\documentclass[11pt]{article}
\usepackage{amsmath,amssymb}
\usepackage[dvips]{epsfig}

\makeatletter\@addtoreset {equation}{section}\makeatother

\newtheorem{theo}{Theorem}[section]
\newtheorem{lem}[theo]{Lemma}

\newtheorem{cor}[theo]{Corollary}
\newtheorem{rem}[theo]{Remark}
\newenvironment{Proof}
{\begin{trivlist} \item[]{\bf Proof. }}%
{\hspace*{\fill}$\rule{.3\baselineskip}{.35\baselineskip}$\end{trivlist}}

\newcommand{\R}{\mathbb{R}}
\newcommand{\C}{\mathbb{C}}

\renewcommand{\geq}{\geqslant}
\renewcommand{\leq}{\leqslant}

\renewcommand{\phi}{\varphi}

\newcommand{\be}{\begin{eqnarray}}
\newcommand{\ee}{\end{eqnarray}}

\newcommand{\eps}{\varepsilon}
\renewcommand{\le}{L_-^\eps}

\begin{document}

\title{\bf Eigenvalues of a nonlinear ground state \\
in the Thomas--Fermi approximation}

\author{Cl\'ement Gallo and Dmitry Pelinovsky \\
{\small Department of Mathematics, McMaster
University, Hamilton, Ontario, Canada, L8S 4K1}  }

\date{\today}
\maketitle

\begin{abstract}
We study a nonlinear ground state of the Gross--Pitaevskii equation with a
parabolic potential in the hydrodynamics limit often referred to
as the Thomas--Fermi approximation. Existence of the energy
minimizer has been known in literature for some time but it was
only recently when the Thomas--Fermi approximation was
rigorously justified. The spectrum of linearization of the
Gross--Pitaevskii equation at the ground state consists of
an unbounded sequence of positive eigenvalues. We analyze
convergence of eigenvalues in the hydrodynamics limit.
Convergence in norm of the resolvent operator is proved and
the convergence rate is estimated. We also study asymptotic and numerical
approximations of eigenfunctions and eigenvalues using Airy functions.
\end{abstract}

\section{Introduction}

Recent experiments in Bose--Einstein condensation has stimulated
an intense research around the Gross--Pitaevskii equation with a
parabolic potential \cite{bose-book}. Considered in a
one-dimensional cigar-shaped geometry and in the limit of a
compact Thomas-Fermi cloud, the repulsive Bose gas is described by
the Gross--Pitaevskii equation in the form
\begin{equation}
\label{GP}
i u_t + \eps^2 u_{xx} + (1 - x^2) u - |u|^2 u = 0,
\end{equation}
where $u = u(x,t)$ is a complex-valued amplitude, the subscripts
denote partial differentiations, $\eps$ is a small parameter,
and all other parameters are normalized to unity.

Existence of the ground state $u = \eta_{\eps}(x)$ for a fixed,
sufficiently small $\eps > 0$, where $\eta_{\eps}$ is a
real-valued, positive-definite, global minimizer of the
Gross--Pitaevskii energy
$$
E_{\eps}(u) =  \int_{\mathbb{R}} \left( \frac{1}{2} \eps^2 |u_x|^2 + \frac{1}{2} (x^2-1) |u|^2 +
\frac{1}{4} |u|^4  \right) dx
$$
in the energy space
$$
\mathcal{H}_1 = \left\{ u \in H^1(\mathbb{R}) : \;\; x u \in L^2(\mathbb{R}) \right\},
$$
has been proved in the literature long ago (see, i.e., Brezis \&
Oswald \cite{BO}). Recent works of Ignat \& Millot \cite{IM} and
Aftalion, Alama, \& Bronsard \cite{AAB} have focused, among other
problems related to existence of vortices in a two-dimensional
rotating Bose--Einstein condensate, on the rigorous justification
of the Thomas-Fermi asymptotic formula
\begin{equation}
\label{Thomas-Fermi}
\eta_0(x) = \left\{ \begin{array}{cl} (1 - x^2)^{1/2}, \;\; & \mbox{for} \;\; |x| < 1, \\
0, \;\; & \mbox{for} \;\; |x| > 1, \end{array} \right.
\end{equation}
which was believed to be a weak limit of $\eta_{\eps}(x)$ as $\eps
\to 0$ since the work of Thomas \cite{Thomas} and Fermi
\cite{Fermi}. To be precise, Proposition 2.1 of \cite{IM} and
Proposition 1 in \cite{AAB} state that $\eta_{\eps}(x)$ converges
to $\eta_0(x)$ as $\eps \to 0$ in the sense that
\begin{equation}
\label{TF} \left\{ \begin{array}{cl} (1 - C \eps^{1/3})  \leq
\frac{\eta_{\eps}(x)}{(1 - x^2)^{1/2}} \leq 1 \;\; & \mbox{for}
\;\; |x| \leq 1 - \eps^{2/3}  \\
0 \leq \eta_{\eps}(x) \leq C \eps^{1/3} \exp\left(\frac{1 - x^2}{4
\eps^{2/3}}\right) \;\; & \mbox{for} \;\; |x| \geq 1 - \eps^{2/3},
\end{array} \right.
\end{equation}
for an $\eps$-independent constant $C > 0$. (The results of
\cite{IM,AAB} are formulated in the space of two dimensions, but
the extension to the one-dimensional case is trivial.) It was
proved in \cite{IM} that $\| \eta_{\eps} - \eta_0 \|_{C^1(K)} \leq
C_K \eps^2$ for any compact subset $K \subset (-1,1)$, which
justified the WKB approximation of the ground state considered
earlier by formal expansions (see, i.e., \cite{BK}).

We are concerned here with the spectrum of linearization of the
Gross--Pitaevskii equation (\ref{GP}) at the ground state
$\eta_{\eps}$, which is defined by the eigenvalue problem
\begin{equation}
\label{gen-eig-prob-0} -\eps^2 u'' + (x^2 - 1 + 3 \eta_{\eps}^2 )
u = -\lambda w, \qquad -\eps^2 w'' + (x^2 - 1 + \eta_{\eps}^2 ) w
= \lambda u,
\end{equation}
where $(u + i w) e^{\lambda t} + (\bar{u} - i \bar{w})
e^{\bar{\lambda} t}$ is a perturbation to $\eta_{\eps}$. The
eigenvalue problem (\ref{gen-eig-prob-0}) determines the spectral
stability of the ground state $\eta_{\eps}$ with respect to the
time evolution of the Gross--Pitaevskii equation (\ref{GP}) and
gives preliminary information for nonlinear analysis of orbital
stability and long-time dynamics of ground states. More complex
phenomena of pinned vortices (dark solitons) on the top of the
ground state can also be understood from the analysis of
eigenvalues of the spectral problem (\ref{gen-eig-prob-0}) (see,
i.e., \cite{PK}).

In what follows, we shall simplify the spectral problem
(\ref{gen-eig-prob-0}) and replace $\eta_{\eps}$ by $\eta_0$. We
do not claim that eigenvalues of these two problems are close to
each other but, given a complexity of the problem, we would like
to deal with a simpler problem in this article. Therefore, we
analyze here solutions of the model eigenvalue problem defined
explicitly by
\begin{eqnarray}
\label{gen-eig-prob} \left\{ \begin{array}{cl} -\eps^2 u'' + 2 (1
- x^2) u = -\lambda w, \;\;
-\eps^2 w'' = \lambda u \;\; & \mbox{for} \;\; |x| < 1, \\
-\eps^2 u'' + (x^2 - 1) u = -\lambda w, \;\; -\eps^2 w'' + (x^2 -
1) w = \lambda u \;\; & \mbox{for} \;\; |x| > 1, \end{array}
\right.
\end{eqnarray}
with appropriate matching conditions at $x=\pm1$. It will be left
for the forthcoming work to study solutions of the original
eigenvalue problem (\ref{gen-eig-prob-0}) with $\eta_{\eps} =
\eta_0 + {\cal O}_{L^{\infty}(\R)}(\eps^{1/3})$, according to the
bound (\ref{TF}) above.

Formal weak solutions of (\ref{gen-eig-prob}) have been
constructed in the pioneer work of Stringari \cite{stringari} and
have been used in a more complex context of three-dimensional
anisotropic repulsive Bose gas in \cite{FCSG,EGD}. To recover
these solutions, let us denote $\lambda = i \eps \gamma^{1/2}$ and
drop $-\eps^2 u''$ term in the first equation of
(\ref{gen-eig-prob}). Then, the model eigenvalue problem is closed
at the singular Sturm--Liouville problem
\begin{equation}
\label{Gegenbauer} -2(1-x^2) w'' = \gamma w, \quad -1 < x < 1,
\end{equation}
which has a ${\cal C}^2$ solution on $[-1,1]$ for $\gamma \neq 0$
if and only if $w(1) = w(-1) = 0$. We will show in Lemma
\ref{lemma-hypergeometric} below that the only solutions of
(\ref{Gegenbauer}) with $w(1) = w(-1) = 0$ are the Gegenbauer
polynomials $w(x) = C_{n+1}^{-1/2}(x)$, which correspond to
eigenvalues at $\gamma = \gamma_n = 2 n (n+1)$, where $n \geq 1$
is an integer. Solutions $w(x) = C_{n+1}^{-1/2}(x)$ of
(\ref{Gegenbauer}) on the interior domain $[-1,1]$ are completed
with the zero function $w = 0$ on the exterior domain $|x| \geq
1$. In this way, we glue together weak solutions of system
(\ref{gen-eig-prob}) in the hydrodynamics limit $\eps = 0$. It is
the main goal of this article to develop a rigorous justification
of persistence of eigenvalues $\{ \gamma_n \}_{n \in \mathbb{N}}$
for small non-zero values of $\eps$. Our main result is the
following theorem.

\medskip

{\bf Main Theorem.} {\em Spectral problem (\ref{gen-eig-prob}) for
$\eps > 0$ has a purely discrete spectrum that
consists of eigenvalues at $\lambda = \pm i \eps
(\gamma_{n,\eps})^{1/2}$, where the set $\{ \gamma_{n,\eps} \}_{n
\in \mathbb{N}}$ is sorted in the increasing order
$$
0 < \gamma_{1,\eps} \leq \gamma_{2,\eps} \leq \gamma_{3,\eps} \leq
\gamma_{4,\eps} \leq ...,
$$
while
$$
\gamma_{n,\eps} \longrightarrow \gamma_n \quad \text{as } \eps \to 0
$$
for every fixed $n \in \mathbb{N}$. Moreover, for any fixed
$\delta > 0$, there exists $C_n > 0$ such that
$$
|\gamma_{n,\eps} - \gamma_n | \leq C_n \eps^{1/3-\delta}
$$
for sufficiently small $\eps > 0$.}

\medskip

{\bf Remark.} {\em The convergence rate of eigenvalues is not
sharp and our numerical results indicate that the convergence rate
is ${\cal O}(\eps^{2})$ for a fixed $n \in \mathbb{N}$.}

\medskip

Before going into technical details of our analysis, we mention
three relevant applications where eigenvalues of the singular
Sturm--Liouville problem (\ref{Gegenbauer}) have appeared
recently.

\begin{itemize}
\item Propagation of self-similar pulses in an amplifying optical
medium is described by the Gross--Pitaevskii equation with a
parabolic potential \cite{BTNN}
$$
i U_{\tau} + \tau^{-2} U_{\xi \xi} + (1 - \xi^2) U - |U|^2 U = 0.
$$
The small parameter $\eps = \tau^{-1}$ changes with the time
$\tau$ due to evolution of the self-similar optical pulse in the
presence of the gain. The decomposition of perturbation to the
optical pulse via Gegenbauer polynomials is used for understanding
the effects of higher-order dispersion and gain terms on the
long-term optical pulse dynamics \cite{BT}.

\item Analysis of radiation from a dark soliton oscillating in a wide parabolic
potential was studied in \cite{PFK} using asymptotic multi-scale expansion methods.
The analysis leaded to the wave equation with a space-dependent speed
$$
U_{\tau \tau} = \left( (1 - \xi^2) U_{\xi} \right)_{\xi}.
$$
Eigenvalues of the wave equation are given by eigenvalues of the
Sturm--Liouville problem (\ref{Gegenbauer}). The corresponding
eigenfunctions are needed to match the dark soliton with its
far-field radiation tail and to predict radiative corrections to
the soliton dynamics \cite{PFK}.

\item Numerical approximations of eigenvalues of the spectral problem
associated with a dark soliton in the Gross--Pitaevskii equation
$$
i U_{\tau} + U_{\xi \xi} + (\mu -  \xi^2) U - |U|^2 U = 0
$$
showed convergence of eigenvalues in the limit $\mu \to \infty$
\cite{PK}. It was observed that the whole spectrum consisted of
eigenvalues associated with the ground state and an additional
pair of pure imaginary eigenvalues. The countable infinite set of
eigenvalues associated with the ground state corresponds to the
set of eigenvalues of the Sturm--Liouville problem
(\ref{Gegenbauer}) after an appropriate rescaling transformation
of $\xi$, $\tau$, and $U$.
\end{itemize}

This article is organized as follows. Section 2 discusses
properties of the two Schr\"{o}dinger operators that define the
spectral problem (\ref{gen-eig-prob}) as well as the properties of
their product. Section 3 gives a proof of the Main Theorem.
Section 4 is devoted to asymptotic and numerical approximations of
eigenvalues of the spectral problem (\ref{gen-eig-prob}). In the
Appendix, we give the proofs of several technical lemmas used in
the article, as well as the description of the numerical method.

\paragraph{Notations.} In what follows, if $A$ and $B$ are two
quantities depending on a parameter $p$ in a set $\mathcal{P}$, the
notation $A(p)\lesssim B(p)$ indicates that there exists a positive
constant $C$ such that
$$
A(p)\leq CB(p) \quad \mbox{for every} \; p\in \mathcal{P}.
$$
The notation $A(p)\approx B(p)$ means that $A(p)\lesssim B(p)$ and
$A(p) \gtrsim B(p)$. We say that a property is satisfied for $0 <
\eps \ll 1$ if there exists $\eps_0 \in (0,1)$ such that the
property is true for every $\eps\in (0,\eps_0)$. If $E$ and $F$
are two Banach spaces, $\mathcal{L}(E,F)$ denotes the space of
bounded linear operators from $E$ into $F$, endowed with its
natural norm
$$
\|u\|_{\mathcal{L}(E,F)}=\underset{x\in E, \; x\neq
  0}{\sup}\frac{\|u(x)\|_F}{\|x\|_E}.
$$
If $E=F$, we simply denote $\mathcal{L}(E)=\mathcal{L}(E,E)$. The dual
space of $E$ is denoted by $E'=\mathcal{L}(E,\R)$. If $S$ is
a subset of $\R$, $\mathbf{1}_S$ denotes the characteristic
function of $S$:
$$
\mathbf{1}_S(x) = \left\{\begin{array}{ccc} 1 & \text{if}
    & x\in S,\\ 0 & \text{if} & x\notin S.\end{array}\right.
$$
If $f$ is a function defined on some set $D$ and $S\subset D$,
$f_{|S}$ denotes the restriction of $f$ to the set $S$. Finally,
$B_{L^2}$ denotes the unit ball of $L^2(\R)$.

\section{Preliminaries}
\subsection{The operator $L_-^\eps$ and its inverse}

Let $L_-^\eps$ be the Friedrichs extension of
$-\partial_x^2+p_\eps(x)$ on $L^2(\R)$ for $\eps>0$ and
$$
p_\eps(x) = \frac{1}{\eps^2} (x^2-1) \mathbf{1}_{\{|x|>1\}}.
$$
Since $p_\eps(x) \geq 0$ for any $x \in \mathbb{R}$, $L_-^\eps$ is
a positive self-adjoint operator. Since $p_\eps(x)\to +\infty$ as
$x\to \infty$, $L_-^\eps$ has compact resolvent. The domain of $L_-^\eps$,
$$
D(L_-^\eps)=\{\phi \in L^2(\R) : -\partial_x^2 \phi + p_\eps \phi \in
L^2(\R) \}=\{\phi \in H^2(\R) : x^2 \phi \in
L^2(\R) \}=:\mathcal{H}_2,
$$
is contained in its form domain
$$
Q(L_-^\eps) = \{ \phi \in H^1(\R) : x \phi \in L^2(\R) \}.
$$
If $\phi\in D(L_-^\eps)$ is in the kernel of $L_-^\eps$, then
$\int_\R\left(|\partial_x\phi|^2+p_\eps|\phi|^2\right)dx=0$, which
implies $\phi=0$. Therefore $0\not\in \sigma(L_-^\eps)$ and
$L_-^\eps$ is invertible. In the following lemma, we state that
the inverse of $L_-^\eps$ is uniformly bounded in
$\mathcal{L}(L^2)$ as $\eps \to 0$.

\begin{lem}
\label{lemma-resolvent}
For $0<\eps\ll 1$,
$$
\|(L_-^\eps)^{-1}\|_{\mathcal{L}(L^2)}\approx 1.
$$
\end{lem}

\begin{Proof}
See Appendix \ref{AL-}.
\end{Proof}

Using Lemma \ref{lemma-resolvent}, we give estimates on various
norms of $(L_-^\eps)^{-1}$ for sufficiently small $\eps > 0$.

\begin{lem}
\label{lemma-space} For $0<\eps\ll 1$,
\be\label{bounds-quotient-0} \|
\partial_x(L_-^\eps)^{-1}\|_{\mathcal{L}(L^2(\mathbb{R}))}
& \lesssim & 1, \\
\label{bounds-quotient-01}
\| \mathbf{1}_{\{|x| > 1\}}
  \partial_x(L_-^\eps)^{-1}\|_{\mathcal{L}(L^2(\mathbb{R}))}
   & \lesssim & \eps^{1/3}, \\
\label{bounds-quotient}
 \| \mathbf{1}_{\{|x| > 1\}}(L_-^\eps)^{-1}\|_{\mathcal{L}(L^2(\mathbb{R}))} & \lesssim
        & \eps \\
\label{bounds-quotient-4} \|
\partial_x(L_-^\eps)^{-1}\|_{\mathcal{L}(L^2(\mathbb{R}),L^\infty(\R))}
& \lesssim & 1, \\
\label{bounds-quotient-41} \|\mathbf{1}_{\{|x| >
  1\}}(L_-^\eps)^{-1}\|_{\mathcal{L}(L^2(\mathbb{R}),L^\infty(\R))}
& \lesssim & \eps^{2/3}. \ee
\end{lem}

\begin{Proof}
Let us take $\eps>0$ sufficiently small, $f\in B_{L^2}$, and
denote $\phi=(L_-^\eps)^{-1}f$. By Lemma \ref{lemma-resolvent},
\be\label{reslem} \|\phi\|_{L^2(\R)} \lesssim 1. \ee Moreover,
$\phi$ satisfies the second--order differential equation \be
\label{eqdiff} -\phi'' + p_\eps \phi = f, \qquad x \in \mathbb{R}.
\ee Multiplying (\ref{eqdiff}) by $\phi$, integrating over $\R$,
using the Cauchy-Schwarz inequality and (\ref{reslem}), we get
\be\label{id} \int_\R|\phi'|^2dx+\int_{|x|> 1}p_\eps|\phi|^2dx =
\int_\R f\phi dx\leq \|f\|_{L^2(\R)}\|\phi\|_{L^2(\R)}\lesssim 1,
\ee which directly proves (\ref{bounds-quotient-0}). Proceeding
like for (\ref{id}), but integrating on $[1,+\infty)$ instead of
$\R$, we obtain \be \label{starting-bound} \int_1^{+\infty}
|\phi'|^2 dx + \int_1^{+\infty} p_\eps |\phi|^2dx \leq |\phi(1)|
|\phi'(1)| + \| \phi \|_{L^2( 1,+\infty)}. \ee Then, we observe
\be\label{phixg}
\|\phi\|_{L^2(1+\eps^{2/3},+\infty)}^2 & = & \eps^2\int_{1+\eps^{2/3}}^{+\infty}
\frac{1}{x^2-1}p_\eps|\phi|^2dx\nonumber\\
& \leq & \frac{\eps^2}{(1+\eps^{2/3})^2-1}
\int_{1+\eps^{2/3}}^{+\infty}p_\eps|\phi|^2dx\nonumber\\
& \lesssim &  \eps^{4/3}\int_1^{+\infty} p_\eps |\phi|^2dx. \ee
Since $\phi''=-f$ on $(-1,1)$ and thanks to bound
(\ref{bounds-quotient-0}), Sobolev's embedding of $H^1(-1,1)$ into
$L^\infty(-1,1)$ yields \be \label{derivative-bound}
\|\phi'\|_{L^\infty(-1, 1)} \lesssim \| \phi' \|_{H^1(-1,1)}
\lesssim \| \phi' \|_{L^2(-1,1)}+\| f \|_{L^2(-1,1)}\lesssim 1.
\ee The triangle inequality yields \be\label{triangle}
\|\phi\|_{L^2(1,+\infty)} & \leq &
\|\phi\|_{L^2(1+\eps^{2/3},+\infty)}+\eps^{1/3}\|\phi\|_{L^\infty(1,1+\eps^{2/3})}.
\ee By the Taylor formula and the Cauchy-Schwarz inequality,
\be\label{phi1-bound} \|\phi\|_{L^\infty(1, 1+\eps^{2/3})}\leq
|\phi(1 + \eps^{2/3})| + \eps^{1/3} \| \phi' \|_{L^2(1,+\infty)}.
\ee Let us introduce the new variable $\xi = (x-1)/\eps^{2/3}$ and
the function $\tilde{\phi}(\xi) = \phi( 1 + \eps^{2/3} \xi)$.
Then, \be \label{H^1-bound} \| \tilde{\phi} \|^2_{H^1(1,+\infty)}
& = & \eps^{2/3} \| \phi' \|^2_{L^2( 1 + \eps^{2/3},+\infty)} +
\eps^{-2/3} \| \phi \|^2_{L^2( 1 + \eps^{2/3},+\infty)} \ee Thus,
by Sobolev's embedding of $H^1(1,+\infty)$ into
$L^{\infty}(1,+\infty)$, (\ref{H^1-bound}) provides the bound
\be\label{28bis} |\phi(1 + \eps^{2/3})| =|\tilde{\phi}(1)|\lesssim
\eps^{1/3} \| \phi' \|_{L^2(1 + \eps^{2/3},+\infty)} + \eps^{-1/3}
\| \phi \|_{L^2( 1 + \eps^{2/3},+\infty)}. \ee Concatenating
(\ref{phixg}), (\ref{starting-bound}), (\ref{derivative-bound}),
(\ref{triangle}), (\ref{phi1-bound}) and (\ref{28bis}), we obtain
\be \label{final-bound} \| \phi' \|^2_{L^2(1,+\infty)} +
\frac{1}{\eps^{4/3}} \| \phi \|^2_{L^2(1 + \eps^{2/3},+\infty)} &
\lesssim &  \eps^{1/3} \| \phi' \|_{L^2(1,+\infty)} + \eps^{-1/3}
\| \phi \|_{L^2( 1 + \eps^{2/3},+\infty)}. \ee There exists $C >
0$ such that (\ref{final-bound}) can be rewritten in the form \be
\nonumber \left( \| \phi' \|_{L^2(1,+\infty)} - C \eps^{1/3}
\right)^2 + \frac{1}{\eps^{4/3}} \left( \| \phi \|_{L^2( 1 +
\eps^{2/3},+\infty)} - C \eps \right)^2 \lesssim  \eps^{2/3} . \ee
Therefore, $\| \phi' \|_{L^2(1,+\infty)} \lesssim \eps^{1/3} $ and
$\| \phi \|_{L^2(1 + \eps^{2/3},+\infty)} \lesssim \eps $. Using also
(\ref{phi1-bound}) and (\ref{28bis}), we deduce
$$\| \phi \|_{L^2(1,1 + \eps^{2/3})}\lesssim\eps^{1/3}\| \phi
\|_{L^\infty(1,1 + \eps^{2/3})}\lesssim \eps,$$ and thus $\| \phi
\|_{L^2(1,+\infty)}\lesssim \eps$. Similar computations on
$(-\infty,-1]$ complete the proof of (\ref{bounds-quotient-01})
and (\ref{bounds-quotient}). Sobolev's embedding of $H^1(\R_+)$
into $L^{\infty}(\R_+)$ for $\tilde{\phi}(\xi) = \phi( 1 +
\eps^{2/3} \xi)$ yields \be \label{phiLinf}
\|\phi\|_{L^\infty(1,+\infty)} & = &
\|\tilde{\phi}\|_{L^\infty(\R_+)} \lesssim
\|\tilde{\phi}\|_{H^1(\R_+)} \lesssim
\|\tilde{\phi}' \|_{L^2(\R_+)}+\|\tilde{\phi}\|_{L^2(\R_+)}\nonumber\\
&\lesssim&
\eps^{1/3}\|\phi'\|_{L^2(1,+\infty)}+\eps^{-1/3}\|\phi\|_{L^2(1,+\infty)}
\lesssim \eps^{2/3}. \ee Combined with a similar estimate for
$\|\phi\|_{L^\infty(-\infty,-1)}$, we get
(\ref{bounds-quotient-41}). Finally, Sobolev's embedding of
$H^1(\R_+)$ into $L^{\infty}(\R_+)$ for $\tilde{\phi}'(\xi) =
\eps^{2/3} \phi'( 1 + \eps^{2/3} \xi)$ similarly yields \be \nonumber
\|\phi'\|_{L^\infty(1,+\infty)}
&\lesssim& \eps^{1/3}\|\phi''\|_{L^2(1,+\infty)}
+\eps^{-1/3}\|\phi'\|_{L^2(1,+\infty)}. \ee Therefore, the bound
(\ref{bounds-quotient-4}) holds if $\| \phi'' \|_{L^2(1,\infty)}
\lesssim \eps^{-1/3}$ since $\| \phi' \|_{L^{\infty}(-\infty,-1)}$
is estimated similarly and $\| \phi' \|_{L^{\infty}(-1,1)}$ is
given by the bound (\ref{derivative-bound}). Since $\phi \in D(L_-^{\eps})=\mathcal{H}_2$, $\lim_{x \to
\infty} p_{\eps} \phi \phi' = 0$, and the bound $\| \phi'' \|_{L^2(1,\infty)}
\lesssim \eps^{-1/3}$ follows from integration by parts: \be
\nonumber 1\geq \| f \|^2_{L^2(1,+\infty)} = \| L_-^\eps \phi \|_{L^2( 1,+\infty)}^2 & = &
\int_1^{+\infty} (\phi'')^2 dx - 2 \int_1^{+\infty} p_{\eps} \phi
\phi'' dx +
\int_1^{+\infty} p_{\eps}^2 \phi^2 dx \\
\nonumber    &
 = & \int_1^{+\infty} (\phi'')^2 dx + 2
\int_1^{+\infty} p_{\eps} (\phi')^2 dx +
\int_1^{+\infty} p_{\eps}^2 \phi^2 dx \\
\label{squared-bound} & \phantom{t} & \phantom{texttexttext} -
\frac{2}{\eps^2} \int_1^{+\infty} \phi^2 dx -\frac{2}{\eps^2}
\phi^2(1), \ee where the
second and third terms in the right-hand-side are positive and the
last two terms are estimated from
(\ref{bounds-quotient}) and (\ref{bounds-quotient-41}).
\end{Proof}

\subsection{The operator $L_+^\eps$ and its inverse}

Let $L_+^\eps$ be defined similarly to $L_-^\eps$ as the
Friedrichs extension of $- \partial_x^2 + q_\eps(x)$ on $L^2(\R)$
for $\eps > 0$, where
$$
q_\eps(x) = \frac{1}{\eps^2} \left[  2(1-x^2)
\mathbf{1}_{\{|x|<1\}}+(x^2-1)\mathbf{1}_{\{|x|>1\}} \right].
$$
The domain of $L_+^\eps$ is $\mathcal{H}_2$ and $L_+^\eps$ is a
positive self-adjoint invertible operator with a
compact resolvent. Similarly as for $(L_-^\eps)^{-1}$, we estimate the
size of $(L_+^\eps)^{-1}$ in $\mathcal{L}(L^2(\R))$.

\begin{lem}\label{22}
For $0<\eps\ll 1$,
$$
\|(L_+^\eps)^{-1}\|_{\mathcal{L}(L^2(\R))}\approx
\eps^{4/3}.
$$
\end{lem}

\begin{Proof}
See Appendix \ref{AL+}.
\end{Proof}

Using Lemma \ref{22}, we give estimates on various norms of
$(L_+^\eps)^{-1}$ for sufficiently small $\eps > 0$.

\begin{lem}
\label{lemma-remainder-plus} For $0<\eps\ll 1$, \be
\label{bounds-quotient-plus} \| \partial_x^2(L_+^\eps)^{-1}
\|_{\mathcal{L}(L^2(\R))} & \lesssim & 1, \\
\label{bounds-quotient-plus-1} \|
\partial_x(L_+^\eps)^{-1} \|_{\mathcal{L}(L^2(\R))} & \lesssim &
\eps^{2/3} \\  \label{bounds-quotient-4-plus} \|
\partial_x(L_+^\eps)^{-1} \|_{\mathcal{L}(L^2(\R),L^\infty(\R))}
& \lesssim & \eps^{1/3}, \\      \label{bounds-quotient-4-plus-1}
 \| (L_+^\eps)^{-1} \|_{\mathcal{L}(L^2(\R),L^\infty(\R))} & \lesssim & \eps.
\ee
\end{lem}

\begin{Proof}
Let $f\in B_{L^2}$ and $\psi=(L_+^\eps)^{-1}f$. The bound
(\ref{bounds-quotient-plus-1}) is obtained by taking an inner
product of $L_+^{\eps} \psi = f$ with $\psi$ and using Lemma
\ref{22}:
$$\|\psi'\|_{L^2(\R)}^2+\int_\R q_\eps|\psi|^2dx\leq
\|f\|_{L^2(\R)}\|\psi\|_{L^2(\R)}\lesssim\eps^{4/3}.$$ The bound
(\ref{bounds-quotient-4-plus-1}) is a consequence of the bound
(\ref{bounds-quotient-plus-1}) and Lemma \ref{22}, applying
Sobolev's embedding of $H^1(\R)$ into $L^\infty(\R)$ to the
function $\tilde{\psi}(\xi)=\psi(\eps^{2/3} \xi)$. To get the
bound (\ref{bounds-quotient-plus}), we compute 
\be \nonumber 1\geq \| f
\|^2_{L^2(\R)} = \|
L_+^{\eps} \psi \|^2_{L^2(\R)} & = & \int_{\R} (\psi'')^2 dx - 2
\int_{\R} q_{\eps} \psi \psi'' dx +
\int_{\R} q_{\eps}^2 \psi^2 dx \\
\nonumber    & = & \int_{\R} (\psi'')^2 dx + 2 \int_{\mathbb{R}}
q_{\eps} (\psi')^2 dx + \int_{\R} q_{\eps}^2 \psi^2 dx \\
\nonumber & \phantom{t} & \phantom{t} + \frac{4}{\eps^2} \int_{|x|
< 1} \psi^2 dx - \frac{2}{\eps^2} \int_{|x| > 1} \psi^2 dx -
\frac{6}{\eps^2} \left( \psi^2(1) + \psi^2(-1) \right), \ee 
where we
have used that $\lim_{|x| \to \infty} q_{\eps} \psi \psi' = 0$, which
is true because $\psi \in
D(L_+^{\eps})=\mathcal{H}_2$. The bound (\ref{bounds-quotient-plus})
holds with the use of the bound (\ref{bounds-quotient-4-plus-1}) and Lemma \ref{22}. The bound
(\ref{bounds-quotient-4-plus}) follows from Sobolev's embedding of
$H^1(\R)$ into $L^\infty(\R)$ applied to $\tilde{\psi}'(\xi) =
\eps^{2/3} \psi'(\eps^{2/3} \xi)$ and from bounds
(\ref{bounds-quotient-plus}) and (\ref{bounds-quotient-plus-1}).
\end{Proof}

\subsection{The operator $(L_+^\eps)^{-1}(L_-^\eps)^{-1}$}

From the results in the two previous sections, we can deduce
easily some estimates on norms of
$(L_+^\eps)^{-1}(L_-^\eps)^{-1}$. For instance,
$$\|(L_+^\eps)^{-1}(L_-^\eps)^{-1}\|_{\mathcal{L}(L^2(\R))}\leq
\|(L_+^\eps)^{-1}\|_{\mathcal{L}(L^2(\R))}\|(L_-^\eps)^{-1}\|_{\mathcal{L}(L^2(\R))}\lesssim
\eps^{4/3}.$$ However, it turns out that these estimates are not
sufficient for the proof of the Main Theorem. To improve the
estimates, we use the fact that if $v \in B_{L^2}$ maximizes
$\left((L_+^\eps)^{-1}v,v\right)\approx\eps^{4/3}$, then
$(L_+^\eps)^{-1} v$ has its $L^2$-norm concentrated about the
points $\pm 1$ (where $q_\eps$ vanishes), whereas if $u\in
B_{L^2}$ maximizes $\left((L_-^\eps)^{-1}u,u\right) \approx 1$,
then $(L_-^\eps)^{-1}u$ has its $L^2$-norm concentrated in the
interval $(-1,1)$, away from the points $\pm 1$. Figure 1 shows
potentials $p_{\eps}$ and $q_{\eps}$ versus $x$. Figure 2
shows schematic shapes of $(L_-^\eps)^{-1} f$ and $(L_+^\eps)^{-1} f$
for a $f \in L^2(\R)$. The precise estimates on norms of
$(L_+^\eps)^{-1}(L_-^\eps)^{-1}$ are summarized in the following
lemma.

\begin{figure}[!h]
\begin{center}
\includegraphics[width=6cm]{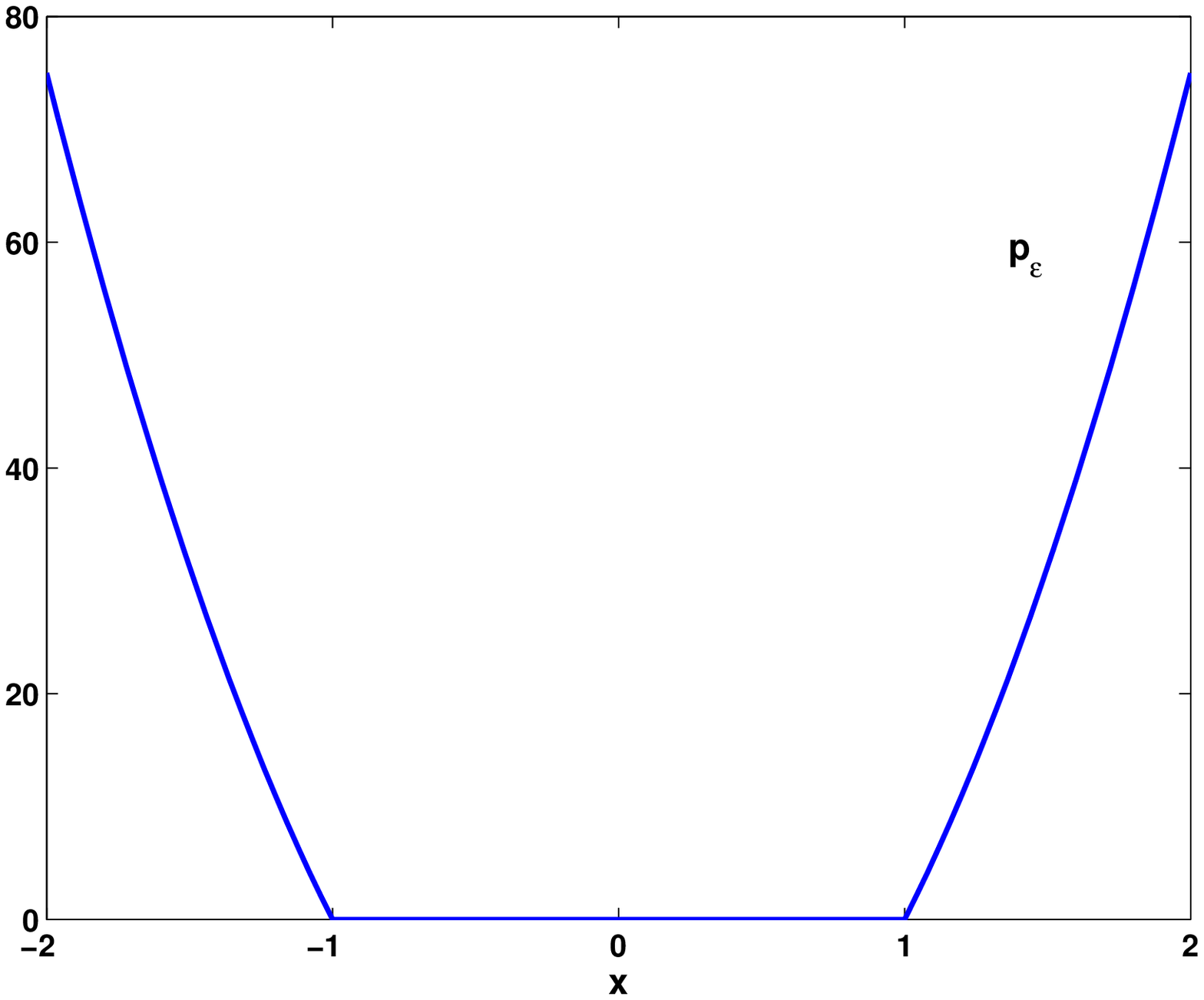}
\includegraphics[width=6cm]{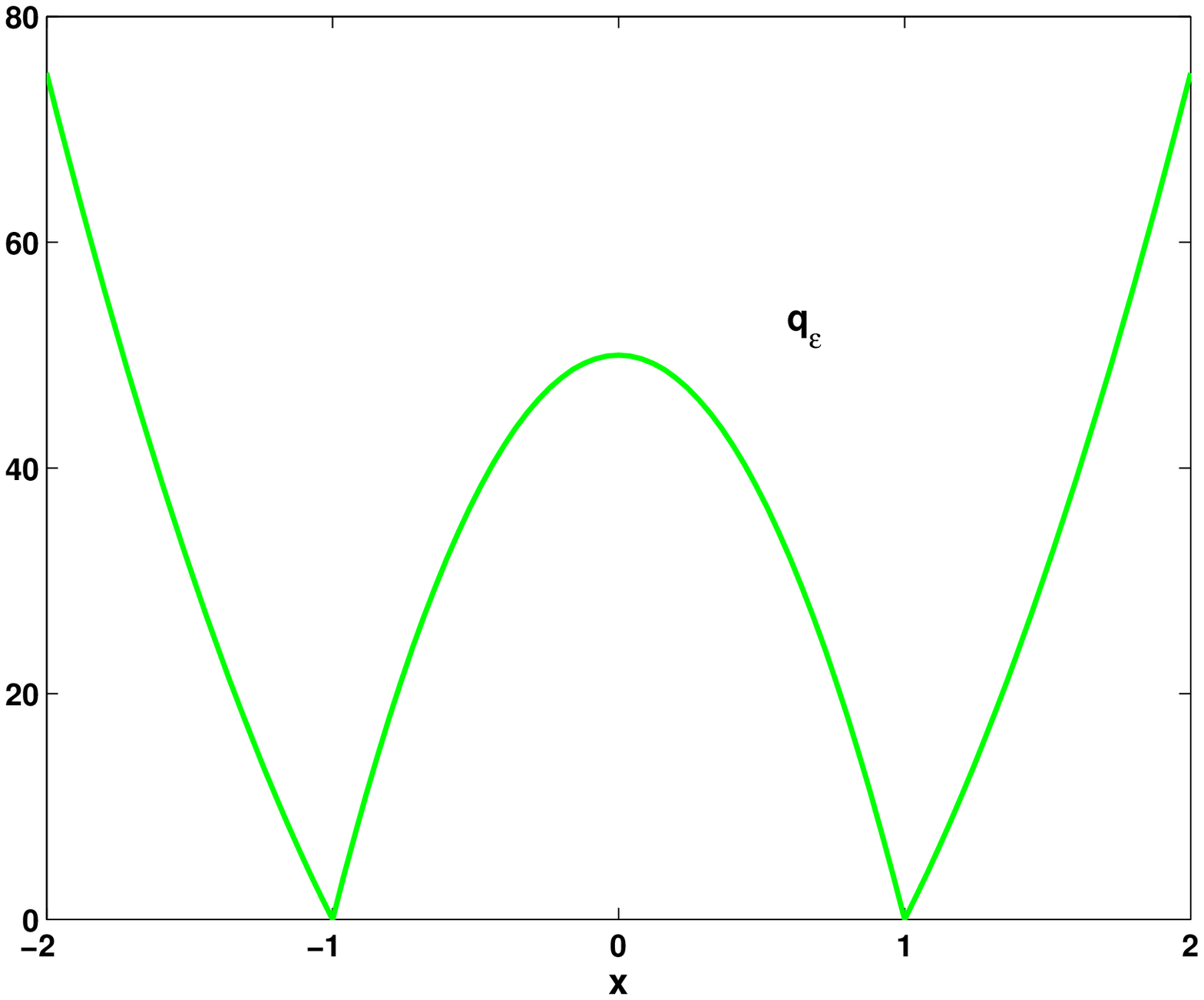}
\caption{Profiles of potentials $p_{\epsilon}$ (left) and
$q_{\epsilon}$ (right) versus $x$.}
\end{center}
\label{figure-0pot}
\end{figure}

\begin{figure}[!h]
\begin{center}
\includegraphics[width=6cm]{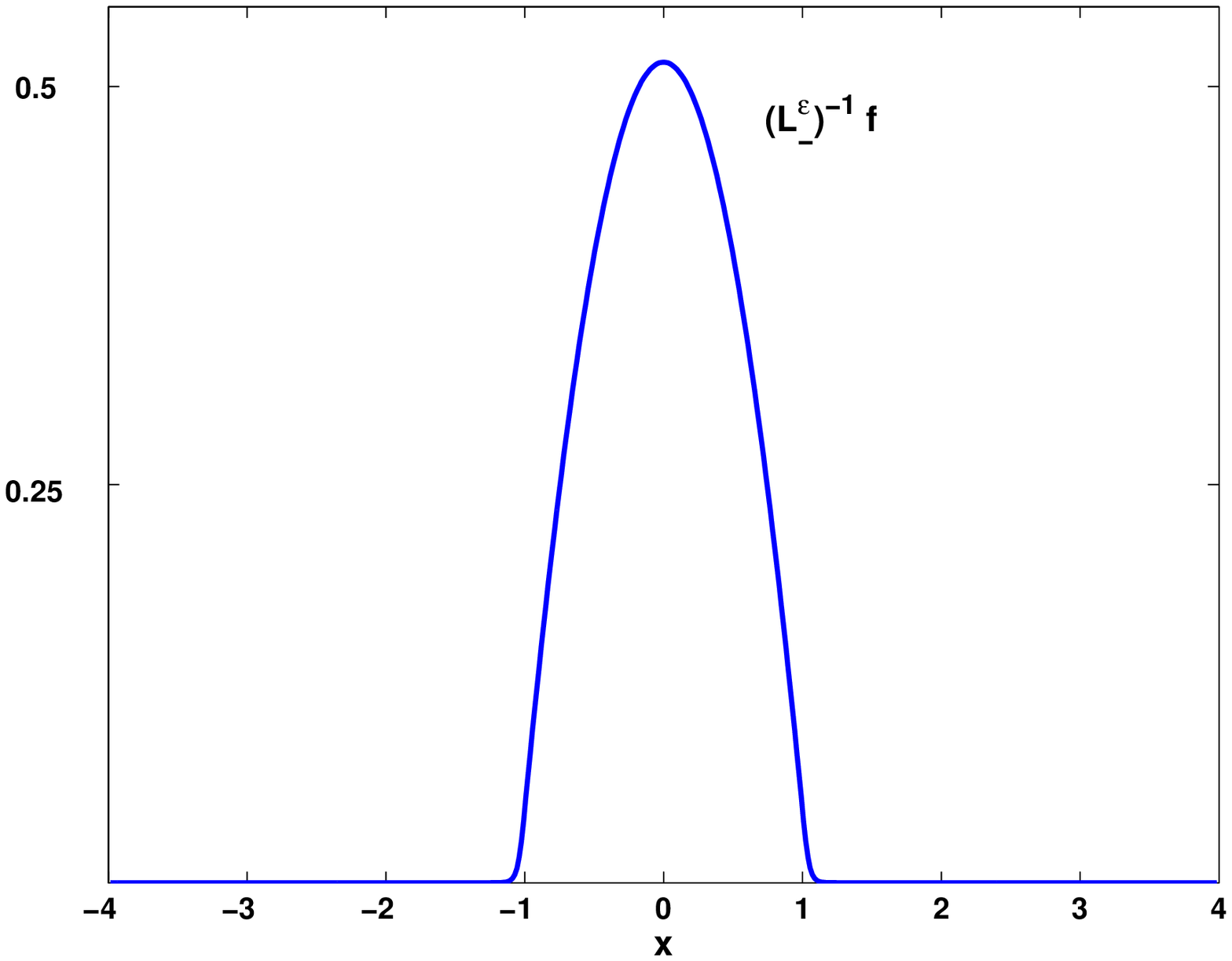}
\includegraphics[width=6.1cm]{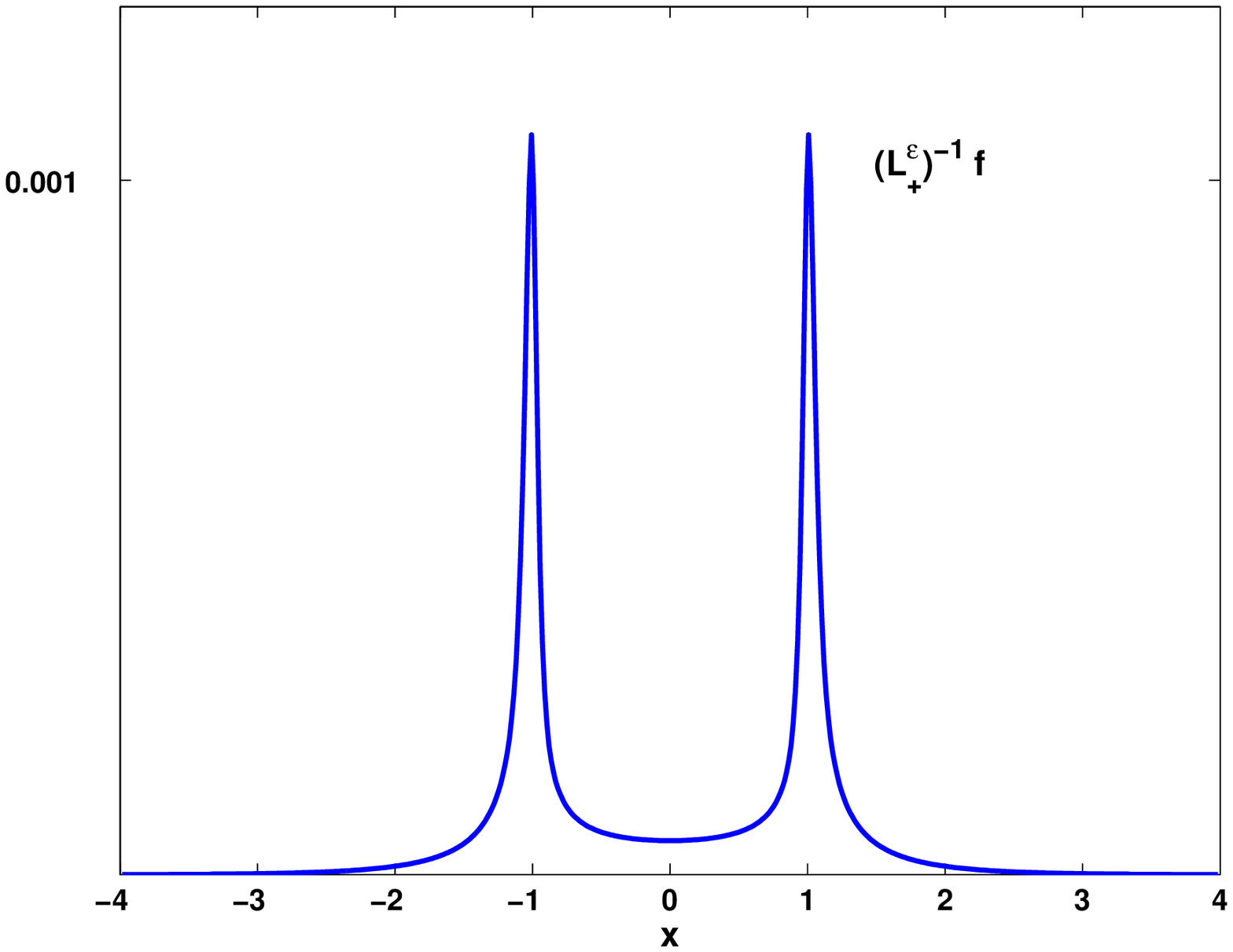}
\caption{Schematic shapes of $(L_-^\eps)^{-1} f$ and $(L_+^\eps)^{-1} f$
for $f(x) = \exp(-x^2/4)\in L^2(\R)$.}
\end{center}
\label{figure-0res}
\end{figure}

\begin{lem}\label{L+L-}
Let $\alpha\in(0,+\infty]$ and $\delta>0$. Then for $0<\eps\ll 1$, \be\label{L+L-der}
\|\partial_x(L_+^\eps)^{-1}(L_-^\eps)^{-1}\|_{\mathcal{L}(L^2(\R))}&\lesssim
&\eps^{11/12},\\
\label{L+L-1}
\|(L_+^\eps)^{-1}(L_-^\eps)^{-1}\|_{\mathcal{L}(L^2(\R))} &
\lesssim & \eps^{26/15-\delta}, \\ \label{L+L-12}
\|\mathbf{1}_{\{|x|>1\}}(L_+^\eps)^{-1}(L_-^\eps)^{-1}\|_{\mathcal{L}(L^2(\R))}
& \lesssim & \eps^{7/3-\delta}, \\ \label{L+L-2}\label{comp2}
\|\mathbf{1}_{\{|x|>1-\eps^\alpha\}}\partial_x\left(L_+^\eps\right)^{-1}\left(L_-^\eps\right)^{-1}\|_{\mathcal{L}(L^2(\R),L^\infty(\R))}&\lesssim &\eps^{\min\left(4/3,1/3+3\alpha/2\right)-\delta},\\\label{comp1}
\|\mathbf{1}_{\{|x|>1-\eps^\alpha\}}\left(L_+^\eps\right)^{-1}\left(L_-^\eps\right)^{-1}\|_{\mathcal{L}(L^2(\R),L^\infty(\R))}
&\lesssim &\eps^{\min\left(2,1+3\alpha/2\right)-\delta},
\ee
where if $\alpha=+\infty$, we use the convention $\eps^\alpha=0$.
\end{lem}

\begin{Proof}
Let $f\in B_{L^2}$, $S =(L_-^\eps)^{-1}f$ and $R =(L_+^\eps)^{-1}
S$. We choose $\gamma\in (0,2/3)$ (in the sequel, we will make
different explicit choices of such $\gamma$), and we split $R$
into three pieces: $R=R_1+R_2+R_3$, where \be
R_1&=&(L_+^\eps)^{-1}\mathbf{1}_{\{|x|>1\}}(L_-^\eps)^{-1}f,\nonumber\\
R_2&=&(L_+^\eps)^{-1}\mathbf{1}_{\{1-\eps^\gamma<|x|<1\}}(L_-^\eps)^{-1}f,\nonumber\\
R_3&=&(L_+^\eps)^{-1}\mathbf{1}_{(-1+\eps^\gamma,1-\eps^\gamma)}(L_-^\eps)^{-1}f.\nonumber
\ee Notice that $R_2$ and $R_3$ depend on $\gamma$. According to
Lemmas \ref{lemma-space}, \ref{22} and \ref{lemma-remainder-plus},
\be\label{R1} &\|R_1'\|_{L^2(\R)}\lesssim \eps^{5/3},\quad
\|R_1\|_{L^2(\R)}\lesssim \eps^{7/3},\quad
\|R_1'\|_{L^\infty(\R)}\lesssim \eps^{4/3},\quad
\|R_1\|_{L^\infty(\R)}\lesssim \eps^2.& \ee Thanks to Lemma
\ref{lemma-space}, the Taylor formula provides
\be\label{S<>}
\| S \|_{L^2(1-\eps^\gamma, 1)}\lesssim
\eps^{\gamma/2}(|S(1)|+\eps^{\gamma}\|S'\|_{L^\infty(-1,1)})\lesssim
\eps^{\gamma/2}(\eps^{2/3}+\eps^{\gamma})\lesssim
\eps^{3\gamma/2},
\ee
because $\gamma<2/3$. Thus, using Lemmas
\ref{22} and \ref{lemma-remainder-plus}, we obtain \be\label{R2}
\|R_2'\|_{L^2(\R)}\lesssim \eps^{2/3+3\gamma/2},
\|R_2\|_{L^2(\R)}\lesssim \eps^{4/3+3\gamma/2},
\|R_2'\|_{L^\infty(\R)}\lesssim \eps^{1/3+3\gamma/2},
\|R_2\|_{L^\infty(\R)}\lesssim \eps^{1+3\gamma/2}.\ \ \ee
The last
component $R_3$ solves the differential equation \be\label{eqR3}
L_+^\eps R_3=\mathbf{1}_{(-1+\eps^\gamma,1-\eps^\gamma)}S, \qquad
x \in \mathbb{R}. \ee We multiply this equality by $R_3$,
integrate over $\R$ and use the Cauchy-Schwarz inequality. Since
$\|S\|_{L^2(\R)}\lesssim 1$, we get \be \label{229}
\|R_3'\|_{L^2(\R)}^2+\int_\R q_\eps |R_3|^2dx\lesssim
\|R_3\|_{L^2(-1+\eps^\gamma,1-\eps^\gamma)}.\ee Thus, since
$\|R_3\|_{L^2(-1+\eps^\gamma,1-\eps^\gamma)}^2\lesssim
\eps^{2-\gamma}\int_\R q_\eps |R_3|^2dx$,
\be
\|R_3'\|_{L^2(\R)}^2+\frac{1}{\eps^{2-\gamma}}\left(\|R_3\|_{L^2(-1+\eps^\gamma,1-\eps^\gamma)}-C\eps^{2-\gamma}\right)^2\lesssim \eps^{2-\gamma}
\ee
for some $C>0$. We deduce
\be\label{82bis} \|R_3'\|_{L^2(\R)}\lesssim \eps^{1-\gamma/2},
\qquad \|R_3\|_{L^2(-1+\eps^\gamma,1-\eps^\gamma)}\lesssim
\eps^{2-\gamma}. \ee Next, we will establish an estimate on
$\|R_3\|_{L^2(1-\eps^\gamma<|x|<1)}$. We first estimate the
$L^\infty(\R)$ norm of $R_3$. Let $\chi$ be a $\mathcal{C}^\infty$
function on $\R$ with values in $[0,1]$ such that
$\chi(x)\equiv 0$ for $x<-1/2$ and $\chi(x)\equiv 1$ for $x>0$.
We denote $\widetilde{\chi R_3}$ the function defined by
$$\widetilde{\chi R_3}(x):=\chi R_3\left(1-\eps^\gamma+(x-1+\eps^\gamma)\eps^{1-\gamma/2}\right).$$
Then, using Sobolev's embedding of $H^1(-\infty,1-\eps^\gamma)$
into $L^\infty(-\infty,1-\eps^\gamma)$ (notice that the norm of
this embedding is the same that the norm of $H^1(\R_+)\subset
L^\infty(\R_+)$, and therefore does not depend on $\eps$), we
obtain \be \|R_3\|_{L^\infty(0,1-\eps^\gamma)} &\leq &\|\chi
R_3\|_{L^\infty(-\infty,1-\eps^\gamma)}
\quad=\quad\|\widetilde{\chi
R_3}\|_{L^\infty(-\infty,1-\eps^\gamma)}
\quad\lesssim \quad \|\widetilde{\chi R_3}\|_{H^1(-\infty,1-\eps^\gamma)}\nonumber\\
&\lesssim &  \eps^{-1/2+\gamma/4}\|\chi
R_3\|_{L^2(-\infty,1-\eps^\gamma)}+\eps^{1/2-\gamma/4}\|(\chi
R_3)'\|_{L^2(-\infty,1-\eps^\gamma)} \nonumber \\ & \lesssim &
\eps^{3/2-3\gamma/4}. \ee Similarly,
$\|R_3\|_{L^\infty(-1+\eps^\gamma,0)}\lesssim
\eps^{3/2-3\gamma/4}$. Since $R_3$ solves
$$
-\partial_x^2 R_3 + q_{\eps} R_3 = 0, \quad |x| > 1 -
\eps^{\gamma},
$$
where $q_\eps\geq 0$ and $R_3\in L^2(\R)$, we infer from the
maximum principle that \be\label{R3inf}
\|R_3\|_{L^\infty(\R)}\lesssim\eps^{3/2-3\gamma/4} . \ee On the
interval $(1-\eps^\gamma,1)$, there exists constants $C_A^\eps$
and $C_B^\eps$ such that $R_3$ is given by the linear combination
$$
R_3=C_A^\eps\psi_A^\eps+C_B^\eps\psi_B^\eps,
$$
where $\psi_A^\eps$ and $\psi_B^\eps$ are defined in Lemma
\ref{airy} below.

\begin{lem}\label{airy}
There exists a constant $C>0$ such that for $\eps>0$ sufficiently
small, the equation
\be\label{eqinn}
 -\psi''(x)+\frac{2(1-x^2)}{\eps^2}\psi(x)=0,\quad -\frac{1}{2} <
 x < 1
\ee has two linearly independent solutions $\psi_A^\eps$ and
$\psi_B^\eps$ in the form
$$\psi_A^\eps(x)=a(1-x){\rm Ai}\left(\frac{\xi(1-x)}{\eps^{2/3}}\right)
\left(1+Q_A^\eps(x)\right),$$
$$\psi_B^\eps(x)=a(1-x){\rm Bi}\left(\frac{\xi(1-x)}{\eps^{2/3}}\right)\left(1+Q_B^\eps(x)\right),$$
where
$\xi(x):=\left(\frac{3}{2}\int_0^x\sqrt{2t(2-t)}dt\right)^{2/3}$,
$a(x):= \left(\xi'(x) \right)^{-1/2}$, ${\rm Ai}$, ${\rm Bi}$ are
the Airy functions, and $Q_A^{\eps}$, $Q_B^{\eps}$ satisfy the
bound
$$\|Q_A^\eps\|_{L^\infty(-1/2,1)} + \|Q_B^\eps\|_{L^\infty(-1/2,1)}\leq
C\eps^{2/3}.$$
\end{lem}

\begin{Proof}
See Appendix \ref{Aairy}.
\end{Proof}

According to 10.4.59 and 10.4.63 in \cite{AS}, the Airy functions
satisfy the following asymptotic behaviour at infinity
\cite[Section 10.4]{AS}: \be\label{daaibi} {\rm Ai}(z)\sim
\frac{1}{2 \pi^{1/2} z^{1/4}} e^{-\frac{2}{3}z^{3/2}} \quad
\text{and} \quad {\rm Bi}(z)\sim \frac{1}{\pi^{1/2} z^{1/4}}
e^{\frac{2}{3}z^{3/2}} \quad \text{as }z\to+\infty. \ee At the
point $x=1$, we deduce from (\ref{R3inf}) that
$$\left| C_A^\eps a(0){\rm Ai}(0)(1+Q_A^\eps(1))+C_B^\eps
  a(0){\rm Bi}(0)(1+Q_B^\eps(1))\right|\lesssim \eps^{3/2-3\gamma/4}.$$
Thus,
\be\label{preCA0}
|C_A^\eps|\lesssim \eps^{3/2-3\gamma/4}+|C_B^\eps|.
\ee
At the point $x=1-\eps^\gamma$, provided that $\gamma<2/3$, we
similarly have
$$
\left| C_A^\eps
  a(\eps^\gamma){\rm Ai}\left(\frac{\xi(\eps^\gamma)}{\eps^{2/3}}\right)(1+Q_A^\eps(1-\eps^\gamma))+C_B^\eps
  a(\eps^\gamma){\rm
    Bi}\left(\frac{\xi(\eps^\gamma)}{\eps^{2/3}}\right)(1+Q_B^\eps(1-\eps^\gamma))\right|\lesssim \eps^{3/2-3\gamma/4}.
$$
Since \be \label{expansion-xi} \xi(x) \sim 2^{2/3} x
\quad \mbox{as} \quad x\to 0 \ee and thanks to (\ref{daaibi}) and
(\ref{preCA0}), we obtain
\be\label{estCACB}
|C_B^\eps|\lesssim \frac{\eps^{3/2-3\gamma/4}}{{\rm
Bi}\left(\frac{\xi(\eps^\gamma)}{\eps^{2/3}}\right)}\quad
  \text{and}\quad |C_A^\eps|\lesssim \eps^{3/2-3\gamma/4},
\ee
where ${\rm
Bi}\left(\frac{\xi(\eps^\gamma)}{\eps^{2/3}}\right) \to \infty$ as $\eps \to 0$.
Since $\gamma<2/3$, one can choose $\beta\in
(\gamma,1-\gamma/2)$. Using again the maximum principle, we get
$$
|R_3(x)|\leq |R_3(1-\eps^\gamma+\eps^\beta)|, \quad x > 1-\eps^\gamma+\eps^\beta.
$$
Moreover, thanks to (\ref{estCACB}), we have \be
|R_3(1-\eps^\gamma+\eps^\beta)|&\lesssim &\eps^{3/2-3\gamma/4}{\rm
Ai}\left(\frac{\xi(\eps^\gamma-\eps^\beta)}{\eps^{2/3}}\right)+\eps^{3/2-3\gamma/4}\frac{{\rm
Bi}\left(\frac{\xi(\eps^\gamma-\eps^\beta)}{\eps^{2/3}}\right)}{{\rm
Bi}\left(\frac{\xi(\eps^\gamma)}{\eps^{2/3}}\right)}.\nonumber \ee
Using (\ref{expansion-xi}) again, we deduce from (\ref{daaibi})
that there exist a constant $c_0 >0$ such that
\begin{eqnarray*}
\eps^{3/2-3\gamma/4} {\rm Ai}\left(\frac{\xi(\eps^\gamma-\eps^\beta)}{\eps^{2/3}}\right)
\lesssim \exp\left(-c_0 \eps^{3\gamma/2-1}\right), \\
\eps^{3/2-3\gamma/4}\frac{{\rm Bi}
\left(\frac{\xi(\eps^\gamma-\eps^\beta)}{\eps^{2/3}}\right)}{{\rm
Bi} \left(\frac{\xi(\eps^\gamma)}{\eps^{2/3}}\right)} \lesssim
\exp\left(-c_0\eps^{\beta + \gamma/2-1}\right),
\end{eqnarray*}
where we have used
$$
\frac{\xi(\eps^\gamma-\eps^\beta)^{3/2}-\xi(\eps^\gamma)^{3/2}}{\eps}\sim
-3\eps^{\beta+\gamma/2-1}\quad \text{as}\quad \eps\to 0,
$$
which holds because $\beta \in (\gamma,1-\gamma/2)$. Therefore, we find
\be\label{R3Linf}
\|R_3\|_{L^\infty(1-\eps^\gamma+\eps^\beta, +\infty)}\lesssim
|R_3(1-\eps^\gamma+\eps^\beta)| \lesssim
\exp\left(-c_0\eps^{\beta+\gamma/2-1}\right),
\ee
which shows that $R_3(1)$ and $C_A^{\eps}$ are actually exponentially decaying as
$\eps \to 0$. Then, we infer from (\ref{R3inf}) and (\ref{R3Linf})
\be\label{R3L2<>}
\|R_3\|_{L^2(1-\eps^\gamma,1)}&\lesssim
&\|R_3\|_{L^2(1-\eps^\gamma,1-\eps^\gamma+\eps^\beta)} +
\|R_3\|_{L^2(1-\eps^\gamma+\eps^\beta,1)}\nonumber\\
&\lesssim &\eps^{\beta/2}\eps^{3/2-3\gamma/4}
+\eps^{\gamma/2} \exp\left(-c_0\eps^{\beta + \gamma/2 - 1}\right) \lesssim
\eps^{3/2+\beta/2-3\gamma/4}. \ee The $L^2$ norm of $R_3$ on the
interval $(-1,-1+\eps^\gamma)$ is estimated in the same way. Next,
we estimate the $L^2$ norm of $R_3$ on the interval $(1,\infty)$.
We multiply (\ref{eqR3}) by $R_3$ and integrate over
$(1,+\infty)$. Since $p_\eps\geq 1$ for $x\geq 2$ and $\eps\leq 1$, we
obtain \be\label{R3L2>1}
\|R_3\|_{L^2(2,+\infty)}^2 \leq
\int_1^{+\infty} (R_3')^2dx + \int_1^{+\infty} p_{\eps} R_3^2dx
& = & -R_3(1)R_3'(1)\nonumber\\
&\lesssim & \exp\left(-c_0\eps^{\beta+\gamma/2-1}\right)\eps^{1/3}, \ee
where $R_3(1)$ has been
estimated with (\ref{R3Linf}) and the bound for $R_3'(1)$ comes from Lemmas
\ref{lemma-remainder-plus} and \ref{lemma-resolvent}. The $L^2$ norm of $R_3$ on $(1,2)$ is estimated thanks to (\ref{R3Linf}). Together with
(\ref{R3L2>1}), we deduce that
$$\|R_3\|_{L^2(1,+\infty)}\lesssim
\exp\left(-c\eps^{\beta+\gamma/2-1}\right),
$$
where $c=c_0/2$. The $L^2$ norm of $R_3$ on
$(-\infty,-1)$ is estimated similarly, thus \be\label{R3L2>1conc}
\|R_3\|_{L^2(|x|>1)}\lesssim \exp\left(-c\eps^{\beta+\gamma/2-1}\right). \ee
Since $R_3$ solves
$$-R_3''+q_\eps R_3=0$$
on $(1-\eps^\gamma,+\infty)$ and $R_3\in L^2(\R)$, we deduce from the
maximum principle that if $R_3$ does not identically vanish on
$(1-\eps^\gamma,+\infty)$, then $R_3$ has a constant sign on that
interval. For instance, $R_3>0$ (the argument is similar in the other
case). Then, $R_3''(x)\geq 0$ for every $x\geq
1-\eps^\gamma$. Therefore $R_3'$ is a negative increasing function on
$(1-\eps^\gamma,+\infty)$. Let us assume by contradiction that
$\left|R_3'(1-\eps^\gamma+\eps^\beta)\right|>\exp(-c_0\eps^{\beta+\gamma/2-1})/\eps^2$.
Then, for $x\geq 0$, it follows
from the Taylor formula and (\ref{R3Linf}) that for $\eps$ sufficiently small,
\be
R_3(1-\eps^\gamma+\eps^\beta+\eps) & = & R_3(1-\eps^\gamma+\eps^\beta)+\eps R_3'(1-\eps^\gamma+\eps^\beta)+\int_0^\eps\int_0^s R_3''(1-\eps^\gamma+\eps^\beta+t)dt ds\nonumber\\
& \leq &\exp(-c_0\eps^{\beta+\gamma/2-1})\left(C-\frac{\eps}{\eps^2}+C\frac{\eps^2}{2}\frac{1}{\eps^{2-\gamma}}\right)<0,\nonumber
\ee
for some $C>0$, which is a contradiction with the positiveness of $R_3$.
As a result,
\be\label{R'3inf}
\|R_3'\|_{L^\infty(1-\eps^\gamma+\eps^\beta,+\infty)}=\left|R_3'(1-\eps^\gamma+\eps^\beta)\right|\lesssim \exp(-c\eps^{\beta+\gamma/2-1}).
\ee
At this stage, we have established all the estimates required to prove
the lemma. First,
(\ref{R1}), (\ref{R2}) and
(\ref{82bis}) yield
\be\label{R'L2}
\|R'\|_{L^2(\R)}&\leq & \|R'_1\|_{L^2(\R)}+\|R'_2\|_{L^2(\R)}+\|R'_3\|_{L^2(\R)}\nonumber\\
&\lesssim &
\eps^{5/3}+\eps^{2/3+3\gamma/2}+\eps^{1-\gamma/2}.
\ee
The choice $\gamma=1/6$ provides (\ref{L+L-der}). From (\ref{R1}),
(\ref{R2}), (\ref{82bis}),
(\ref{R3L2<>}) and (\ref{R3L2>1conc}), we obtain \be\label{RL2}
\|R\|_{L^2(\R)}&\leq &
\|R_1\|_{L^2(\R)}+\|R_2\|_{L^2(\R)}+\|R_3\|_{L^2(-1+\eps^\gamma,1-\eps^\gamma)}
+\|R_3\|_{L^2(1-\eps^\gamma<|x|<1)}+\|R_3\|_{L^2(|x|>1)}\nonumber\\
&\lesssim&
\eps^{7/3}+\eps^{4/3+3\gamma/2}+\eps^{2-\gamma}+\eps^{3/2-3\gamma/4+
\beta/2}+\exp\left(-c\eps^{\beta+\gamma/2-1}\right). \ee The choice
$\gamma=4/15$, $\beta=13/15-2\delta$, for sufficiently small positive
number $\delta$, provides the bound (\ref{L+L-1}). Similarly, we
have \be\label{RL2>1} \|R\|_{L^2(|x|>1)}&\leq &
\|R_1\|_{L^2(|x|>1)}+\|R_2\|_{L^2(|x|>1)}+\|R_3\|_{L^2(|x|>1)}\nonumber\\
&\lesssim&
\eps^{7/3}+\eps^{4/3+3\gamma/2}+\exp\left(-c\eps^{\beta+\gamma/2-1}\right).
\ee The choice $\gamma=2(1-\delta)/3$, $\beta=2/3$, for any small positive number
$\delta$, provides the bound
(\ref{L+L-12}). If $\alpha>0$, $\gamma<\min(\alpha,2/3)$ and if $\eps$ is
sufficiently small, we also obtain from (\ref{R1}), (\ref{R2}) and
(\ref{R'3inf}),
\be\label{R'at1}
\|R'\|_{L^\infty(1-\eps^\alpha,+\infty)} \leq
\|R'\|_{L^\infty(1-\eps^\gamma+\eps^\beta,+\infty)} &\leq &\|R'_1
\|_{L^\infty(\R)}+\|R'_2\|_{L^\infty(\R)}
+\|R'_3\|_{L^\infty(1-\eps^\gamma+\eps^\beta,+\infty)}\nonumber\\
&\lesssim &\eps^{4/3}+\eps^{1/3+3\gamma/2}+
\exp\left(-c\eps^{\beta+\gamma/2-1}\right). \ee A similar argument
on $(-\infty,-1+\eps^\alpha)$ gives (\ref{L+L-2}), for the choice
$\gamma=\min(\alpha,2/3)-2\delta/3$, $\beta=(1+\gamma)/4$. If
$\gamma<\min(\alpha,2/3)$, thanks to (\ref{R1}), (\ref{R2}),
(\ref{R3Linf}) and its twin estimate on
$(-\infty,-1+\eps^\alpha)$, we get similarly, for $\eps$
sufficiently small, \be\label{precomp1}
\|R\|_{L^\infty(|x|>1-\eps^\alpha)}\leq
\|R\|_{L^\infty(|x|>1-\eps^\gamma+\eps^\beta)}& \lesssim
&\|R_1\|_{L^\infty(\R)}+ \|R_2\|_{L^\infty(\R)}+\|R_3\|_{L^\infty(|x|>1-\eps^\gamma+\eps^\beta)}\nonumber\\
& \lesssim &
\eps^2+\eps^{1+3\gamma/2}+\exp\left(-c_0\eps^{\beta+\gamma/2-1}\right).
\ee
The bound (\ref{comp1}) follows from (\ref{precomp1}), again with the
choice $\gamma=\min(\alpha,2/3)-2\delta/3$, $\beta=(1+\gamma)/4$.
\end{Proof}

\section{Proof of the Main Theorem}
\subsection{The operator $A_\eps$ for $\eps>0$}
We consider here the operator \be\label{defAeps} A_\eps :=
\eps^{-2} (-\partial_x^2+p_\eps(x))^{-1}
(-\partial_x^2+q_{\eps}(x))^{-1} =
\eps^{-2}(L_-^\eps)^{-1}(L_+^\eps)^{-1}. \ee As we have seen
before, if $\eps>0$, both operators $L_-^\eps$ and $L_+^\eps$ on
$L^2(\R)$ are invertible with compact resolvent. As a result,
$A_\eps$ is a compact operator on $L^2(\R)$ for any fixed $\eps >
0$. Thus, its spectrum consists of a sequence of eigenvalues which
converges to zero. Moreover, these eigenvalues are all strictly
positive. Indeed, if $\mu$ is an eigenvalue of $A_\eps$ and $\phi$
is an associated eigenvector, $\zeta := (L_+^\eps)^{-1/2} \phi$
satisfies
$$
(L_+^\eps)^{-1/2}(L_-^\eps)^{-1}(L_+^\eps)^{-1/2} \zeta = \mu
\zeta.
$$
Therefore, $\mu$ is an eigenvalue of the self adjoint positive
operator $(L_+^\eps)^{-1/2}(L_-^\eps)^{-1}(L_+^\eps)^{-1/2}$,
which implies $\mu>0$. We order eigenvalues of $A_{\eps}$ as
$$
0 < \cdots \leq \mu_{n,\eps} \leq \cdots \leq \mu_{2,\eps} \leq
\mu_{1,eps} < \infty.
$$

\subsection{The operator $A_0$}\label{a0}
As $\eps\to 0$, we can formally expect that $A_\eps$ converges in some
sense to the operator
$$A_0=(-\partial_x^2+p_0)^{-1}\frac{1}{2(1-x^2)},$$
where
$$p_0(x)=\left\{\begin{array}{ll}0&\text{if } |x|<1,\\+\infty &\text{if
    } |x|>1.\end{array}\right.$$
Let us describe more precisely the action of the operator $A_0$ on
$L^2(\R)$. The following lemma is helpful for that purpose.
\begin{lem}\label{domain1}
If $u\in L^2(\R)$, then $\left(\frac{u}{1-x^2}\right)_{|(-1,1)}\in
\left(H^2\cap H_0^1\right)'(-1,1)$, where
  $\left(H^2\cap H_0^1\right)(-1,1)$ is endowed with the $H^2$
  norm. Moreover,
  the map $u\mapsto \left(\frac{u}{1-x^2}\right)_{|(-1,1)}$ is
  continuous from $L^2(\R)$ into
  $\left(H^2\cap H_0^1\right)'(-1,1)$.
\end{lem}

\begin{Proof} By Sobolev's embedding theorem, $H^2(-1,1)$ is
  continuously embedded into
  $\mathcal{C}^1([-1,1])$. Therefore, if $g\in (H^2\cap H_0^1)(-1,1)$,
  then
$$|g(x)|=|g(x)-g(\pm1)|\leq\|g'\|_{L^\infty}(1-|x|),$$
with $+1$ for $x>0$ and $-1$ for $x<0$. It follows that for every
$x\in (-1,1)$,
$$\left|\frac{g(x)}{1-x^2}\right|\leq
\frac{\|g'\|_{L^\infty}}{1+|x|}\lesssim \|g\|_{H^2}.$$
As a result, using the Cauchy-Schwarz inequality, we obtain
$$\left|\int_{-1}^1\frac{u(x)}{1-x^2}g(x)dx\right|\lesssim
\|u\|_{L^2(\R)}\|g\|_{H^2(-1,1)},$$
which completes the proof.
\end{Proof}

\noindent Let us denote the Dirichlet realization of the Laplacian
$\Delta=\partial_x^2$ on
the interval $(-1,1)$ by $\Delta_D$. It is well known that $(-\Delta_D)^{-1}$
maps continuously $L^2(-1,1)$
into $\left(H^2\cap H_0^1\right)(-1,1)$. By duality, it also
continuously maps $\left(H^2\cap H_0^1\right)'(-1,1)$ into
$L^2(-1,1)$. For $u\in L^2(\R)$, $A_0 u\in L^2(\R)$ is defined by
\be\label{defA0}
\left\{\begin{array}{l}
(A_0u)_{|\{|x|>1\}}  \equiv 0,\\
(A_0u)_{|(-1,1)} =
(-\Delta_D)^{-1}\left(\left(\frac{u}{2(1-x^2)}\right)_{|(-1,1)}\right).
\end{array}\right.
\ee Thanks to Lemma \ref{domain1} and the continuity of
$(-\Delta_D)^{-1}:\left(H^2\cap H_0^1\right)'(-1,1)\mapsto
L^2(-1,1)$, $A_0$ is a bounded operator on $L^2(\R)$. Moreover, we
have the following lemma.

\begin{lem}\label{continuity}
For any $u\in L^2(\R)$ and any $s\in [-1,1]$, \be\label{expA0} A_0u(s) &
= &
\int_s^1\left(\int_{-1}^y\frac{u(x)}{4(1-x)}dx-\int_y^1\frac{u(x)}{4(1+x)}dx\right)dy
+\frac{s-1}{2}I(u), \ee where \be\label{defI} I(u) & := &
\int_{-1}^1\left(\int_{-1}^y\frac{u(x)}{4(1-x)}dx-\int_y^1\frac{u(x)}{4(1+x)}dx\right)dy.
\ee In particular, $A_0u$ is continuous on $\R$.
\end{lem}

\begin{Proof}
For any $u\in L^2(\R)$ and any $y\in (-1,1]$, we have
\be\label{integL1}
\left|\int_y^1\frac{u(x)}{(1+x)}dx\right|
\leq\left(\int_y^1|u(x)|^2dx\right)^{1/2}\left(\int_y^1\frac{1}{(1+x)^2}dx\right)^{1/2}
\leq\frac{\|u\|_{L^2(\R)}}{\sqrt{1+y}}, \ee which implies that the
map $u\mapsto \int_y^1\frac{u(x)}{1+x}dx$ is continuous from
$L^2(\R)$ into $L^1(-1,1)$. Similarly, one can see that the map
$u\mapsto \int_{-1}^y\frac{u(x)}{1-x}dx$ has the same property. As
a result, $u\mapsto I(u)$ is a continuous linear form on
$L^2(\R)$, and the map which assigns to $u$ the right hand side in
(\ref{expA0}) is continuous from $L^2(\R)$ into
$L^\infty(-1,1)\subset L^2(-1,1)$. As we have seen before, so is
$u\mapsto (A_0 u)_{|(-1,1)}$. Actually, both sides in
(\ref{expA0}) only depend on the restriction of $u$ to $(-1,1)$,
so that they can be considered as continuous from $L^2(-1,1)$ into
itself. Therefore, using the principle of extension for uniformly
continuous functions, it suffices to check (\ref{expA0}) for $u$
in a dense subset of $L^2(-1,1)$. This can be done for $u\in
\mathcal{C}_c^\infty(-1,1)$. Indeed, in this case
$\left(\frac{u}{1-x^2}\right)_{|(-1,1)}\in L^2(-1,1)$, therefore
$(A_0u)_{|(-1,1)}\in \left(H^2\cap H_0^1\right)(-1,1)$. In
particular, $\underset{s\to\pm 1 \mp 0}{\lim}(A_0u)(s)=0$. On the
other side, we can easily check that the right hand side in
(\ref{expA0}) also vanishes at $s=\pm 1$ and its second derivative
is $-\frac{u(x)}{2(1-x^2)}$, which completes the proof of
(\ref{expA0}). It remains to prove that $\underset{s\to\pm
1 \mp 0}{\lim}(A_0u)(s)=0$ is true for any $u\in L^2(\R)$. This
follows from the fact that the maps $y\mapsto
\int_y^1\frac{u(x)}{1+x}dx$ and $y\mapsto
\int_{-1}^y\frac{u(x)}{1-x}dx$ are in $L^1(-1,1)$.
\end{Proof}

\begin{lem}
$A_0$ is a compact operator on $L^2(\R)$.
\end{lem}

\begin{Proof}
By Lemma \ref{continuity}, $A_0$ is continuous. Thus, according to
a standard criterion of relative compactness for a subset of
$L^2(\R)$ (see, for instance, Corollary IV.26 in \cite{B}), it is
sufficient to check the following two conditions
\begin{enumerate}
\item for every $\eta>0$, there exists a compact subset
  $\omega\subset \R$ such that for every $u\in B_{L^2}$,
  $$
  \|A_0u\|_{L^2(\R\string\ \omega)}<\eta
  $$
\item for every $\eta>0$ and for every compact subset
  $\omega\subset \R$, there exists $\delta>0$ such that for every $u\in B_{L^2}$
  and for every $h$ with $|h|<\delta$,
$$\|A_0u(\cdot+h)-A_0u\|_{L^2(\omega)}<\eta.$$
\end{enumerate}
In our case, condition (i) is trivially satisfied: we choose
$\omega=[-1,1]$ and then $\|A_0u\|_{L^2(\R\string\ \omega)}=0$ for
every $u\in B_{L^2}$. To check condition (ii), we note
that if $-1\leq s, s+h\leq 1$, then \be
\left|A_0u(s+h)-A_0u(s)\right| & = &
\left|-\int_s^{s+h}\left(\int_{-1}^y\frac{u(x)}{4(1-x)}dx-\int_y^1\frac{u(x)}{4(1+x)}dx\right)dy
+\frac{h}{2}I(u)\right|\nonumber\\
&\leq &
\left|\int_s^{s+h}\frac{\|u\|_{L^2(\R)}}{4}\left(\frac{1}{\sqrt{1+y}}+\frac{1}{\sqrt{1-y}}\right)dy\right|+\frac{|h|}{2}C\|u\|_{L^2(\R)}\nonumber\\
&\leq & \frac{\sqrt{|h|}}{4}+\frac{C|h|}{2},\nonumber \ee for some
constant $C > 0$. A similar estimate holds if either $+1$ or $-1$ lies
between $s$ and $s+h$ (which can only happen if
$|s|<1+|h|$), whereas if both $s$ and $s+h$ are outside of
$(-1,1)$, then $A_0u(s+h)-A_0u(s)=0$. Therefore,
$$
\| A_0u(\cdot+h)-A_0u\|_{L^2(\R)}\leq
\left(2(1+|h|)\right)^{1/2}\left(\frac{\sqrt{|h|}}{4}+\frac{C|h|}{2}\right),
$$
and condition (ii) follows.
\end{Proof}

Since $A_0$ is compact, its spectrum is purely discrete. Clearly,
$0$ is an eigenvalue of $A_0$ and the associated infinite-dimensional eigenspace is
made of the set of functions in $L^2(\R)$ supported in the exterior domain
$\{x\in \R : \; |x|\geq 1\}$. If $\mu\neq 0$ is an eigenvalue of $A_0$ and
$w$ an associated eigenvector, it follows from the definition of
$A_0$ that $w\equiv 0$ on $\{x \in \R: \; |x|\geq 1\}$, whereas
on $\{x \in \R: \; |x| < 1\}$, $w$ solves
\begin{equation}
\label{2-ODE} -2 (1 - x^2) w''(x) = \gamma w(x), \quad -1 < x < 1,
\end{equation}
where $\gamma = 1/\mu$. Moreover, thanks to Lemma
\ref{continuity}, $w = \gamma A_0w$ is continuous so that $w(-1) = w(1) = 0$. We
shall now prove that the only solutions of (\ref{2-ODE}) vanishing
at the endpoints $\pm1$ are the Gegenbauer polynomials
$C_{n+1}^{-1/2}(x)$ for $\gamma_n = 2n(n+1)$, where $n\geq 1$ is
integer. Thus, the spectrum of operator $A_0$ is given by
$$
\sigma(A_0) = \left\{\mu_n:=\frac{1}{2n(n+1)}, \;\; n\geq 1\right\} \cup \{0 \}.
$$

\begin{lem}
\label{lemma-hypergeometric} Equation (\ref{2-ODE}) admits a
family of solutions $(\gamma,w)=(\gamma_n,C_{n+1}^{-1/2})$, for
$n\geq -1$, where $\gamma_n = 2n(n+1)$ and $C_m^{\lambda}$ is a
Gegenbauer polynomial with degree $m$. If $(\gamma,w)\not\in
\{(\gamma_n,\alpha C_{n+1}^{-1/2})\ |\ n\geq -1, \alpha\in\R\}$ is a
solution of (\ref{2-ODE}), then it  satisfies \be
\label{solution-behavior}  \lim_{x \to 1-0} \left( |w(x)| + |w(-x)|
\right) \neq 0, \quad \lim_{x \to 1-0} \left( |w'(x)| + |w'(-x)|
\right) = \infty. \ee The only solutions $(\gamma,w)$ of (\ref{2-ODE})
such that $w(1)=w(-1)=0$ are $(\gamma_n,\alpha C_{n+1}^{-1/2})$, for
$n\geq 1$ and $\alpha\in \R$.
\end{lem}

\begin{Proof}
Explicit computations show that Gegenbauer polynomials
$C_{n+1}^{-1/2}(x)$ from Section 8.93 in \cite{Grad} are solutions
of (\ref{2-ODE}) for $\gamma_n$, for any $n\geq -1$. In particular, for $n\geq 1$, by equation 8.935 in \cite{Grad},
we have
$$
C_{n+1}^{-1/2}(x) =  -\frac{(1-x^2)}{n(n+1)}\frac{d^2}{dx^2}C_{n+1}^{-1/2}(x) = \frac{(1-x^2)}{n(n+1)} C_{n-1}^{3/2}(x),
$$
which proves that $C_{n+1}^{-1/2}(1) = C_{n+1}^{-1/2}(-1) = 0$ for
$n\geq 1$, whereas $C_0^{-1/2}(x)=1$ and $C_1^{-1/2}(x)=-x$. We
next prove that if $(\gamma,w)$ solves (\ref{2-ODE}) and $w$ is
not proportional to $C_{n+1}^{-1/2}$ with $n\geq -1$, then $w$
satisfies (\ref{solution-behavior}). We introduce the new variable
$z = x^2$ for $0<x<1$, and the function $u(z):=w(x)$. It is
equivalent for $w(x)$ to solve (\ref{2-ODE}) on $(0,1)$ or for
$u(z)$ to solve the hypergeometric equation: \be\label{hypergeom}
z(1-z) u''(z) + \frac{1}{2}(1-z) u'(z) + \frac{\gamma}{8} u(z) =
0, \quad 0 < z < 1. \ee This equation admits a general solution
given by 9.152 in \cite{Grad} \be\label{solhypergeom} u(z) = c_1
F(a,b,c;z) + c_2 z^{1/2}
F\left(a+\frac{1}{2},b+\frac{1}{2},\frac{3}{2};z\right), \ee where
$$
a + b = -\frac{1}{2}, \;\; ab = -\frac{\gamma}{8}, \;\; c =
\frac{1}{2}
$$
and $F(a,b,c;z)$ is a hypergeometric function. Clearly, the
function $x\mapsto u(x^2)=w(x)$ defined by (\ref{solhypergeom}) is
analytic for $0<x<1$ and can be extended into an function
$\tilde{w}$ which is analytic for $-1<x<1$, given by \be\nonumber
\tilde{w}(x) & := & c_1 F(a,b,c;x^2) + c_2 x
F\left(a+\frac{1}{2},b+\frac{1}{2},\frac{3}{2};x^2\right), \ee
where the first term is even in $x$ and the second term is odd in
$x$. Since $\tilde{w}$ solves (\ref{hypergeom}), the uniqueness in
the Cauchy-Lipshitz Theorem ensures that $w=\tilde{w}$. In order
to prove the Lemma, it is sufficient to consider one component of
the solution at one boundary point, e.g. $F(a,b,c;x^2)$ at $x = 1$
($z = 1$). Since ${\rm Re}(c-a-b) = 1 > 0$, the function
$F(a,b,c;z)$, which is analytic on $\{z: |z| < 1\}$, is also
bounded as $z \to 1$ (see 15.1.1 in \cite{AS}). Using 15.1.20 in
\cite{AS}, that is
$$
F(a,b,c;1) = \frac{\Gamma(c) \Gamma(c-a-b)}{\Gamma(c-a)
\Gamma(c-b)},
$$
we find that
$$
F(a,b,c;1) = \frac{\pi^{1/2}}{\Gamma(1+a) \Gamma(1/2-a)} =
-\frac{\sin(\pi a) \Gamma(-a)}{\pi^{1/2} \Gamma(1/2-a)} =
\frac{\cos(\pi a) \Gamma(1/2+a)}{\pi^{1/2} \Gamma(1+a)}.
$$
Parameters $a$ and $\gamma$ are related by $\gamma = 4a(1+2a)$. If
$\gamma = \gamma_{2m-1} = 4m(2m-1)$ for $m \geq 1$, then either $a
= -m$ or $a = -1/2+m$, both give $F(a,b,c;1) = 0$, corresponding
to even polynomial solutions $C_{2m}^{-1/2}$. For all other values
of $\gamma$ and $a$, $F(a,b,c;1)$ is bounded but non-zero. On the
other hand, using 15.2.1 in \cite{AS}, that is
$$
\frac{d}{dz} F(a,b,c;z) = \frac{ab}{c} F(a+1,b+1,c+1;z),
$$
since ${\rm Re}(c+1-a-1-b-1) = 0$, we obtain that $\frac{d}{dx}
F(a,b,c;z)=2x\frac{d}{dz} F(a,b,c;z)$ diverges as $z \to 1$ (see
15.1.1 in \cite{AS}), unless the series for $F(a,b,c,z)$ is
truncated into a polynomial function, which happens precisely when
$a$ or $b$ is a negative integer, which implies that $\gamma$
equals one of the $\gamma_{2m-1}$'s for some $m\geq 0$. Therefore,
$\lim_{x \to 1} |w'(x)| = \infty$ if $w(x)$ is an even solution of
(\ref{2-ODE}) and $\gamma \neq \gamma_{2m-1}$ for $m \geq 0$.
Similarly, the statement is proved for an odd solution of
(\ref{2-ODE}), given by $x F\left(a+1/2,b+1/2,3/2;x^2\right)$ for
$\gamma \neq \gamma_{2m}$ with $m \geq 0$, where $\gamma =
\gamma_{2m} = 4m (2m+1)$ correspond to odd polynomial solutions
$C_{2m-1}^{-1/2}$.
\end{Proof}

\subsection{Convergence in norm of $A_\eps$ to $A_0$ as $\eps \to 0$}

Our goal in this section is to prove the following result.
\begin{theo}\label{convinnorm}
It is true that
$$
A_\eps\longrightarrow A_0 \quad \text{in $\mathcal{L}(L^2(\R))$ as } \eps\to 0.
$$
\end{theo}
Once this result has been proved, we immediately have the
corollary.

\begin{cor}\label{coro}
For every integer $n\geq 1$,
$$\mu_{n,\eps}\longrightarrow \mu_n \quad \text{as }\eps\to 0.$$
Moreover, if $w_n$ is an eigenvector of $A_0$ associated to the
eigenvalue $\mu_n$, there exists a set
$(w_{n,\eps})_{\eps>0}\subset L^2(\R)$ of eigenvectors of $A_\eps$
associated to the eigenvalues $\mu_{n,\eps}$ for
$\eps>0$, such that 
$$w_{n,\eps}\longrightarrow w_n\quad \text{in } L^2(\R)
\text{ as }\eps\to 0.$$
\end{cor}

\begin{Proof}
Since convergence in norm in $\mathcal{L}(L^2)$ implies generalized
convergence, it follows from Theorem 3.16 on p.212 in \cite{K}
that for every integer $N\geq 1$ and for $0<\eps\ll 1$,
$$
\left|\left(\frac{\mu_N+\mu_{N+1}}{2},+\infty\right)\cap
  \sigma(A_\eps)\right|=N.
$$
Moreover, $\mu_{n,\eps}\to \mu_n$ as $\eps\to 0$, for any $1\leq n\leq
N$, which proves the convergence of the eigenvalues. For the
eigenvectors, let us fix $n\geq 1$, and let $\Omega_n\subset \C$ be a
neighborhood of $\mu_n$ such that $\overline{\Omega_n}$ does not
contain 0 nor any other eigenvalue of $A_0$. From the convergence of
the eigenvalues, it follows that for $\eps$
sufficiently small, $A_\eps$ has a unique eigenvalue in $\Omega_n$,
which is $\mu_{n,\eps}$. For any integer $m\geq 1$, we
denote by $E_m$ (resp. $E_m^\eps$) the  eigenspace of $A_0$ (resp
$A_\eps$) associated to the eigenvalue $\mu_m$ (resp
$\mu_{m,\eps}$). We also define 
$$F_n:=\left(\underset{m\neq
  n}{\oplus}E_m\right){\oplus}{\rm Ker}A_0\quad \text{ and }\quad F_{n,\eps}:=\underset{m\neq
  n}{\oplus}E_m^\eps,$$
as well as $P_n\in \mathcal{L}(L^2(\R))$
(resp. $P_{n,\eps}$) the projector on $E_n$ (resp $E_{n,\eps}$) along
$F_n$ (resp. $F_{n,\eps}$). Then, Theorem 3.16 in \cite{K} also ensures
that $P_{n,\eps}\longrightarrow P_n$ in $\mathcal{L}(L^2)$ as $\eps\to
0$. Thus, $w_{n,\eps}:=P_{n,\eps} w_n$ is an eigenvector of $A_\eps$ for
the eigenvalue $\mu_{n,\eps}$, and we have
$$\|w_{n,\eps}-w_n\|_{L^2(\R)}=\|(P_{n,\eps}-P_n)w_n\|_{L^2(\R)}\leq
\|P_{n,\eps}-P_n\|_{\mathcal{L}(L^2(\R))}\|w_n\|_{L^2(\R)}\underset{\eps\to
  0}{\longrightarrow} 0,$$
which completes the proof.
\end{Proof}

\begin{rem}\label{eigenvector*}
A straightforward consequence of Theorem \ref{convinnorm} is that
$A_\eps^*\to A_0^*$ in $\mathcal{L}(L^2(\R))$ as $\eps\to 0$. Thus,
an analogous result to Corollary \ref{coro} holds for the eigenvalues
and eigenvectors of $A_\eps^*$ and $A_0^*$.
\end{rem}

The convergence statement of the Main Theorem directly follows from
Corollary \ref{coro}, since the spectrum of system
(\ref{gen-eig-prob}) is made is made of the eigenvalues $\lambda = \pm
i\eps/\sqrt{\mu}$, where $\mu$ describes the spectrum $\sigma(A_\eps)$ of
$A_\eps$. Indeed, if
$(\lambda,u,w)\in\C\times L^2(\R) \times L^2(\R)$ solves
(\ref{gen-eig-prob}), a straightforward
computation shows that
$$A_\eps w=-\frac{\eps^2}{\lambda^2}w,$$
thus $\lambda=\pm \frac{i\eps}{\sqrt{\mu}}$ for some $\mu \in
\sigma(A_\eps)$. Conversely, if $A_\eps w=\mu w,$ with $w\in
L^2(\R)$, then $(i\eps/\sqrt{\mu}, u,w)\in\C\times L^2(\R) \times
L^2(\R)$ solves system (\ref{gen-eig-prob}) with
$$
u:=-\frac{i}{\eps\sqrt{\mu}}(L_+^\eps)^{-1}w.
$$

Let us now turn to the proof of Theorem \ref{convinnorm}.
In order to compare $A_0 u$ and $A_\eps u$ for $\eps>0$ and $u\in
L^2(\R)$, we would like
first to express $A_0 u$ as $A_0 u=A_\eps(A_\eps)^{-1}A_0 u$. This
can be done with the help of the following lemma.

\begin{lem}\label{extension}
Let $H$ be a Hilbert space and $L$ be a self-adjoint operator on $H$ with
domain $D(L)$ endowed with the graph-norm
$\|\cdot\|_{D(L)}=\left(\|\cdot\|_H^2+ \|L\cdot\|_H^2\right)^{1/2}$. Assume that
$L$ is continuously invertible and $X$ is a Banach space continuously
embedded in $H$. $L$ induces an operator $L_X$ on $X$, defined by
$$D(L_X)=\{x\in X,\ \ L_Xx\in X\}, \quad L_Xx=Lx\quad \text{for any } x\in
D(L_X).$$
$D(L_X)$ is endowed with the graph-norm
$\|\cdot\|_{D(L_X)}=\left(\|\cdot\|_X^2+
  \|L_X\cdot\|_X^2\right)^{1/2}$. Assume further
that $D(L_X)$ is dense in $H$ and that $D(L)$ is continuously
embedded in $X$.
Then $L$ is extended to $X'$ as a bicontinuous map $L_{X'}:X'\mapsto
D(L_X)'$ defined by
$$\left<L_{X'}f,\phi\right>_{D(L_X)',D(L_X)}:=\left<f,L_X\phi\right>_{X',X}\quad
\text{for any }f\in X' \text{ and } \phi\in D(L_X).$$
\end{lem}

\begin{Proof}
See Appendix \ref{Aextension}.
\end{Proof}

To prove that $A_0 u=A_\eps(A_\eps)^{-1}A_0 u$ for any $\eps > 0$
and $u \in L^2(\R)$, we apply Lemma \ref{extension} twice.
For the first application, $H=X=L^2(\R)$ and $L=L_-^\eps$, such
that $L_-^\eps$ is extended as a bicontinuous map (also denoted
$L_-^\eps$ for convenience) from $L^2(\R)$ into $D(L_-^\eps)'$.
Thus, $A_0u=(L_-^\eps)^{-1}L_-^\eps A_0 u$. For the second
application, $H=L^2(\R)$, $X=D(L_-^\eps)$ and $L=L_+^\eps$ such
that $L_+^\eps$ is extended as a bicontinuous map (that we will
also denote $L_+^\eps$) from $D(L_-^\eps)'$ into
$$
D_{D(L_-^\eps)}(L_+^\eps):=\{v\in D(L_-^\eps),\ \ L_+^\eps v\in
D(L_-^\eps)\}.
$$
Note here that $D(L_+^\eps)$ is continuously
embedded in $X=D(L_-^\eps)$, since
$L_+^\eps-L_-^\eps=\frac{2(1-x^2)}{\eps^2}\mathbf{1}_{(-1,1)}\in
\mathcal{L}(L^2(\R))$ (actually, $D(L_+^\eps)=D(L_-^\eps)$ and the
norms $\|\cdot\|_{D(L_-^\eps)}$ and $\|\cdot\|_{D(L_+^\eps)}$ are
equivalent). As a result,
$$A_0u=(L_-^\eps)^{-1}(L_+^\eps)^{-1}L_+^\eps L_-^\eps A_0 u=A_\eps
\eps^2L_+^\eps L_-^\eps A_0 u=A_\eps(A_\eps)^{-1}A_0u,$$ where
$(A_\eps)^{-1}$ maps $D_{D(L_-^\eps)}(L_+^\eps)$ into $L^2(\R)$.

The identity (\ref{expA0}) provides an explicit expression of $A_0u$
for any $u\in L^2(\R)$. Let us next use this identity to express $L_-^\eps A_0 u\in
D(L_-^\eps)'$. If $\phi\in
D(L_-^\eps)$ and $u\in L^2(\R)$, then direct computations involving
integration by parts give
\be\label{actionL-}
\lefteqn{\left<L_-^\eps A_0
    u,\phi\right>_{D(L_-^\eps)',D(L_-^\eps)}}\nonumber\\
 & = &
\left<A_0 u,L_-^\eps \phi\right>_{L^2,L^2}
\ =\  -\int_{-1}^1(A_0u)(s)\phi''(s)ds\nonumber\\
& = &
\int_{-1}^1\left(\int_s^1\left(\int_y^1\frac{u(x)}{4(1+x)}dx
    -\int_{-1}^y\frac{u(x)}{4(1-x)}dx\right)dy
  +\frac{s-1}{2}I(u)\right)\phi''(s)ds\nonumber\\
& = &
\int_{-1}^1\left(\int_s^1\frac{u(x)}{4(1+x)}dx-
  \int_{-1}^s\frac{u(x)}{4(1-x)}dx\right)\phi'(s)ds
-\frac{I(u)}{2}(\phi(1)-\phi(-1)).
\ee
Performing another integration by parts, the first term in the right
hand side of (\ref{actionL-}) can be expressed as
\be\label{actionL-2}
\lefteqn{\int_{-1}^1\left(\int_s^1\frac{u(x)}{4(1+x)}dx-
  \int_{-1}^s\frac{u(x)}{4(1-x)}dx\right)\phi'(s)ds}\nonumber\\
 & = &
\underset{\delta\to 0}{\lim}\int_{-1+\delta}^{1-\delta}\left(\int_s^1\frac{u(x)}{4(1+x)}dx-
  \int_{-1}^s\frac{u(x)}{4(1-x)}dx\right)\phi'(s)ds\nonumber\\
 & = &
\underset{\delta\to
  0}{\lim}\left(\int_{-1+\delta}^1\frac{u(x)}{4(1+x)}\left(\phi(x)-\phi(-1+\delta)\right)dx+\int^{1-\delta}_{-1}\frac{u(x)}{4(1-x)}\left(\phi(x)-\phi(1-\delta)\right)dx\right)\nonumber\\
& =
&\int_{-1}^1\frac{u(x)}{4(1+x)}\left(\phi(x)-\phi(-1)\right)dx+\int^1_{-1}\frac{u(x)}{4(1-x)}\left(\phi(x)-\phi(1)\right)dx.
\ee The first limit in the right hand side of (\ref{actionL-2}) is
evaluated as follows. (The second limit is evaluated similarly.)
We write \be\label{actionL-3}
\lefteqn{\left|\int_{-1+\delta}^1\frac{u(x)}{4(1+x)}
\left(\phi(x)-\phi(-1+\delta)\right)dx-\int_{-1}^1\frac{u(x)}{4(1+x)}
\left(\phi(x)-\phi(-1)\right)dx\right|}\nonumber\\
& = & \left|\int_{-1+\delta}^1\frac{u(x)}{4(1+x)}
\left(\phi(-1)-\phi(-1+\delta)\right)dx-
\int_{-1}^{-1+\delta}\frac{u(x)}{4(1+x)}
\left(\phi(x)-\phi(-1)\right)dx\right|. \ee The two terms in
the right hand side of (\ref{actionL-3}) converge to $0$ as
$\delta$ goes to 0 thanks to Lebesgue's dominated convergence
theorem. For the first term, the integrand is dominated by
\begin{eqnarray*}
\left|\frac{u(x)}{4(1+x)}\left(\phi(-1)-\phi(-1+\delta)\right)
\mathbf{1}_{(-1+\delta,1)} \right| & \leq & \left|\frac{\delta
    u(x)\|\phi'\|_{L^\infty}}{4(1+x)}\mathbf{1}_{(-1+\delta,1)}\right|\\
& \leq & \frac{|u(x)|\|\phi'\|_{L^\infty}}{4}\in
L^1(-1,1).
\end{eqnarray*}
The integrand of the second term is dominated by the same integrable
majorant. Then, from (\ref{actionL-}) and (\ref{actionL-2}) we deduce that
\be\label{actionL-fin}
\lefteqn{\left<L_-^\eps A_0
    u,\phi\right>_{D(L_-^\eps)',D(L_-^\eps)}}\\
 & = &
 \int_{-1}^1\frac{u(x)}{4}\frac{\phi(x)-\phi(-1)}{1+x}dx+\int^1_{-1}\frac{u(x)}{4}\frac{\phi(x)-\phi(1)}{1-x}dx-\frac{I(u)}{2}(\phi(1)-\phi(-1)).\nonumber
\ee
Thus, if $u\in L^2(\R)$ and $\phi\in D_{D(L_-^\eps)}(L_+^\eps)$, then
\be\label{actionL+L-}
\lefteqn{\left<\eps^2 L_+^\eps L_-^\eps A_0
    u-u,\phi\right>_{D_{D(L_-^\eps)}(L_+^\eps)',D_{D(L_-^\eps)}(L_+^\eps)}}\nonumber\\
& = &\left<L_-^\eps A_0
    u,\eps^2 L_+^\eps \phi\right>_{D(L_-^\eps),D(L_-^\eps)}-\int_{\R}u(x)\phi(x)dx\nonumber\\
 & = &
 -\eps^2\int_{-1}^1\frac{u(x)}{4(1+x)}
 \left(\phi''(x)-\phi''(-1)\right)dx-\eps^2
 \int^1_{-1}\frac{u(x)}{4(1-x)}\left(\phi''(x)
 -\phi''(1)\right)dx\nonumber\\
&& \phantom{texttexttext}+\frac{\eps^2I(u)}{2}
(\phi''(1)-\phi''(-1))-\int_{|x|>1}u(x)\phi(x)dx.\nonumber
\ee
Finally, if we introduce the adjoint operator of $A_\eps$,
$$A_\eps^*:=\frac{1}{\eps^2}(L_+^\eps)^{-1}(L_-^\eps)^{-1}\in
\mathcal{L}(L^2(\R),D_{D(L_-^\eps)}(L_+^\eps)),$$
we get for any $u,\phi\in L^2(\R)$
\be\label{action}
\lefteqn{\left< A_0 u-A_\eps u,\phi\right>_{L^2,L^2}\quad = \quad\left<A_\eps(\eps^2 L_+^\eps L_-^\eps A_0
    u-u),\phi\right>_{L^2,L^2}}\nonumber\\ & = & \left<\eps^2 L_+^\eps L_-^\eps A_0
    u-u,A_\eps^*\phi\right>_{D_{D(L_-^\eps)}(L_+^\eps)',D_{D(L_-^\eps)}(L_+^\eps)}\nonumber\\
&=& -\eps^2\int_{-1}^1\frac{u(x)}{4}\frac{(A_\eps^*\phi)''(x)-(A_\eps^*\phi)''(-1)}{1+x}dx-\eps^2\int^1_{-1}\frac{u(x)}{4}\frac{(A_\eps^*\phi)''(x)-(A_\eps^*\phi)''(1)}{1-x}dx\nonumber\\
&&
+\frac{\eps^2I(u)}{2}((A_\eps^*\phi)''(1)-(A_\eps^*\phi)''(-1))-\int_{|x|>1}u(x)(A_\eps^*\phi)(x)dx.
\ee In order to prove the convergence of $A_\eps$ to $A_0$ in
$\mathcal{L}(L^2(\R))$, it is sufficient to prove that the right
hand side in (\ref{action}) converges to 0 as $\eps \to 0$
uniformly for $u,\phi\in B_{L^2}$. Up to terms which may be
estimated similarly, it hence suffices to prove that the three
quantities
$$Q^\eps_1(u,\phi):=\left|\eps^2I(u)(A_\eps^*\phi)''(1)\right|,$$
$$Q^\eps_2(u,\phi):=\left|\int_{|x|>1}u(x)(A_\eps^*\phi)(x)dx\right|,$$
$$Q^\eps_3(u,\phi):=\left|\eps^2\int^1_{-1}u(x)
  \frac{(A_\eps^*\phi)''(x)-(A_\eps^*\phi)''(1)}{1-x}dx\right|,$$
defined for $u,\phi\in L^2(\R)$, converge to 0 as $\eps\to 0$,
uniformly for $u,\phi\in B_{L^2}$. In other words, we should
choose $u$ and $\phi$ in $B_{L^2}$ and prove that
\be
\label{bound-Q-terms}
Q^\eps_1(u,\phi) + Q^\eps_2(u,\phi) + Q^\eps_3(u,\phi) \lesssim
C(\eps),
\ee
where $C(\eps)$ does not depend on $u$ or $\phi$ and $C(\eps)\to 0$ as $\eps\to 0$.

\paragraph{Estimate on $Q_1^\eps$.}  We have already seen in the proof
of Lemma \ref{continuity} that
$|I(u)|\lesssim 1$. On the other side,
$$\eps^2\partial_x^2A_\eps^*=
q_\eps(L_+^\eps)^{-1}(L_-^\eps)^{-1}-(L_-^\eps)^{-1}.$$ Since
$q_\eps(1)=0$, it follows from Lemma \ref{lemma-space} that
$$\left|(\eps^2\partial_x^2A_\eps^*\phi)(1)\right|=\left|\left((L_-^\eps)^{-1}\phi\right)(1)\right|\lesssim
\eps^{2/3}.$$ Therefore \be\label{Q1} Q^\eps_1(u,\phi) &\lesssim &
\eps^{2/3}. \ee

\paragraph{Estimate on $Q_2^\eps$.}
It follows from Lemma \ref{L+L-} and from the Cauchy-Schwarz
inequality that \be\label{Q2new} Q^\eps_2(u,\phi) &\lesssim &
\eps^{1/3-\delta}, \ee for any $\delta>0$.

\paragraph{Estimate on $Q_3^\eps$.} Thanks to the Cauchy-Schwarz
inequality, it suffices to prove that
$$
\left\|\frac{(\eps^2\partial_x^2A_\eps^*)\phi(x)
    -(\eps^2\partial_x^2A_\eps^*)\phi(1)}{1-x}\right\|_{L^2(-1,1)} \longrightarrow 0
    \quad \text{as }\eps\to 0,
$$
uniformly for $\phi\in B_{L^2}$. Using a commutator, we first
decompose the operator
$\eps^2\mathbf{1}_{(-1,1)}\partial_x^2A_\eps^*$ as
\be\label{decop}
\lefteqn{\eps^2\mathbf{1}_{(-1,1)}\partial_x^2A_\eps^* =
-\mathbf{1}_{(-1,1)} L_-^\eps(L_+^\eps)^{-1}(L_-^\eps)^{-1}}\nonumber\\
& = &
-\mathbf{1}_{(-1,1)}(L_+^\eps)^{-1}+\mathbf{1}_{(-1,1)}
\partial_x^2\left[(L_+^\eps)^{-1},(L_-^\eps)^{-1}\right].
\ee We introduce the functions $r:=(L_+^\eps)^{-1}\phi$,
$s:=(L_-^\eps)^{-1}\phi$, $R:=(L_+^\eps)^{-1}s$,
$S:=(L_-^\eps)^{-1}r$ and $\omega:=\partial_x^2(R-S)$. Then,
$$
\left\|\frac{(\eps^2\partial_x^2A_\eps^*)\phi(x)-(\eps^2\partial_x^2A_\eps^*)\phi(1)}{1-x}\right\|_{L^2(-1,1)}
\leq
\left\|\frac{r(x)-r(1)}{1-x}\right\|_{L^2(-1,1)}+\left\|\frac{\omega(x)-\omega(1)}{1-x}\right\|_{L^2(-1,1)}.
$$
According to Lemma \ref{lemma-remainder-plus},
$\|r'\|_{L^\infty(\R)}\lesssim \eps^{1/3}$ and the first term is hence estimated by
\be\label{estimr} \left\|\frac{r(x)-r(1)}{1-x}\right\|_{L^2(-1,1)}
&\lesssim & \eps^{1/3}. \ee Let us now estimate the second term in
the inequality above. If we make the difference
of the two fourth--order differential equations satisfied by $R$
and $S$ on $(-1,1)$, we find that $\omega$ solves the differential
equation \be\label{odeomega} -\partial_x^2
\omega+\frac{2(1-x^2)}{\eps^2}\omega & = &
\frac{4}{\eps^2}R+\frac{8x}{\eps^2}R', \quad -1 < x < 1. \ee Let
$\alpha\in (0,2)$ (different explicit choices of $\alpha$ will be
made later), $\beta=23/30-\delta$ and $\gamma=7/15+\delta$, where
$0<\delta<1/45$. Thanks to the triangle inequality,
\be\label{splitQ3}
\left\|\frac{\omega(x)-\omega(1)}{1-x}\right\|_{L^2(-1,1)}
\!\!\!\!\lesssim\|\omega\|_{L^2(-1,0)}+
\eps^{-\gamma}\|\omega\|_{L^2(0,
    1-\eps^{\gamma})}
+\eps^{-\gamma}|\omega(1)|
+\left\|\frac{\omega(x)-\omega(1)}{1-x}\right\|_{L^2(1-\eps^{\gamma},1)}\!\!\!\!\!.
\ee
Next, for $x\in (-1,1)$, we have
\be\label{omegaexpl}
\omega(x)=\partial_x^2(R-S)(x)=r(x)-s(x)+\frac{2(1-x^2)}{\eps^2}R(x)
\ee
and
\be\label{omega'expl}
\omega'(x)=r'(x)-s'(x)+\frac{2(1-x^2)}{\eps^2}R'(x)-\frac{4x}{\eps^2}R(x).
\ee
Thanks to Lemmas \ref{lemma-space},
\ref{lemma-remainder-plus}, and \ref{L+L-}, we obtain
\be\label{om1}
|\omega(\pm 1)|=|r(\pm 1)-s(\pm 1)|\lesssim \eps^{2/3}
\ee
and
\be\label{om1'}
|\omega'(\pm 1)|=\left|r'(\pm 1)-s'(\pm 1)\mp \frac{2}{\eps^2}R(\pm
1)\right|\lesssim 1+\frac{|R(\pm1)|}{\eps^2}\lesssim\eps^{-\delta}.
\ee
If we multiply (\ref{odeomega}) by $\omega$, integrate over $(-1,1)$
and use the Cauchy-Schwarz
inequality, we get
\be\label{estimomega}
\lefteqn{\|\omega'\|_{L^2(-1,1)}^2+\frac{1}{\eps^2}\int_{-1}^1(1-x^2)\omega^2dx}\nonumber\\
&   \lesssim &
\frac{1}{\eps^2}\|R\|_{L^2(-1,1)}\|\omega\|_{L^2(-1,1)}+\frac{1}{\eps^2}
      \|R\|_{L^2(-1,1)}\|\omega'\|_{L^2(-1,1)}
\nonumber\\
&&
      +|\omega(1)||\omega'(1)|+|\omega(-1)||\omega'(-1)|
      +\frac{|\omega(1)||R(1)|+|\omega(-1)||R(-1)|}{\eps^2}.
\ee
Decomposing $(-1,1)$ into $(-1+\eps^\alpha,1-\eps^\alpha)$,
$(-1,-1+\eps^\alpha)$ and $(1-\eps^\alpha,1)$ and using the Taylor
formula and the Cauchy-Schwarz inequality on the last two intervals, we get
thanks to (\ref{om1})
\be\label{omegaL2}
\|\omega\|_{L^2(-1,1)}&\lesssim
&\|\omega\|_{L^2(-1+\eps^\alpha,1-\eps^{\alpha})}+\eps^{\alpha/2}
\left(|\omega(1)|+|\omega(-1)|+\eps^{\alpha/2} \|\omega'\|_{L^2(-1,1)}\right)\nonumber\\
&\lesssim
&\|\omega\|_{L^2(-1+\eps^\alpha,1-\eps^{\alpha})}+\eps^{\alpha/2+2/3}+\eps^{\alpha}\|\omega'\|_{L^2(-1,1)}.
\ee From (\ref{estimomega}), (\ref{om1}), (\ref{om1'}),
(\ref{omegaL2}) and Lemma \ref{L+L-} we deduce, for sufficiently
small $\delta > 0$, \be
\lefteqn{\|\omega'\|_{L^2(-1,1)}^2+\eps^{\alpha-2}\|\omega\|_{L^2(-1+\eps^\alpha,1-\eps^{\alpha})}^2}\nonumber\\
&\lesssim &
\eps^{26/15 -\delta-2}\left(\|\omega\|_{L^2(-1+\eps^\alpha,1-\eps^{\alpha})}
+\eps^{\alpha/2+2/3}+\eps^{\alpha}\|\omega'\|_{L^2(-1,1)}
+\|\omega'\|_{L^2(-1,1)}\right)+\eps^{2/3-\delta}+\eps^{2/3-\delta}\nonumber\\
&\lesssim &
\eps^{2/3-\delta}+\eps^{\alpha/2+2/5-\delta}+\eps^{-4/15-\delta}
\|\omega\|_{L^2(-1+\eps^\alpha,1-\eps^{\alpha})}+\eps^{-4/15-\delta}\|\omega'\|_{L^2(-1,1)}.
\ee Therefore there exists a positive constant $C$ such that
\be\label{quadrat}
\left(\|\omega'\|_{L^2(-1,1)}-C\eps^{-4/15-\delta}\right)^2+\eps^{\alpha-2}
\left(\|\omega\|_{L^2(-1+\eps^{\alpha},1-\eps^\alpha)}-
C\eps^{26/15-\alpha-\delta}\right)^2 \nonumber\\
\lesssim \quad \eps^{2/3-\delta}+
\eps^{\alpha/2+2/5-\delta}+\eps^{-8/15-2\delta}+
\eps^{22/15-\alpha-2\delta}. \ee We deduce that for any $\alpha\in
(0,2)$, \be\label{omegaL2intint}
\|\omega\|_{L^2(-1+\eps^{\alpha},1-\eps^\alpha)} &\lesssim &
\eps^{26/15-\alpha-\delta}+\eps^{4/3-\alpha/2-\delta/2}
+\eps^{6/5-\alpha/4-\delta/2}+\eps^{11/15-\alpha/2-\delta}
\nonumber \\
& \lesssim & \eps^{11/15-\alpha/2-\delta} \ee and
\be\label{omega'L2}
\|\omega'\|_{L^2(-1,1)} &\lesssim &
\eps^{-4/15-\delta}+\eps^{1/3-\delta}
+\eps^{1/5+\alpha/4-\delta/2}+\eps^{-4/15-\delta}+
\eps^{11/5-\alpha/2-\delta}
\nonumber \\
&\lesssim &\eps^{-4/15-\delta}. \ee Using (\ref{omegaL2}),
(\ref{omegaL2intint}), and (\ref{omega'L2}), we obtain \be
\|\omega\|_{L^2(-1,1)} &\lesssim & \eps^{11/15-\alpha/2-\delta}+
\eps^{\alpha/2+2/3}+\eps^{-4/15+\alpha-\delta}.\nonumber \ee For
$\alpha=2/3$, we get \be\label{omegaL2f} \|\omega\|_{L^2(-1,1)}
&\lesssim & \eps^{2/5-\delta}. \ee Coming back to (\ref{splitQ3}),
thanks to (\ref{om1}), (\ref{omegaL2intint}) with $\alpha=\gamma$,
and (\ref{omegaL2f}), we obtain \be\label{splitQ3bis}
\left\|\frac{\omega(x)-\omega(1)}{1-x}\right\|_{L^2(-1,1)}
&\lesssim &
\eps^{2/5-\delta}+\eps^{11/15-3\gamma/2-\delta}+\eps^{2/3-\gamma}
+\left\|\frac{\omega(x)-\omega(1)}{1-x}\right\|_{L^2\left(1-\eps^{\gamma},1\right)}.\ \ \ \ \
\ee
If $\gamma = 7/15+\delta$ and $\beta = 23/30-\delta$, we have
$$
1-\eps^{7/15}+\eps^{23/30-\delta}<1-\eps^{\gamma}
$$
for sufficiently small $\eps > 0$ and therefore
$$
\left\|\frac{\omega(x)-\omega(1)}{1-x}\right\|_{L^2\left(1-\eps^{\gamma},1\right)} \leq
\left\|\frac{\omega(x)-\omega(1)}{1-x}\right\|_{L^2\left(1-\eps^{7/15}+\eps^{23/30-\delta},1\right)}.
$$
From (\ref{omegaexpl}) we
infer, for $x\in (-1,1)$, \be\label{omegafracexpl}
\frac{\omega(x)-\omega(1)}{1-x}=
\frac{r(x)-r(1)}{1-x}+\frac{s(x)-s(1)}{1-x}+\frac{2(1+x)}{\eps^2}R(x).
\ee Like in (\ref{estimr}), it follows from Lemmas
\ref{lemma-space} and \ref{lemma-remainder-plus} that
\be\label{estimrs}
\left\|\frac{r(x)-r(1)}{1-x}\right\|_{L^2\left(1-\eps^{7/15}+\eps^{23/30-\delta},1\right)}
& \lesssim & \eps^{17/30}, \\\label{estimrs-extra}
\left\|\frac{s(x)-s(1)}{1-x}\right\|_{L^2\left(1-\eps^{7/15}+\eps^{23/30-\delta},1\right)}
& \lesssim & \eps^{7/30}. \ee Splitting $R$ as $R_1+R_2+R_3$ as
in the proof of Lemma \ref{L+L-}, and using (\ref{R1}),
(\ref{R2}) and (\ref{R3Linf}), we deduce that
\be\label{Rclose1}
\|R\|_{L^2(1-\eps^{7/15}+\eps^{23/30-\delta},1)}\lesssim
\eps^{7/3}+\eps^{61/30+3\delta/2}+ \exp\left(-c\eps^{23/30-\delta+7/30-1}\right)
\lesssim \eps^{61/30}, \ee
for some $c>0$, since $7/15<2/3$ and $7/15<23/30-\delta<1-7/30$.
As a result, combining
(\ref{splitQ3bis}), (\ref{omegafracexpl}), (\ref{estimrs}), (\ref{estimrs-extra}),
and (\ref{Rclose1}), we obtain \be
\left\|\frac{\omega(x)-\omega(1)}{1-x}\right\|_{L^2(-1,1)}
&\lesssim &
\eps^{2/5-\delta}+\eps^{1/30-5\delta/2}+\eps^{1/5-\delta}+
\eps^{7/30}+\eps^{1/30} \nonumber \\
& \lesssim & \eps^{1/30-5\delta/2}, \nonumber \ee which
provides the required result for $\delta<1/45$. Combining all together,
we proved that $C(\eps) \to 0$ as $\eps \to 0$ in bound (\ref{bound-Q-terms}).
According to the previous construction, this finishes the proof of Theorem \ref{convinnorm}.

\subsection{Convergence rate of eigenvalues of $A_{\eps}$}

To prove the convergence rate of the Main Theorem, we write the
eigenvalue problem $ A_{\eps} w=\mu w$ as the generalized
eigenvalue problem \be \label{eig-prob} L_-^\eps w = \gamma
\eps^{-2} (L_+^\eps)^{-1} w, \ee
where $\gamma=1/\mu$. Let us first introduce some notations. For any integer $n\geq 1$, let
$w_n$ be an eigenvector of $A_0$ for the eigenvalue $\mu_n =
\frac{1}{2n(n+1)}$, and let $u_n=\frac{w_n}{2(1-x^2)}$. According to
the results of section
\ref{a0}, $w_n$
is identically equal to 0 outside of the interval $(-1,1)$ and its
restriction to $(-1,1)$ is a polynomial which vanishes at the
endpoints $\pm 1$. In particular, $u_n\in L^2(\R)$. Moreover, $u_n$
solves the equation 
$$\frac{1}{2(1-x^2)}\left(-\partial_x^2+p_0\right)^{-1}u_n=\mu_n u_n,$$
which means that $\mu_n$ is an eigenvalue of $A_0^*$, with associated
eigenvector $u_n$. Conversely, if $u\in L^2$ is an eigenvector of $A_0^*$
for an eigenvalue $\mu$, then $w=2(1-x^2)u$ defines an eigenvector of $A_0$
for the same eigenvalue $\mu$. Therefore $A_0$ and $A_0^*$ have
the same eigenvalues $\{\mu_n\}_{n\geq 1}$. Similarly, for $\eps>0$,
$A_\eps$ and
$A_\eps^*$ have the same eigenvalues $\{\mu_{n,\eps}\}_{n\geq 1}$, and
$w_{n,\eps}\in L^2$ is an eigenvector of $A_\eps$ for an eigenvalue $\mu_{n,\eps}$ if
and only if $u_{n,\eps}=L_-^\eps w_{n,\eps}$ is an eigenvector of $A_\eps^*$ for the
same eigenvalue $\mu_{n,\eps}$. For convenience, $w_n$ and $u_n$ are normalized
by
$$\|u_n\|_{L^2(\R)}=1.$$ 
Then, according to Remark \ref{eigenvector*},
for any $n\geq 1$ and any $\eps>0$, we can define an eigenvector
$u_{n,\eps}$ of $A_\eps^*$ for the eigenvalue $\mu_{n,\eps}$, in such a
way that
$$u_{n,\eps}\to u_n\quad \text{in } L^2(\R) \text{ as } \eps\to 0.$$
We also define 
$$w_{n,\eps}:=\mu_{n,\eps}^{-1}(L_-^\eps)^{-1}u_{n,\eps}=\eps^2 L_+^\eps
u_{n,\eps}.$$ 
Then, we have the following lemma, which gives directly
the rate of convergence of $\gamma_{n,\eps}=1/\mu_{n,\eps}$ to
$\gamma_n=1/\mu_n$ in the Main Theorem.

\begin{lem}
\label{lemma-correspondence} Let $m,n\geq 1$ be two integers and fix
$\delta > 0$ small. The following alternative is true:
\begin{itemize}
\item If  $m \neq n$, then $|\int_{-1}^1 w_n u_{m,\eps} dx |
\lesssim \eps^{1/3 -\delta}$. 
\item If $m=n$, then $|\int_{-1}^1
w_n u_{m,\eps} dx | \gtrsim 1$ and $|\mu_m^\eps - \mu_n| \lesssim
\eps^{1/3-\delta}$.
\end{itemize}
\end{lem}

\begin{Proof}
We prefer to work with $\gamma_{n,\eps} = 1/\mu_{n,\eps}$ and
$\gamma_n = 1/\mu_n$. The eigenvector of $A_\eps$, $w_{m,\eps} = \gamma_m^\eps
A_{\eps} w_{m,\eps}$ solves the problem
$$
-w_{m,\eps}''(x) = \gamma_m^\eps u_{m,\eps}, \quad -1 < x < 1,
$$
while the eigenvector $w_n = \gamma_n A_0 w_n$ solves the
second--order differential equation
$$
-2(1-x^2) w_n''(x) = \gamma_n w_n(x), \quad -1 < x < 1.
$$
Multiplying the first equation by $w_n$ and integrating by parts
on $[-1+\eps^{2/3},1-\eps^{2/3}]$, we
obtain
\begin{eqnarray}
(\gamma_m^\eps - \gamma_n) \int_{|x| < 1 - \eps^{2/3}} w_n
u_{m,\eps} dx = \left[ w_n' w_{m,\eps} - w_n w_{m,\eps}' \right] |_{x =
-1+\eps^{2/3}}^{x=1-\eps^{2/3}} - \gamma_n \int_{|x| < 1 -
\eps^{2/3}} w_n \theta_{m,\eps} dx,\ \ \  \label{computations-estimate}
\end{eqnarray}
where
$$
\theta_{m,\eps}(x) = u_{m,\eps}(x) - \frac{w_{m,\eps}(x)}{2(1-x^2)}
$$
By Lemma \ref{lemma-space}, since $\| L_-^{\eps} w_{m,\eps} \|_{L^2}
= \gamma_m^\eps\|u_{m,\eps}\|_{L^2}\to\gamma_m$ as $\eps\to 0$, we obtain
\begin{eqnarray}
\| w_{m,\eps}' \|_{L^{\infty}(1-\eps^{2/3} < |x| < 1)} & \leq &
\| w_{m,\eps}'\|_{L^{\infty}(\R)} \lesssim 1,\label{rhs3391}\\
\| w_{m,\eps} \|_{L^{\infty}(1-\eps^{2/3} < |x| < 1)} & \leq &
|w_{m,\eps}(- 1)|+|w_{m,\eps}(1)| + \eps^{2/3} \| w_{m,\eps}'
\|_{L^{\infty}(1-\eps^{2/3} < |x| < 1)} \lesssim \eps^{2/3}.\ \ \ \ \label{rhs3392}
\end{eqnarray}
The last term in the right-hand-side of
(\ref{computations-estimate}) is estimated by
\be\label{rhs3393}
\left| \int_{|x| < 1 - \eps^{2/3}} w_n \theta_{m,\eps} dx \right|
\lesssim \| \theta_{m,\eps} \|_{L^2(|x| < 1 - \eps^{2/3})}.
\ee
The function $\theta_{m,\eps}(x)$
solves the second--order differential equation for $|x| < 1 -
\eps^{2/3}$: \be \label{ode-theta} -\eps^2 \theta_{m,\eps}''(x) +
2(1-x^2) \theta_{m,\eps}(x) & = & \eps^2 g_{m,\eps}''(x), \qquad
\text{where }g_{m,\eps}(x) = \frac{w_{m,\eps}(x)}{2(1-x^2)}. \ee
We infer that
\begin{eqnarray}
|g_{m,\eps}(\pm(1-\eps^{2/3}))| \lesssim
1, \quad |g_{m,\eps}'(\pm(1-\eps^{2/3}))|
\lesssim \eps^{-2/3}.\label{boundg'}
\end{eqnarray}
We take a scalar product of (\ref{ode-theta}) with $\theta_{m,\eps}$
and obtain the bound \be \nonumber  
\lefteqn{\eps^2 \|
\theta_{m,\eps}' \|^2_{L^2(|x| < 1 - \eps^{2/3})} + \eps^{2/3}
\| \theta_{m,\eps} \|^2_{L^2(|x| < 1 - \eps^{2/3})} \lesssim \eps^2
|\theta_{m,\eps}(1-\eps^{2/3})|
|\theta_{m,\eps}'(1-\eps^{2/3})|}\nonumber  \\
& &  + \eps^2
|\theta_{m,\eps}(-1+\eps^{2/3})|
|\theta_{m,\eps}'(-1+\eps^{2/3})| + \eps^2 \| \theta_{m,\eps}
\|_{L^2(|x| < 1 - \eps^{2/3})} \| g_{m,\eps}'' \|_{L^2(|x| < 1 -
\eps^{2/3})}.\ \ \ \ \ \  \label{theta-bounds} \ee 
By Lemma \ref{L+L-} for $\alpha=2/3$, we
have for any small $\delta >0$
\begin{eqnarray}
|u_{m,\eps}(\pm(1-\eps^{2/3}))| = \eps^{-2} |((L_+^{\eps})^{-1}w_{m,\eps})(\pm(1-\eps^{2/3}))|
&\lesssim &\eps^{-\delta},\label{boundpsi} \\
|u_{m,\eps}'(\pm(1-\eps^{2/3}))| = \eps^{-2} |((L_+^{\eps})^{-1}
w_{m,\eps})'(\pm(1-\eps^{2/3}))| &\lesssim &\eps^{-2/3-\delta}.\label{boundpsi'}
\end{eqnarray}
The bounds (\ref{boundg'}), (\ref{boundpsi}), and
(\ref{boundpsi'}), induce, if $\delta<1$,
\begin{eqnarray}
|\theta_{m,\eps}(\pm(1-\eps^{2/3}))| \leq
|u_{m,\eps}(\pm(1-\eps^{2/3}))|
+ |g_{m,\eps}(\pm(1-\eps^{2/3}))| & \lesssim &
\eps^{-\delta}, \label{boundtheta}\\
|\theta_{m,\eps}'(\pm(1-\eps^{2/3}))| \leq
|u_{m,\eps}'(\pm(1-\eps^{2/3}))| +
|g_{m,\eps}'(\pm(1-\eps^{2/3}))| & \lesssim &
\eps^{-2/3-\delta}.\label{boundtheta'}
\end{eqnarray}
On the other hand, it follows from the definition of $g_{m,\eps}$ in
(\ref{ode-theta}) that for $x\in(-1+\eps^{2/3},1-\eps^{2/3})$,
$$
w_{m,\eps}''(x) = 2 (1-x^2) g_{m,\eps}''(x) - 8 x
g_{m,\eps}'(x) - 4 g_{m,\eps}(x).
$$
We multiply this identity by $g_{m,\eps}''$ and integrate over
$(-1+\eps^{2/3},1-\eps^{2/3})$. We get
\be
\lefteqn{2\int_{-1+\eps^{2/3}}^{1-\eps^{2/3}}(1-x^2)|g_{m,\eps}''|^2dx
  +8\int_{-1+\eps^{2/3}}^{1-\eps^{2/3}}|g_{m,\eps}'|^2dx}\nonumber\\ 
&=&\int_{-1+\eps^{2/3}}^{1-\eps^{2/3}}w_{m,\eps}g_{m,\eps}''dx+4\left[
  xg_{m,\eps}'(x)^2
  +g_{m,\eps}(x)g_{m,\eps}'(x)\right]_{-1+\eps^{2/3}}^{1-\eps^{2/3}},\nonumber 
\ee
which implies thanks to Lemma \ref{lemma-resolvent}, (\ref{boundg'}) and the Cauchy-Schwarz
inequality \be
\eps^{2/3}\|g_{m,\eps}''\|_{L^2(-1+\eps^{2/3},1-\eps^{2/3})}^2
+\|g_{m,\eps}'\|_{L^2(-1+\eps^{2/3},1-\eps^{2/3})}^2 
&\lesssim
&\|g_{m,\eps}''\|_{L^2(-1+\eps^{2/3},1-\eps^{2/3})}+\eps^{-4/3}.\
\ \ \ \ \ \ 
\ee 
It follows that there exists $C>0$ such that
\be\label{preg''L2}
\eps^{2/3}\left(\|g_{m,\eps}''\|_{L^2(-1+\eps^{2/3},1-\eps^{2/3})}-C\eps^{-2/3}\right)^2
+\|g_{m,\eps}'\|_{L^2(-1+\eps^{2/3},1-\eps^{2/3})}^2 &\lesssim
&\eps^{-4/3}. \ee 
As a result, 
\be \|g_{m,\eps}' \|_{L^2(|x| < 1 - \eps^{2/3})} \lesssim
\eps^{-2/3}, \quad \| g_{m,\eps}'' \|_{L^2(|x|
< 1 - \eps^{2/3})} \lesssim \eps^{-1}.\label{g'g''l2}
\ee Then, thanks to (\ref{theta-bounds}), (\ref{boundtheta}),
(\ref{boundtheta'}) and (\ref{g'g''l2}), we obtain 
\be\nonumber
\eps^2 \| \theta_{m,\eps}' \|^2_{L^2(|x| < 1 -
    \eps^{2/3})}+\eps^{2/3} \| \theta_{m,\eps} \|^2_{L^2(|x| < 1 -
    \eps^{2/3})}
& \lesssim & \eps \|
\theta_{m,\eps} \|_{L^2(|x| < 1 - \eps^{2/3})} +
\eps^{4/3-2\delta}. \ee
Therefore, there exists $\eps$-independent constant $C > 0$ such
that \be \eps^2 \| \theta_{m,\eps}' \|^2_{L^2(|x| < 1 -
    \eps^{2/3})}+\eps^{2/3}\left( \| \theta_{m,\eps} \|_{L^2(|x| < 1 -
      \eps^{2/3})} - C \eps^{1/3}
  \right)^2
& \lesssim & \eps^{4/3-2\delta}\nonumber
\ee Thus, \be\label{351} \| \theta_{m,\eps} \|_{L^2(|x| < 1 -
      \eps^{2/3})}&\lesssim &
    \eps^{1/3-\delta}
\ee 
We deduce from
(\ref{computations-estimate}), (\ref{rhs3391}), (\ref{rhs3392}),
(\ref{rhs3393}) and (\ref{351}) that
\be\label{star} 
\left| (\gamma_m^\eps - \gamma_n) \int_{-1+\eps^{2/3}}^{1-\eps^{2/3}} w_n
u_{m,\eps} dx \right| \lesssim  \eps^{1/3-\delta}. \ee If $m \neq
n$, then $|\gamma_m^\eps - \gamma_n| \gtrsim 1$ and therefore $\left|
\int_{-1+\eps^{2/3}}^{1-\eps^{2/3}} w_n u_{m,\eps} dx \right| \lesssim
\eps^{1/3-\delta}$. Since $u_{m,\eps}\to u_m$ in $L^2(\R)$, using the
Cauchy-Schwarz inequality, we obtain
$$\left|\int_{-1}^1w_n u_{m,\eps} dx\right|\leq \left|
\int_{-1+\eps^{2/3}}^{1-\eps^{2/3}} w_n u_{m,\eps} dx
\right|+\left|\int_{1-\eps^{2/3}<|x|<1} w_n u_{m,\eps} dx
\right|\lesssim
\eps^{1/3-\delta}+\eps^{1/3}\|u_{m,\eps}\|_{L^2(\R)}\lesssim
\eps^{1/3-\delta},$$
which is the estimate of the first alternative. If $m = n$, since
$u_{n,\eps}\to u_n$ in $L^2(\R)$, we also have 
$\mathbf{1}_{[-1+\eps^{2/3},1-\eps^{2/3}]}u_{n,\eps}\to u_n$ in $L^2(\R)$,
and thus
$$\int_{-1+\eps^{2/3}}^{1-\eps^{2/3}} w_n u_{n,\eps} dx\underset{\eps\to 0}{\longrightarrow}
 \int_{-1}^1w_n u_n dx=\int_{-1}^1\frac{w_n^2}{2(1-x^2)}dx>0.$$
Combined with (\ref{star}), it gives $|\gamma_{n,\eps}-\gamma_n|\lesssim
\eps^{1/3-\delta}$, which is the second alternative.
\end{Proof}

\section{Eigenvalues of the spectral problem (\ref{gen-eig-prob})}
\label{section-asymptotics}

As we have seen before, if $(u,w)\in L^2(\R)\times L^2(\R)$ solves system
(\ref{gen-eig-prob}), then $w$ is an eigenvector of $A_\eps$ associated to the eigenvalue $1/\gamma$, where $\gamma = -\lambda^2/\eps^2$. In other words, $w$ solves the two
fourth--order differential equations \be
\label{4-ode}
\left\{ \begin{array}{cl} \eps^2 \left( -
\partial_x^2 + \frac{1}{\eps^2} (x^2-1) \right)^2 w(x) = \gamma
w(x), & \;\; \mbox{for} \;\; |x| > 1, \\ -2 (1 - x^2) w''(x) +
\eps^2 w''''(x) = \gamma w(x), & \;\; \mbox{for} \;\; |x| < 1,
\end{array} \right. \ee which also means that $w$ solves
the generalized eigenvalue problem (\ref{eig-prob}). Since $w \in
L^2(\R)$, we have
$(L_+^{\eps})^{-1} w \in H_{\text{loc}}^2(\R) \subset {\cal
C}^1(\R)$ for any fixed $\eps > 0$. From the generalized
eigenvalue problem (\ref{eig-prob}), we infer that $w$ is twice
continuously differentiable on $\mathbb{R}$ and $w'''(x)$ has jump
discontinuities at $x = \pm 1$: \be \label{jump-conditions} w'''
|_{x=1-0}^{x=1+0} = \frac{2}{\eps^2} w(1), \qquad w''' |_{x =
-(1+0)}^{x=-(1-0)} = \frac{2}{\eps^2} w(-1). \ee Solutions of the
first equation of system (\ref{4-ode}) on the outer intervals
$\{|x| > 1\}$ can be constructed analytically. Solutions of the
second equation of system (\ref{4-ode}) on the inner interval
$(-1,1)$ can be approximated numerically. Following to a
classical shooting method, we shall find numerically an estimate on
the convergence rate of $\gamma_{n,\eps}$ to $\gamma_n$ as $\eps
\to 0$, for a fixed $n \geq 1$. The convergence rate we observe
numerically is faster that the one  in the Main Theorem.

For convenience, we will only consider even eigenfunctions
$w(x)$ near $\gamma_{2m-1} = 4m(2m-1)$ for an integer $m \geq 1$.
A similar analysis can be developed for odd eigenfunctions near
$\gamma_{2m} = 4m(2m+1)$ for an integer $m \geq 1$.

\subsection{Asymptotic solutions on the outer interval}
For a fixed value of $\gamma>0$, $w$ solves the first equation of
system (\ref{4-ode}) on $[1,+\infty)$ if and only if
\be\label{fourth-order-ode}
0 & = &
\left(-\partial_x^2+\frac{x^2-(1+\eps\sqrt{\gamma})}{\eps^2}\right)\left(-\partial_x^2+\frac{x^2-(1-\eps\sqrt{\gamma})}{\eps^2}\right)w\nonumber\\
& = &
\left(-\partial_x^2+\frac{x^2-(1-\eps\sqrt{\gamma})}{\eps^2}\right)\left(-\partial_x^2+\frac{x^2-(1+\eps\sqrt{\gamma})}{\eps^2}\right)w.
\ee
Thus, linear combinations of solutions of the second--order
differential equations 
\be\label{sec-order-ode}
0 & = & \left(-\partial_x^2+\frac{x^2-(1+\eps\nu)}{\eps^2}\right)w
\ee
for $\nu=\pm\sqrt{\gamma}$ provide solutions of the fourth--order
differential equation (\ref{fourth-order-ode}). We shall see that they
are the only solutions of (\ref{fourth-order-ode}). First, the
following lemma gives a set of
two linearly independent solutions of (\ref{sec-order-ode}). 
\begin{lem}\label{airy-back}
Fix $\nu\in \R$. There exists a constant $C>0$ such that
for $\eps>0$ sufficiently small, the equation 
\be\label{eqext}
 -\psi''(x)+\frac{(x^2-1)}{\eps^2}\psi(x)=\frac{\nu}{\eps}\psi(x),\quad x\geq 1
\ee has two linearly independent solutions $\psi_A^{\nu,\eps}$ and
$\psi_B^{\nu,\eps}$ such that for $x\geq 0$
$$\psi_A^{\nu,\eps}(\sqrt{1+\eps\nu}(1+x))=a(x){\rm Ai}\left(\frac{(1+\eps\nu)^{1/3}\xi(x)}{\eps^{2/3}}\right)
\left(1+Q_A^{\nu,\eps}(\xi(x))\right),$$
$$\psi_B^{\nu,\eps}(\sqrt{1+\eps\nu}(1+x))=a(x){\rm Bi}\left(\frac{(1+\eps\nu)^{1/3}\xi(x)}{\eps^{2/3}}\right)\left(1+Q_B^{\nu,\eps}(x)\right),$$
where
$\xi(x):=\left(\frac{3}{2}\int_0^x\sqrt{t(2+t)}dt\right)^{2/3}$,
$a(x):= (\xi'(x))^{-1/2}$ and $Q_A^{\nu,\eps}$, $Q_B^{\nu,\eps}$ satisfy the
bound
$$\|Q_A^{\nu,\eps}\|_{L^\infty(\R^+)} + \|Q_B^{\nu,\eps}\|_{L^\infty(\R^+)}\leq
C\eps^{2/3}.$$
Moreover,
\begin{eqnarray}
\frac{(\psi_A^{\nu,\eps})'(1)}{\psi_A^{\nu,\eps}(1)} = \frac{2^{1/3}
{\rm Ai}'(\eps^{1/3} 2^{-2/3} \nu)}{\eps^{2/3}{\rm Ai}(\eps^{1/3} 2^{- 2/3}\nu)} \left( 1 + {\cal O}(\eps^{2/3})
\right) = -\frac{6^{1/3} \Gamma(2/3)}{\eps^{2/3}\Gamma(1/3)} \left( 1 +
{\cal O}(\eps^{1/3}) \right), \label{asymptotic-limit}
\end{eqnarray}
where ${\cal O}(\eps^{1/3})$ and ${\cal O}(\eps^{2/3})$ in (\ref{asymptotic-limit}) are uniform in $\nu\in K$, for any compact set $K\subset \R$.
\end{lem}

\begin{Proof}
See Appendix \ref{Aairy}.
\end{Proof}

\begin{rem}
Note that solutions of (\ref{sec-order-ode}) can be expressed in terms
of the Whittaker's functions of the parabolic cylinder equation. The
connection of these functions with Airy functions, similarly as in
Lemma \ref{airy-back}, was studied by Olver \cite{O} using asymptotic
formal methods.
\end{rem}

\begin{cor}
\label{corollary-space-improvement} Let $n\geq 1$ and $w_\eps \in L^2(\R)$ be an
eigenvector of the generalized eigenvalue problem (\ref{eig-prob}) for
the eigenvalue $\gamma_{n,\eps}$. Then, 
there exists constants $c_+$ and $c_-$ such that
\begin{equation}
\label{solution-exterior}
w_\eps(x) = c_+\psi_A^{\sqrt{\gamma_{n,\eps}},\eps}(x)  + c_-\psi_A^{-\sqrt{\gamma_{n,\eps}},\eps}(x) , \qquad x > 1.
\end{equation}
Moreover,
\be
\label{bounds-improvement} 
w_\eps(1)=\frac{-\Gamma(1/3)\eps^{2/3}w_\eps'(1)}{6^{1/3} \Gamma(2/3)}\left( 1 +
{\cal O}(\eps^{1/3}) \right), \  w_\eps''(1)=\frac{-\Gamma(1/3)\eps^{2/3}w_\eps'''(1-0)}{6^{1/3} \Gamma(2/3)}\left( 1 +
{\cal O}(\eps^{1/3}) \right).
\ee
\end{cor}

\begin{Proof}
First, we remark that if $\gamma>0$, then
$\psi_A^{\sqrt{\gamma},\eps},\psi_B^{\sqrt{\gamma},\eps},\psi_A^{-\sqrt{\gamma},\eps}$
and $\psi_B^{-\sqrt{\gamma},\eps}$ are four linearly independent
solutions of the fourth--order equation
(\ref{fourth-order-ode}). Indeed, if $C_A^\pm, C_B^\pm$ are constants
such that
\be\label{lin-indep}
C_A^+\psi_A^{\sqrt{\gamma},\eps}+C_B^+\psi_B^{\sqrt{\gamma},\eps}+C_A^-\psi_A^{-\sqrt{\gamma},\eps}+C_B^-\psi_B^{-\sqrt{\gamma},\eps}=0,
\ee
applying the operator $-\partial_x^2+\frac{x^2-1}{\eps^2}$ to (\ref{lin-indep}), we obtain 
\be\nonumber
C_A^+\psi_A^{\sqrt{\gamma},\eps}+C_B^+\psi_B^{\sqrt{\gamma},\eps}-C_A^-\psi_A^{-\sqrt{\gamma},\eps}-C_B^-\psi_B^{-\sqrt{\gamma},\eps}=0.
\ee
Combined with (\ref{lin-indep}), it gives
$$C_A^+\psi_A^{\sqrt{\gamma},\eps}+C_B^+\psi_B^{\sqrt{\gamma},\eps}=0\quad\text{and}\quad
C_A^-\psi_A^{-\sqrt{\gamma},\eps}+C_B^-\psi_B^{-\sqrt{\gamma},\eps}=0.$$
From Lemma \ref{airy-back} and from the asymptotic behaviour
(\ref{daaibi})
of Ai and Bi, we deduce that for any $\nu\in \R$, $\psi_A^{\nu,\eps}$
and $\psi_B^{\nu,\eps}$ are linearly independent. As a result, $C_A^+=
C_B^+=C_A^-= C_B^-=0$. It follows that the only solutions of
(\ref{fourth-order-ode}) which vanish at infinity, are the linear
combinations of $\psi_A^{\sqrt{\gamma},\eps}$ and
$\psi_A^{-\sqrt{\gamma},\eps}$. It results in the decomposition (\ref{solution-exterior}). Since $\gamma_{n,\eps}\to \gamma_n$ as $\eps\to 0$,
the asymptotic expansions (\ref{bounds-improvement}) come from
(\ref{asymptotic-limit}) and the identities
\begin{eqnarray*}
w_\eps(1) & = & c_+ \psi_A^{\sqrt{\gamma_{n,\eps}},\eps}(1) +
c_-\psi_A^{-\sqrt{\gamma_{n,\eps}},\eps}(1) , \\ 
w_\eps'(1) & = &   c_+(\psi_A^{\sqrt{\gamma_{n,\eps}},\eps})'(1)  + c_-
(\psi_A^{-\sqrt{\gamma_{n,\eps}},\eps})'(1) , \\ 
w_\eps''(1) & = &  \eps^{-1} (\gamma_{n,\eps})^{1/2} \left[ -c_+
  \psi_A^{\sqrt{\gamma_{n,\eps}},\eps}(1) + c_-
  \psi_A^{-\sqrt{\gamma_{n,\eps}},\eps}(1) \right] \\ 
w_\eps'''(1+0) & = &  \eps^{-1} (\gamma_{n,\eps})^{1/2} \left[ -c_+
  (\psi_A^{\sqrt{\gamma_{n,\eps}},\eps})'(1) + c_-
  (\psi_A^{-\sqrt{\gamma_{n,\eps}},\eps})'(1) \right] \\ 
& \phantom{t} & 
+ 2 \eps^{-2} \left[ c_+ \psi_A^{\sqrt{\gamma_{n,\eps}},\eps}(1) + c_-
  \psi_A^{-\sqrt{\gamma_{n,\eps}},\eps}(1) \right]\\
& = &w_\eps'''(1-0)+ 2 \eps^{-2} \left[ c_+ \psi_A^{\sqrt{\gamma_{n,\eps}},\eps}(1) + c_-
  \psi_A^{-\sqrt{\gamma_{n,\eps}},\eps}(1) \right].
\end{eqnarray*}
\end{Proof}

\begin{rem}
\label{remark-self-adjoint} Asymptotic limit (\ref{asymptotic-limit}) 
implies that for $0 < \eps \ll 1$, the eigenvalue
$\lambda_n^{\eps}$ of the self-adjoint problem $L_-^{\eps} w_{\eps}
= \lambda_n^{\eps} w_{\eps}$ satisfies a sharp bound
\begin{equation}
\label{bound-self-adjoint}
C_n^- \eps^{2/3} \leq |\lambda_n^{\eps} - \lambda_n | \leq C_n^+ \eps^{2/3}
\end{equation}
for a fixed integer $n \geq 1$, where $\lambda_n = \frac{\pi^2
  n^2}{4}$, $\lambda_n^{\eps}$ is the $n^\text{th}$
eigenvalue of $L_-^\eps$ and $0 < C_n^- < C_n^+ < \infty$ are
some constants. Indeed,
differential equation $L_-^{\eps} w = \lambda w$ has analytic
solutions for even eigenfunctions
$$
w = \left\{ \begin{array}{cl} \cos(\sqrt{\lambda} x) & \;\; \mbox{for} \;\; |x| < 1, \\
c \psi_A^{\eps\lambda,\eps}(|x|) & \;\; \mbox{for} \;\; |x| > 1, \end{array} \right.
$$
where $c$ is a constant. Notice that for $\lambda>0$ fixed,
$\nu=\eps\lambda$ stays in a compact subset of $\R$ when $\eps$ goes
to 0. Continuity of $w(x)$ and $w'(x)$ across $1$ leads to an
algebraic system, where $c$ can be eliminated and $\lambda$ is found
from the transcendental equation
$$
\frac{\cos(\sqrt{\lambda})}{\sqrt{\lambda} \sin(\sqrt{\lambda})} =
-\frac{\psi_A^{\eps\lambda,\eps}(1)}{(\psi_A^{\eps\lambda,\eps})'(1)}\underset{\eps\to 0}{\sim}\eps^{2/3}\frac{\Gamma(1/3)}{6^{1/3} \Gamma(2/3)},
$$
where we have used (\ref{asymptotic-limit}). We deduce that for some
integer $m\geq 1$,
$\sqrt{\lambda}=\sqrt{\lambda_{2m-1}^\eps}=\sqrt{\lambda_{2m-1}}-\delta_m(\eps)$,
where $\sqrt{\lambda_{2m-1}^\eps}=\frac{\pi(2m-1)}{2}$ for ${m\geq 1}$
are the roots of ${\rm cos}\sqrt{\lambda}$, and $\delta_m(\eps)\underset{\eps\to 0}{\sim}
\eps^{2/3}\frac{(2m-1)\pi\Gamma(1/3)}{2\cdot 6^{1/3} \Gamma(2/3)}$. It
proves (\ref{bound-self-adjoint}) for $n$ odd. For odd eigenfunctions
($n$ even), the analysis is similar.
\end{rem}

\subsection{Numerical solutions on the inner interval}

Unfortunately, Remark \ref{remark-self-adjoint} is not useful in the context
of the non-self-adjoint system (\ref{4-ode}) because we do not know explicit
analytic solutions of the second equation of system (\ref{4-ode}).
Therefore, we use a numerical method to approximate
these solutions on the inner interval $[-1,1]$.

Considering even eigenfunctions of (\ref{eig-prob}) we let
$w_1(x)$ and $w_2(x)$ be two particular solutions of the second
equation in (\ref{4-ode})
on $[0,1]$ subject to the boundary conditions \be \nonumber
\left\{ \begin{array}{cc} w_1(1)
= 1, \quad w_1''(1) = 0, \quad w_1'(0) = 0,  \quad w_1'''(0) = 0, \\
w_2(1) = 0, \quad w_2''(1) = 1,  \quad w_2'(0) = 0, \quad
w_2'''(0) = 0. \end{array} \right. \ee Then, a general even solution of the
second equation of system (\ref{4-ode}) writes
\begin{equation}
\label{solution-interior}
w(x) = a_1 w_1(x) + a_2 w_2(x), \quad 0 < x < 1,
\end{equation}
for some constants $a_1$, $a_2$. The continuity of $w(x)$ and $w''(x)$ across $x = 1$ leads to the
scattering map from $(a_1,a_2)$ to $(c_+,c_-)$ in the solutions (\ref{solution-exterior})
and (\ref{solution-interior}), which is solved uniquely by
$$
c_{\pm} = \frac{a_1 \mp \eps \gamma^{-1/2} a_2}{2\psi_A^{\pm\sqrt{\gamma},\eps}(1)},
$$
where for conciseness, $\gamma_{n,\eps}$ is simply denoted $\gamma$. The continuity of $w'(x)$ and the jump condition
(\ref{jump-conditions}) on $w'''(x)$ across $x = 1$ lead to a
linear system on $(a_1,a_2)$ in the form
\begin{eqnarray*}
\left[ U_p - \eps^{2/3} w_1'(1) \right] a_1
+ \left[ \eps \gamma^{-1/2} U_m - \eps^{2/3} w_2'(1) \right] a_2 & = & 0,\\
\left[ \gamma^{1/2} U_m - \eps^{5/3} w_1'''(1) \right] a_1 +
\left[ \eps U_p - \eps^{5/3} w_2'''(1) \right] a_2 & = & 0,
\end{eqnarray*}
where
$$
U_p = \frac{\eps^{2/3}(\psi_A^{\sqrt{\gamma},\eps})'(1)}{2 \psi_A^{\sqrt{\gamma},\eps}(1)} + \frac{\eps^{2/3}(\psi_A^{-\sqrt{\gamma},\eps})'(1)}{2 \psi_A^{-\sqrt{\gamma},\eps}(1)}, \qquad
U_m = -\frac{\eps^{2/3}(\psi_A^{\sqrt{\gamma},\eps})'(1)}{2 \psi_A^{\sqrt{\gamma},\eps}(1)} + \frac{\eps^{2/3}(\psi_A^{-\sqrt{\gamma},\eps})'(1)}{2 \psi_A^{-\sqrt{\gamma},\eps}(1)},
$$
By the ODE theory, unique classical solutions $w_1(x)$ and $w_2(x)$ exist for any $\eps > 0$
and the dependence of $w_{1,2}(x)$ on $\eps$ is analytic for $\eps > 0$. If there exists
a simple root of the determinant of the linear system for a particular value $\eps_0 > 0$,
the root persists for other values of $\eps > 0$ near $\eps = \eps_0$. This method is used
for tracing eigenvalues $\gamma(\eps)$ of the spectral problem
(\ref{eig-prob}) as $\eps \to 0$.

To do it numerically, we approximate solutions $w_1(x)$ and
$w_2(x)$ with the second--order central--difference method on a
uniform grid with the grid size $h = 0.005$. The numerical method
is explained in Appendix \ref{Anumerics}. On the other hand,
the values of $U_p$ and $U_m$ can be evaluated from the asymptotic
formula (\ref{asymptotic-limit}) for $\eps \in [10^{-6},10^{-4}]$
with $20$ data points.
Using these approximations, the determinant of the linear system
for $(a_1,a_2)$ is plotted versus $\gamma$ near $\gamma = \gamma_1
= 4$ and $\gamma = \gamma_3 = 24$ and its zero is detected
numerically. Then, the zero is plotted versus $\eps$ and its best
power fit is used to detect the convergence rate of $|\gamma -
\gamma_n| \sim C \eps^p$. The numerical zeros and the best power
fit is shown on Figure 3 for $\gamma_1 = 4$ (left)
and $\gamma_3 = 24$ (right), while the numerical approximations of
the eigenfunctions for $\eps = 10^{-4}$ are shown on Figure
4 (dots) together with the limiting profiles obtained
from the polynomial $C_2^{-1/2}(x)$ and $C_4^{-1/2}(x)$ at $\eps = 0$
(dashed lines). The 
numerical values of the power of the best power fit are found to
be $1.9959$ for $\gamma_1 = 4$ and $1.9662$ for $\gamma_3 = 24$,
which suggests that the sharp asymptotic bound is
$$
|\gamma_{n,\eps} - \gamma_n | \lesssim  \eps^2,
$$
for $n\geq 1$. Finally, Figure 5 shows the ratio $a_1/a_2$ obtained from the linear
system for $\eps = 10^{-6}$ in $\gamma$ near $\gamma_1 = 4$ (left)
and the values of the ratio at the non-zero solution of the linear
system in $\eps$ (right). The power fit was found to be $1.99998$
and it illustrates that $\lim_{\eps \to 0} a_1/a_2 = 0$, such that
$\lim_{\eps \to 0} w(x) = w_2(x)$ (up to renormalization).
\begin{figure}[!h]
\begin{center}
\begin{minipage}{12.5cm}
\includegraphics[width=6cm]{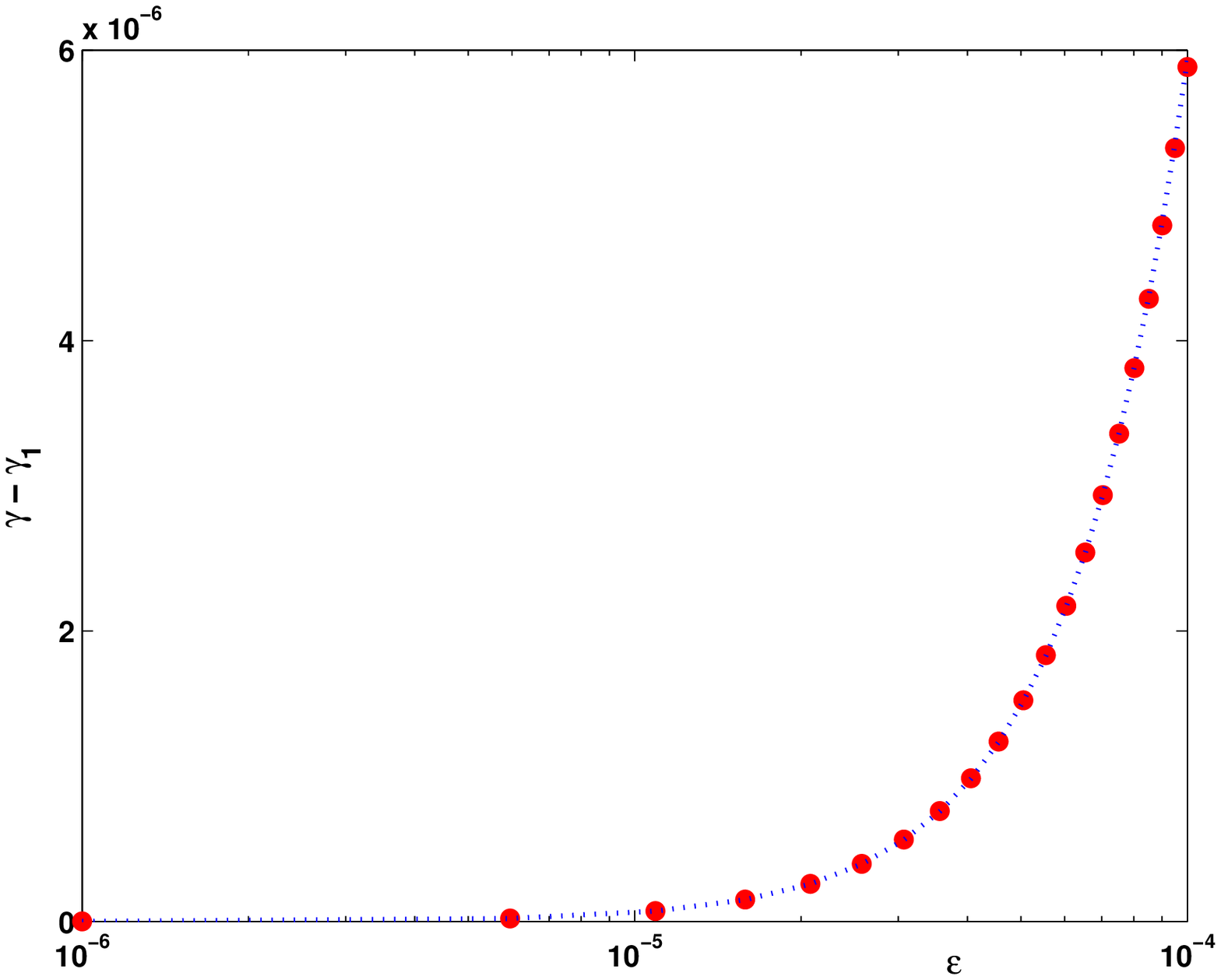}
\includegraphics[width=6cm]{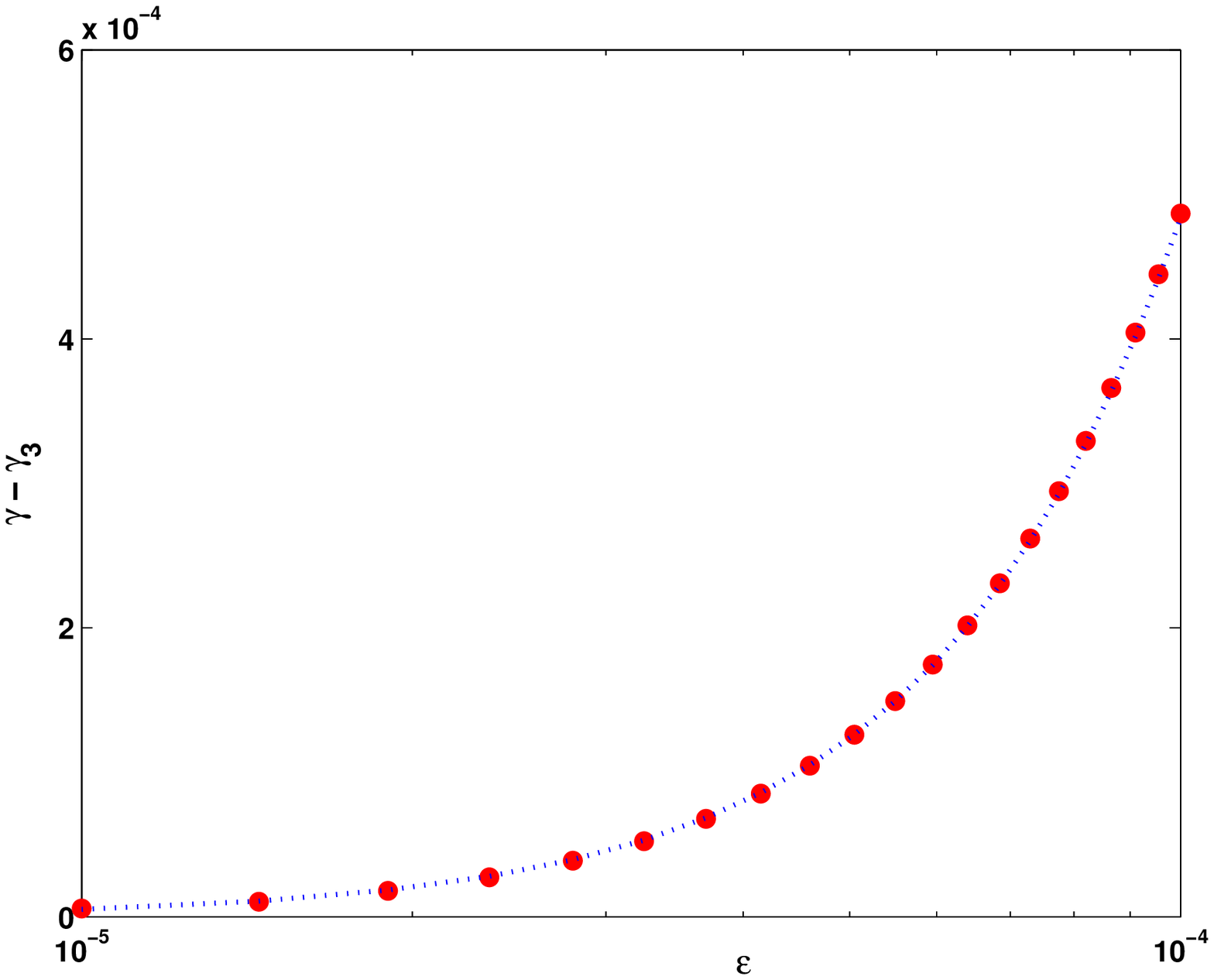}
\caption{The numerical zeros of the determinant of the linear system (dots)
and its best power fit (dashed line)
for $\gamma_1 = 4$ (left) and $\gamma_3 = 24$ (right).}
\label{figure-1}
\end{minipage}\vspace{.55cm}\\
\begin{minipage}{12.5cm}
\includegraphics[width=6cm]{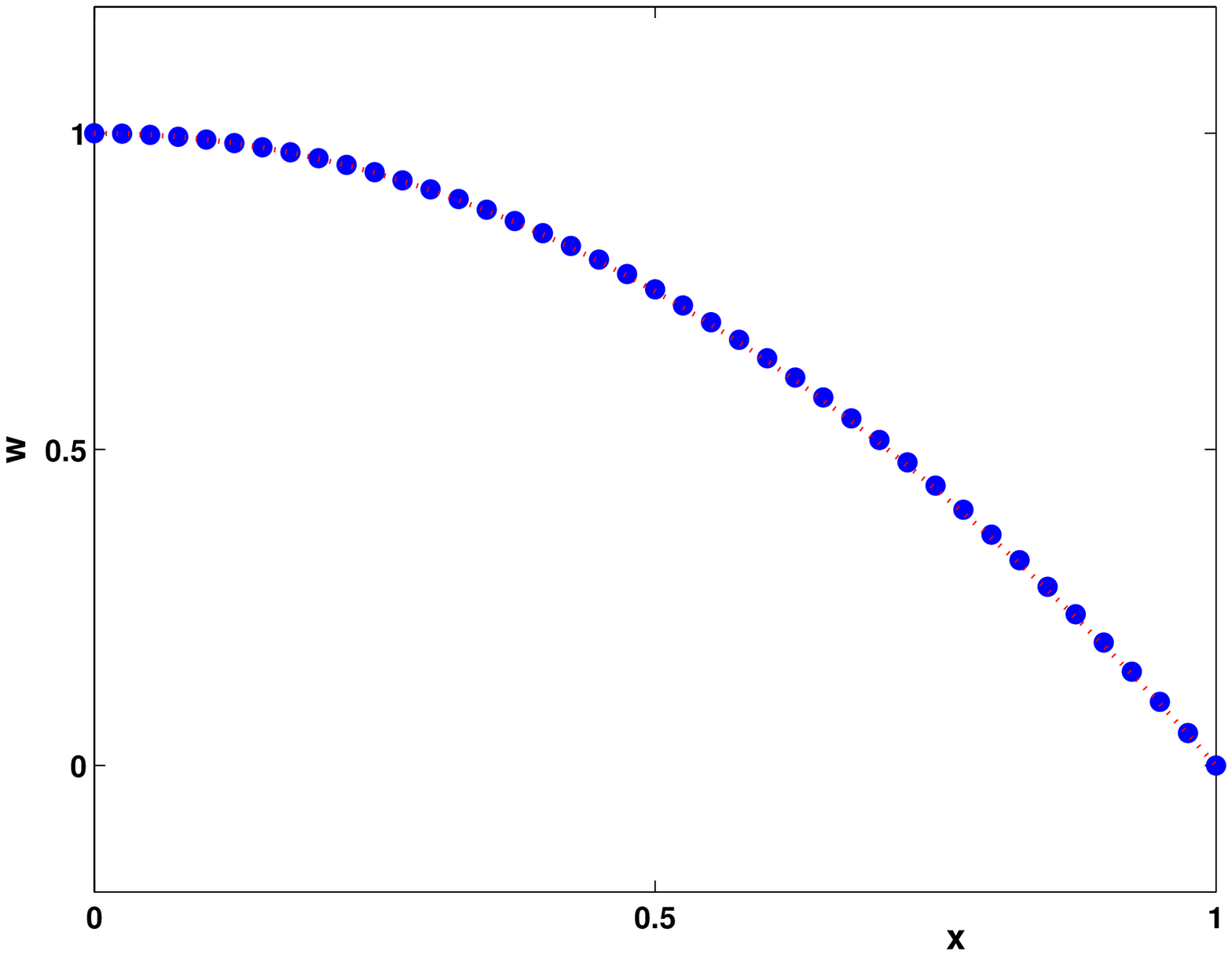}
\includegraphics[width=6cm]{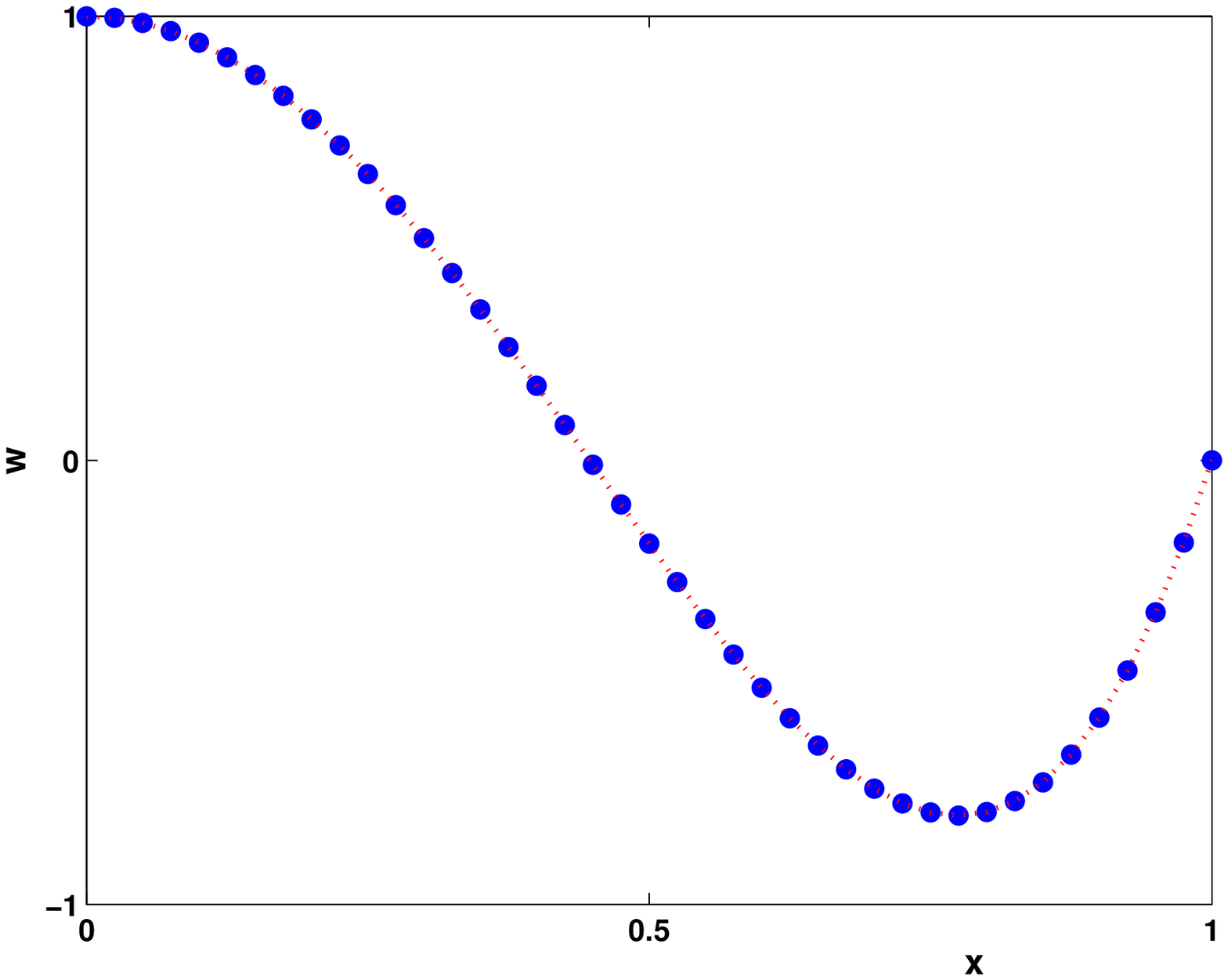}
\caption{The numerical approximation of even eigenfunctions
(dots) for $\eps = 10^{-4}$ near $\gamma_1 = 4$ (left) and
$\gamma_3 = 24$ (right) and the even polynomial solutions for $\eps
= 0$ (dashed line).}
\label{figure-2}
\end{minipage}\vspace{.55cm}\\
\begin{minipage}{12.5cm}
\includegraphics[width=6cm]{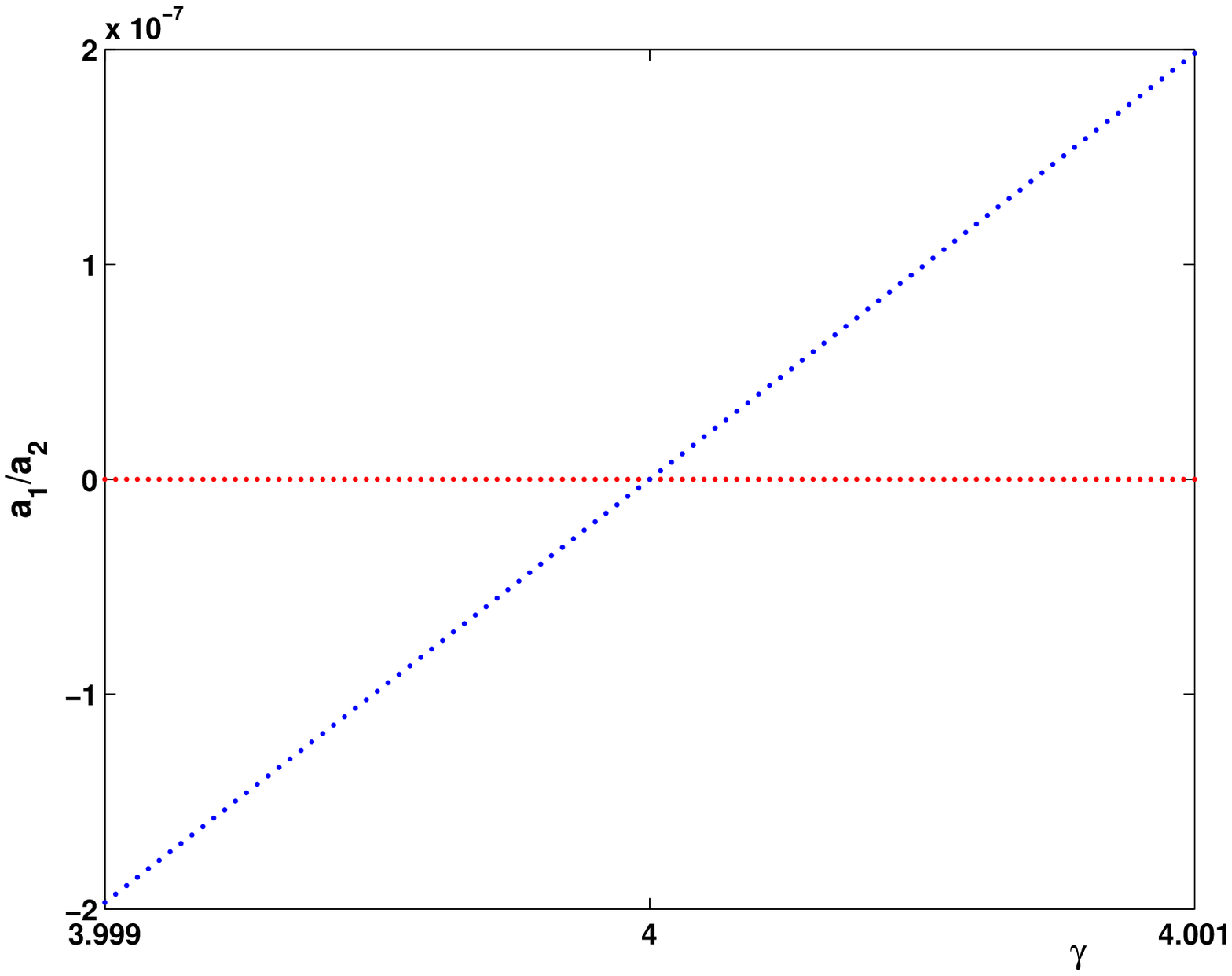}
\includegraphics[width=6cm]{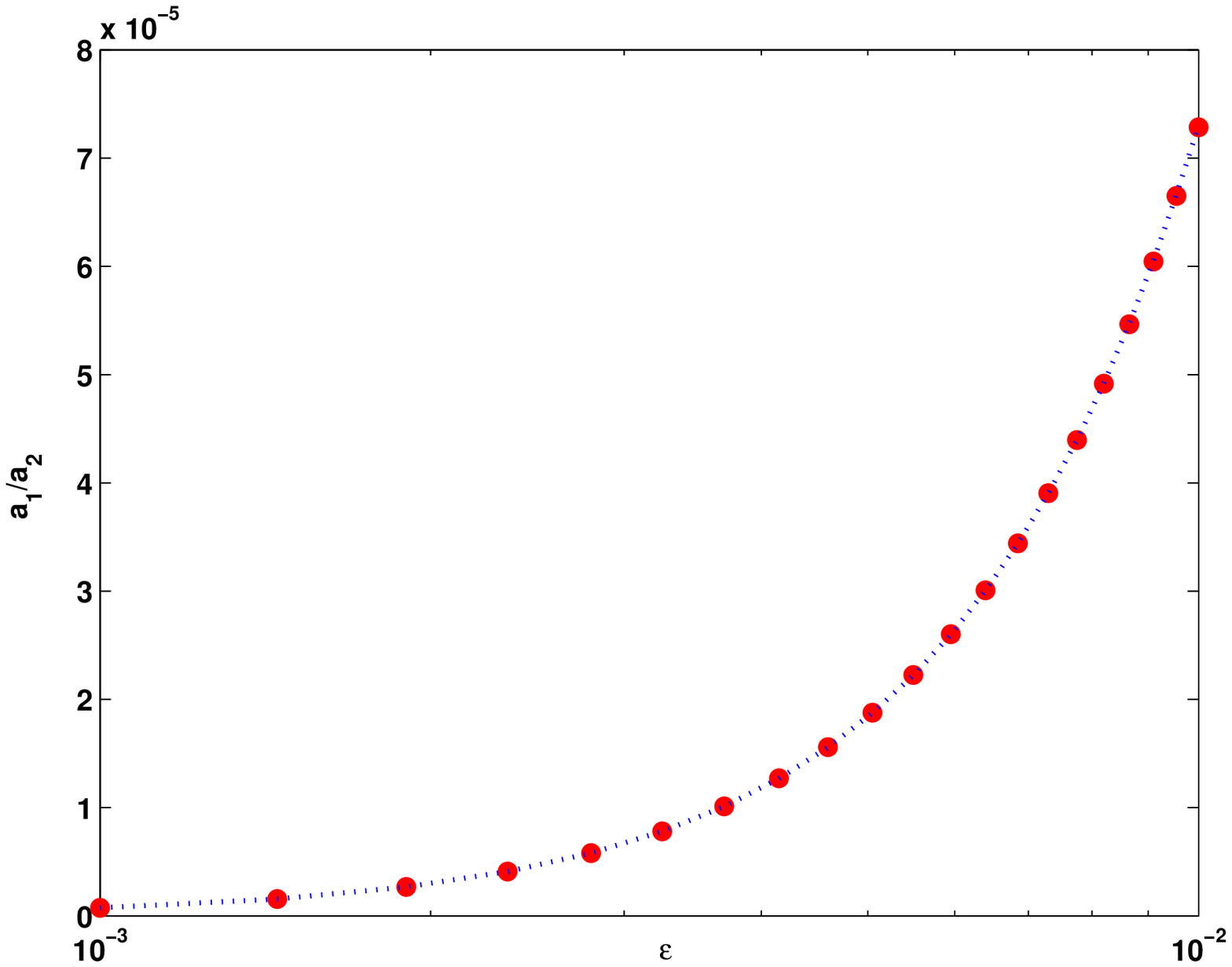}
\caption{The ratio $a_1/a_2$ for the two equations in the linear
  system versus $\gamma$ for $\eps = 10^{-6}$ near $\gamma = \gamma_1
  = 4$
(left) and for the solution of the linear system versus $\eps$
(right). The best
power fit is shown by dashed line.}
\label{figure-3}
\end{minipage}
\end{center}
\end{figure}
\appendix
\section{Appendix}
\subsection{Proof of Lemma \ref{lemma-resolvent}.}\label{AL-}
Let us denote by $\lambda_1(L_-^\eps)$ the smallest eigenvalue of
$L_-^\eps$. We first show that $\lambda_1(L_-^\eps) \gtrsim
1$. Let $\chi\in C_c^\infty(\R)$ be such that $0\leq\chi\leq 1$,
${\rm supp}(\chi) \subset (-3,3)$, and $\chi\equiv 1$ on $(-2,2)$. Let
$\delta>0$ to be fixed later (independently of $\eps$).
The Max-Min principle ensures that
\be\label{bigmin}
\lambda_1(L_-^\eps) & = & \underset{v\in D(L_-^\eps)}{\inf}\frac{<\le
  v,v>}{\|v\|_{L^2}^2}\nonumber\\
& = & \underset{v\in Q(L_-^\eps),
  \|v\|_{L^2}=1}{\inf}\left(\|v'\|_{L^2}^2+\int_{|x|>1}p_{\eps} |v|^2dx\right)
  = \min\{ \Lambda^{(1)},\Lambda^{(2)}\},
\ee
where
\be
\nonumber
\Lambda^{(1)} & = & \underset{\text{\scriptsize$\begin{array}{c}v\in Q(L_-^\eps), \|v\|_{L^2}=1,\\ \int_{|x|> 2}|v|^2dx\geq\delta\end{array}$}}
{\inf}\left(\|v'\|_{L^2}^2+ \int_{|x|>1} p_{\eps} |v|^2dx\right), \\ \nonumber
\Lambda^{(2)} & = & \underset{\text{\scriptsize$\begin{array}{c}v\in
  Q(L_-^\eps),  \|v\|_{L^2}=1,\\ \int_{|x|>
    2}|v|^2dx\leq\delta\end{array}$}}{\inf}\left(\|v'\|_{L^2}^2+
\int_{|x|>1} p_{\eps}|v|^2dx\right).
\ee
If $\|v\|_{L^2}=1$ and $\int_{|x|>
  2}|v|^2dx\geq \delta$, then
\be
&\int_{|x|>
  2}(x^2-1)|v|^2dx\geq 3\int_{|x|>
  2}|v|^2dx\geq3\delta.&\nonumber
\ee
Therefore for $\eps\leq 1$,
\be\label{min1}
\Lambda^{(1)} \geq\frac{3\delta}{\eps^2}\geq
3\delta.
\ee
On the other side, let us now take $v\in Q(L_-^\eps)$ such that $\|v\|_{L^2}=1$ and
$\int_{|x|> 2}|v|^2dx\leq \delta$. Then
\be\label{min2a}
\int_{|x|> 1}(x^2-1)|\chi v|^2dx\leq \int_{|x|>
  1}(x^2-1)|v|^2dx,
\ee
and since $\chi'(x)$ is supported in $\{2\leq |x|\leq 3\}$, we also have
in this case
\be\label{min2b}
\int_\R |(\chi v)'|^2dx & = & \int\left[\chi^2
  |v'|^2+2\chi\chi'vv'+{\chi'}^2|v|^2\right]dx \nonumber\\
&\leq &\|v'\|_{L^2(\R)}^2
+2\|v'\|_{L^2(\R)}\|\chi'\|_{L^\infty(\R)}\|v\|_{L^2(|x|> 2)}
+\|\chi'\|_{L^\infty(\R)}^2\|v\|_{L^2(|x|> 2)}^2\nonumber\\
&\leq & 2 \|v'\|_{L^2(\R)}^2+2\delta\|\chi'\|_{L^\infty(\R)}^2.
\ee
Next, since $\chi\equiv 1$ on $\{|x|\leq 2\}$,
\be\label{min2c}
\int_\R|\chi v|^2dx \geq \int_{-2}^2|v|^2dx \geq 1-\delta.
\ee
Thanks to (\ref{min2a}), (\ref{min2b}) and (\ref{min2c}), it turns out
that
\be\label{vchiv}
\lefteqn{\frac{\int_\R |(\chi v)'|^2dx+ \int_{|x|>1} p_{\eps}|\chi  v|^2dx}{\int_\R|\chi v|^2dx}}\nonumber\\
 &\leq &  \frac{2 \|v'\|_{L^2(\R)}^2+
   2\delta\|\chi'\|_{L^\infty(\R)}^2+ \int_{|x|> 1} p_{\eps}|v|^2dx}{1-\delta}.
\ee
As a result, using (\ref{vchiv}), since $(\chi v)_{|(-3,3)}\in
H_0^1(-3,3)$ for $v\in H^1(\R)$,
\be\label{min2}
\frac{2}{1-\delta}\Lambda^{(2)} &\geq &
 -\frac{2\delta\|\chi'\|_{L^\infty(\R)}^2}{1-\delta}+
 \underset{w\in H_0^1(-3,3)}{\inf}\frac{\int_{-3}^3|w'|^2dx+
\int_{|x|>1}p_{\eps} |w|^2dx}{\int_{-3}^3 |w|^2dx}\nonumber\\
& \geq &
-\frac{2\delta\|\chi'\|_{L^\infty(\R)}^2}{1-\delta}+\underset{w\in
  H_0^1(-3,3)}{\inf}\frac{\|w'\|_{L^2}^2}{\|w\|_{L^2}^2}=:R_\delta.
\ee
Thanks to the Poincar\'e inequality, we can now choose $\delta\in (0,1)$ sufficiently small such that
$R_\delta>0$. Then, according to (\ref{bigmin}), (\ref{min1}) and
(\ref{min2}),
\be\label{C1}
\lambda_1(L_-^\eps) & \geq & \min\left(3\delta,
\frac{(1-\delta)R_\delta}{2}\right),
\ee
which provides the estimate $\lambda_1(L_-^\eps)\gtrsim
1$ for $0 < \eps \leq 1$. The other estimate
$\lambda_1(L_-^\eps)\lesssim 1$ is a direct consequence of (\ref{bigmin}) and
of the Poincar\'e inequality. Indeed, the right hand side in
(\ref{bigmin}) is bounded from above by the infimum of the same
quantity, taken over $v\in L^2(\R)$ such that $v_{|(-1,1)}\in
H_0^1(-1,1)$ and $v_{|\{|x|>1\}}\equiv 0$.\hfill $\blacksquare$

\subsection{Proof of Lemma \ref{22}.}\label{AL+}
To prove Lemma \ref{22}, we use the following lemma.
\begin{lem}\label{lambda1abs}
For $\eps>0$,
$$
L^\eps:=-\partial_x^2+\frac{|x|}{\eps^2}
$$
defines a self-adjoint operator on $L^2(\R)$. The spectrum of $L^\eps$
is made of a sequence of strictly positive eigenvalues increasing to
infinity, and the smallest eigenvalue satisfies
$$
\lambda_1(L^\eps) \approx \eps^{-4/3}.
$$
\end{lem}

\begin{Proof}
The first assertion is straightforward. Thanks to the Max-Min
principle, $\lambda_1(L^\eps)$ is given by
$$
\lambda_1(L^\eps)=\underset{\text{\scriptsize$\begin{array}{c}v\in
      Q(L^\eps)\\
      \|v\|_{L^2}=1\end{array}$}}{\inf}\left(\|v'\|_{L^2}^2
      +\frac{1}{\eps^2}\int_\R |x| v^2 dx\right),
$$
where
$$
Q(L^\eps)=\{v\in H^1(\R) : \; |x|^{1/2}v\in L^2(\R)\}
$$
is the form domain of $L^\eps$.
If $v\in L^2(\R)$ and $\|v\|_{L^2}=1$, $v$ can be rewritten as
$v(x)=hw(h^2x)$, with $h>0$ and $w\in Q(L^\eps)$, with $\|w\|_{L^2}=1$
and $\|w'\|_{L^2}=1$. Moreover, $h$ and $w$ are uniquely defined this
way, and we have
$$
\|v'\|_{L^2}^2=h^4
$$
and
$$
\int_\R|x|v^2dx = h^{-2}\int_\R|x|w^2dx.
$$
Thus,
$$\lambda_1(L^\eps)=\underset{h>0}{\inf}\left(h^4+\eps^{-2}h^{-2}\beta\right)
=\left(\frac{1}{2^{2/3}}+2^{1/3}\right)\beta^{2/3}\eps^{-4/3},$$
where
$$
\beta:=\underset{\text{\scriptsize $\begin{array}{c}w\in Q(L^\eps)\\
    \|w\|_{L^2}=1, \|w'\|_{L^2}=1\end{array}$}}{\inf}\int_\R|x|w^2dx.
$$
The lemma follows if we prove that $\beta>0$. Let us assume by
contradiction that $\beta=0$. Let $\left(w_\delta\right)_{\delta>0}$
be a minimizing sequence, that is
$\|w_\delta\|_{L^2}=\|w_\delta'\|_{L^2}=1$ and
$\int_\R|x|w_\delta^2dx\to 0$ as $\delta\to 0$. Let
$\chi\in\mathcal{C}_c^\infty(\R)$ be such that $0\leq\chi\leq 1$,
$\text{supp}(\chi) \subset [-1,1]$, and $\chi\equiv 1$ on
$[-1/2,1/2]$. For $a>0$, we also define $\chi_a(x)=\chi(x/a)$, as well
as $w_{\delta,a}:=\chi_a w_\delta$. Thanks to the Poincar\'e
inequality, $\alpha:=\underset{v\in
  H_0^1(-1,1)}{\inf}\frac{\|v'\|_{L^2}}{\|v\|_{L^2}}>0$, and then
$\underset{v\in H_0^1(-a,a)}{\inf}\frac{\|v'\|_{L^2}}{\|v\|_{L^2}}=\frac{\alpha}{a}>0$. Thus,
\be\label{minor}
\|w_{\delta,a}'\|_{L^2(\R)}^2 & \geq &
\frac{\alpha^2}{a^2}\|w_{\delta,a}\|_{L^2(\R)}^2\nonumber\\
& \geq & \frac{\alpha^2}{a^2}\|w_{\delta}\|_{L^2(-\frac{a}{2},\frac{a}{2})}^2\nonumber\\
& = &
\frac{\alpha^2}{a^2}\left(\|w_{\delta}\|_{L^2(\R)}^2
  -\|w_{\delta}\|_{L^2(|x|>\frac{a}{2})}^2\right)\nonumber\\
& \geq &
\frac{\alpha^2}{a^2}\left(1-\frac{2}{a}\int_\R|x|w_{\delta}^2dx\right).
\ee
On the other side, since $\chi'(x)$ is supported in $\left\{ \frac{1}{2} \leq |x| \leq 1 \right\}$, we have
\be\label{major}
\|w_{\delta,a}'\|_{L^2}^2 & = &
\int_\R \left( (\chi'_a)^2 w_\delta^2  + 2 \chi_a \chi'_a w_\delta w_\delta'
  + \chi^2_a (w_\delta')^2 \right) dx \\\nonumber
& \leq &
\frac{\|\chi'\|_{L^\infty(\R)}^2}{a^2} \| w_\delta\|^2_{L^2(\frac{a}{2} < |x| <  a)}
+ \frac{2}{a}\|\chi'\|_{L^\infty(\R)}\|w_\delta\|_{L^2(\frac{a}{2}
  < |x| < a)} \|w_\delta'\|_{L^2(\R)}+\|w_\delta'\|_{L^2(\R)}^2.
\ee
According to the assumption, given $a>0$, we can find $\delta(a)$
sufficiently small such that
$$
\int_\R|x|w_{\delta(a)}^2dx\leq a^2.
$$
Then,
\be\label{tcheb}
\int_{\frac{a}{2} < |x| < a}w_\delta^2dx\leq
\int_{|x|>\frac{a}{2}}w_\delta^2dx
\leq\frac{2}{a}\int_\R|x|w_\delta^2dx\leq
2a.
\ee
It follows from (\ref{minor}), (\ref{major}) and (\ref{tcheb}) with
$\delta=\delta(a)$ that
$$\frac{\alpha^2}{a^2}(1-2a)\leq
\frac{2\|\chi'\|_{L^\infty(\R)}^2}{a}
+\frac{2^{3/2}\|\chi'\|_{L^\infty(\R)}}{a^{1/2}}+1.$$
Letting $a$ go to $0$ yields to a contradiction, which completes
the proof of the lemma.
\end{Proof}

Thanks to the Max-Min principle, we know that the lowest eigenvalue of
$L_+^\eps$ is given by
\be\label{minmax}
\lambda_1(L_+^\eps) & = & \underset{v\in
  Q(L_+^\eps)}{\inf}\frac{\|v'\|_{L^2}^2+
\int_\R q_\eps|v|^2dx}{\|v\|_{L^2}^2},
\ee
where
$$
Q(L_+^\eps)=\{v\in H^1(\R) : \; xv\in L^2(\R) \}
$$
is the form domain of $L_+^\eps$. The statement of Lemma \ref{22} is
equivalent to $\lambda_1(L_+^\eps)\approx \eps^{-4/3}$. We first prove
the upper bound on $\lambda_1(L_+^\eps)$. Let us define
$v_\eps$ on $\R$ as
$$v_\eps(x):=\left\{\begin{array}{ll}
x-1+\eps^{2/3} \;\; & \text{for} \;\; 1-\eps^{2/3} <
x < 1,\\
-(x-1-\eps^{2/3}) \;\; & \text{for} \;\; 1 <
x < 1+\eps^{2/3},\\
0 & \text{elsewhere}.
\end{array}\right.$$
and denote $q(x): =\eps^2q_\eps(x) = 2(1-x^2) \mathbf{1}_{\{|x|<1\}} + (x^2-1)\mathbf{1}_{\{|x|>1\}}$.
Then
$$
\|v_\eps'\|_{L^2(\R)}^2=2\eps^{2/3}, \quad \|v_\eps\|_{L^2(\R)}^2=\frac{2\eps^2}{3},
$$
and since $q(x)\leq 4|x-1|$ for $|x-1|\leq 1$,
$$\int_\R q_\eps|v_\eps|^2dx\leq
\frac{4}{\eps^2}\int_{1-\eps^{2/3}}^{1+\eps^{2/3}}|1-x|v_\eps^2dx
=\frac{2\eps^{2/3}}{3}.$$
As a result,
$$\lambda_1(L_+^\eps)  \leq  \frac
{2\eps^{2/3}+2\eps^{2/3}/3}{2\eps^2/3}=4\eps^{-4/3}.$$
It remains to find a bound on $\lambda_1(L_+^\eps)$ from below. Let us first introduce
the two intervals
$$D_+:= \left\{ x\geq 0, \;\; q(x)\leq \frac{1}{2} \right\}=
\left[\frac{\sqrt{3}}{2},\sqrt{\frac{3}{2}}\right], \quad
D_- := \left\{x\leq 0, \;\; q(x)\leq \frac{1}{2} \right\}=-D_+,$$
and denote $D:=D_+\cup D_-$. If $v\in Q(L_+^\eps)$, $\|v\|_{L^2}=1$ and
$\int_{D}|v|^2dx\leq 1-\eps^{1/2}$, then
$$
\int_\R q|v|^2dx\geq\int_{\R\string\ D}q|v|^2dx
\geq\frac{1}{2}\int_{\R\string\ D}|v|^2dx\geq\frac{\eps^{1/2}}{2}>4\eps^{2/3}
$$
for sufficiently small $\eps > 0$.
As a result, thanks to (\ref{minmax}) and the upper bound on
$\lambda_1(L_+^\eps)$, we deduce that
\be\label{minmaxrestricted}
\lambda_1(L_+^\eps) & = &
\underset{\text{\scriptsize$\begin{array}{c}v\in Q(L_+^\eps),\\
  \|v\|_{L^2}=1,\\
  \int_{D}|v|^2dx\geq1-\eps^{1/2}\end{array}$}}{\inf}\left[\|v'\|_{L^2}^2+
\int_\R q_\eps|v|^2dx\right].
\ee
From now on, we assume that $v\in Q(L_+^\eps)$,
 $ \|v\|_{L^2}=1$ and $\int_{D}|v|^2dx\geq1-\eps^{1/2}$.
Let $\chi\in \mathcal{C}_c^\infty (\R)$ be such that $0\leq\chi\leq 1$,
${\rm supp}(\chi) \subset [-1/2,1/2] \subset\R\string\ D$, and $\chi(x)\equiv 1$ for
$x\in[-1/4,1/4]$. We also define $\rho:=1-\chi$. In particular,
$\rho\equiv 1$ on $D$, thus
\be\label{x}
\|\rho v\|_{L^2}^2\geq \int_{D}|v|^2dx\geq 1-\eps^{1/2}, \quad
\int_\R q |\rho v|^2dx\leq\int_\R q|v|^2dx,\ee
and since $\rho'$ is supported in $\R\string\ D$, for some $C>0$, we have
\be\label{xxx}
\int_\R|(\rho v)'|^2dx & \leq & \|\rho'\|_{L^\infty(\R)}^2 \|v\|^2_{L^2(\R \string\ D)}
+\|v'\|_{L^2(\R)}^2
+2\|\rho\|_{L^\infty(\R)}\|\rho'\|_{L^\infty(\R)}\|v'\|_{L^2(\R)} \|v
\|_{L^2(\R\string\ D)}\nonumber\\
&\leq &C\eps^{1/2}+\|v'\|_{L^2(\R)}^2+C\eps^{1/4}\|v'\|_{L^2(\R)}\nonumber\\
&\leq &2(\|v'\|_{L^2(\R)}^2 + C\eps^{1/2}).
\ee
Therefore, combining (\ref{x}) and (\ref{xxx}), we obtain,
for $\eps$ sufficiently small,
\be\label{lambda1rho}
\frac{\|(\rho v)'\|_{L^2}^2+
\int_\R q_\eps|\rho v|^2dx}{\|\rho
v\|_{L^2}^2}
 & \leq &
\frac{2(\|v'\|_{L^2(\R)}^2 + C\eps^{1/2})+
\int_\R q_\eps|v|^2dx}{1-\eps^{1/2}}\nonumber\\
 & \leq & 2(\|v'\|_{L^2}^2+
\int_\R q_\eps|v|^2dx)+2C\eps^{1/2}.
\ee
Taking the infimum in $v$ in (\ref{lambda1rho}), we infer thanks to
(\ref{minmaxrestricted}) that
\be
2\lambda_1(L_+^\eps)+2C\eps^{1/2}
&\geq &\underset{\text{\scriptsize$\begin{array}{c}v\in Q(L_+^\eps),\\
  \|v\|_{L^2}=1,\\
  \int_{D}|v|^2dx\geq1-\eps^{1/2}\end{array}$}}{\inf}\frac{\|(\rho
v)'\|_{L^2}^2+
\int_\R q_\eps|\rho v|^2dx}{\|\rho
v\|_{L^2}^2}.
\ee
Therefore, since $q(x)\geq 2|x-1|$ for $x\geq 0$ and
$q(x)\geq 2|x+1|$ for $x\leq 0$, and decomposing $\rho
v=v_1+v_2$ with $v_1$ supported in $(-\infty,-1/4]$ and  $v_2$
supported in $[1/4,+\infty)$, we have
\be
& \phantom{t} & 2\lambda_1(L_+^\eps)+2C\eps^{1/2}\nonumber\\
 & \geq &
\!\!\!\!\!\underset{\text{\scriptsize$\begin{array}{c}v_1,v_2\in
      Q(L_+^\eps),\\
\text{supp}(v_1) \subset (-\infty,-1/4],\\  \text{supp}(v_2) \subset
[1/4,+\infty)\end{array}$}}{\inf}\frac{\|v_1'\|_{L^2}^2+
\int_\R q_\eps|v_1|^2dx+\|v_2'\|_{L^2}^2+
\int_\R q_\eps|v_2|^2dx}{\|v_1\|_{L^2}^2+\|v_2\|_{L^2}^2}\nonumber\\
& \geq &
\!\!\!\!\!\underset{\text{\scriptsize$\begin{array}{c}v_1,v_2\in
      Q(L_+^\eps),\\
 \text{supp}(v_1) \subset (-\infty,-1/4],\\  \text{supp}(v_2) \subset
 [1/4,+\infty)\end{array}$}}{\inf}\frac{\|v_1'\|_{L^2}^2+
\frac{2}{\eps^2}\int_\R|x+1||v_1|^2dx+\|v_2'\|_{L^2}^2+
\frac{2}{\eps^2}\int_\R|x-1||v_2|^2dx}{\|v_1\|_{L^2}^2+\|v_2\|_{L^2}^2}
\nonumber\\
& \geq & \underset{\text{\scriptsize$\begin{array}{c}v_1,v_2\in
    Q(L_+^\eps)\end{array}$}}{\inf}\frac{\|v_1'\|_{L^2}^2+
\frac{2}{\eps^2}\int_\R|x+1||v_1|^2dx+\|v_2'\|_{L^2}^2+
\frac{2}{\eps^2}\int_\R|x-1||v_2|^2dx}{\|v_1\|_{L^2}^2+\|v_2\|_{L^2}^2}
\nonumber\\
& = & \underset{\text{\scriptsize$\begin{array}{c}v_1,v_2\in
    Q(L_+^\eps)\end{array}$}}{\inf}\frac{\|v_1'\|_{L^2}^2+
\frac{2}{\eps^2}\int_\R|x||v_1|^2dx+\|v_2'\|_{L^2}^2+
\frac{2}{\eps^2}\int_\R|x||v_2|^2dx}{\|v_1\|_{L^2}^2+\|v_2\|_{L^2}^2}
\nonumber\\
& = & \underset{\text{\scriptsize$\begin{array}{c}v_1,v_2\in
    Q(L_+^\eps)\\\|v_1\|_{L^2}\leq\|v_2\|_{L^2}\end{array}$}}{\inf}
\frac{\|v_1'\|_{L^2}^2+
\frac{2}{\eps^2}\int_\R|x||v_1|^2dx+\|v_2'\|_{L^2}^2+
\frac{2}{\eps^2}\int_\R|x||v_2|^2dx}{\|v_1\|_{L^2}^2+\|v_2\|_{L^2}^2}
\nonumber\\
& \geq & \underset{\text{\scriptsize$\begin{array}{c}v_1,v_2\in
    Q(L_+^\eps)\\\|v_1\|_{L^2}\leq\|v_2\|_{L^2}=1\end{array}$}}{\inf}
\left(\frac{\|v_1'\|_{L^2}^2+
\frac{2}{\eps^2}\int_\R|x||v_1|^2dx}{2}+\frac{\|v_2'\|_{L^2}^2+
\frac{2}{\eps^2}\int_\R|x||v_2|^2dx}{2}\right)\nonumber\\
& \geq & \frac{1}{2}\underset{\text{\scriptsize$\begin{array}{c}v_2\in
    Q(L_+^\eps)\\ \|v_2\|_{L^2}=1\end{array}$}}{\inf}\left(\|v_2'\|_{L^2}^2+
\frac{2}{\eps^2}\int_\R|x||v_2|^2dx\right)
\geq\frac{1}{2}\lambda_1(L^\eps)\gtrsim
\eps^{-4/3},
\ee
where we have used Lemma \ref{lambda1abs} in the last
estimation. \hfill $\blacksquare$

\subsection{Proofs of Lemmas \ref{airy} and
  \ref{airy-back}}\label{Aairy}
\paragraph{Proof of Lemma \ref{airy}.\ \ }
The proof of Lemma \ref{airy} relies on WKB approximation techniques,
  explained for instance in \cite{M}.
If we define $w(x):=\psi(1-x)$, it is equivalent for $\psi$ to solve
  (\ref{eqinn}) or for $w$ to solve
\be\label{eqw}
\eps^2 w''-2x(2-x)w=0,\quad x\in \left(0,\frac{3}{2}\right).
\ee
In the new
variable $\xi=\xi(x) :=\left(\frac{3}{2}\int_0^x\sqrt{2t(2-t)}dt\right)^{2/3}$,
it is equivalent for $w$ to solve (\ref{eqw}) or for
$v(\xi):=\frac{w(x)}{a(x)}$ to solve
\be\label{eqv}
\eps^2\frac{d^2v}{d\xi^2}-\xi v=\eps^2\delta(\xi)v, \quad \xi\in
(0,\xi_0),
\ee
where $\xi_0:=\xi(3/2)$, $a(x):= \left(\xi'(x) \right)^{-1/2}$,
and $\delta(\xi):=-a''(x)a^3(x)$. Next, we look
for $v$ in the form
$v(\xi)={\rm Ai}\left(\frac{\xi}{\eps^{2/3}}\right)(1+Q(\xi))$. Using that
${\rm Ai}(\xi/\eps^{2/3})$ solves the homogeneous equation
$$\eps^2\frac{d^2v}{d\xi^2}-\xi v=0,$$
it is equivalent for $v$ to solve (\ref{eqv}) or for $Q$ to solve
\be\label{eqQ}
\frac{d}{d\xi}\left[{\rm Ai}\left(\frac{\xi}{\eps^{2/3}}\right)^2Q'(\xi)\right]
& = & \delta(\xi){\rm Ai}\left(\frac{\xi}{\eps^{2/3}}\right)^2(1+Q(\xi)),
\quad \xi\in(0,\xi_0).
\ee
By integration, (\ref{eqQ}) is equivalent to the integral equation
\be\label{defF}
Q(\xi)=F(Q)(\xi):=\int_\xi^{\xi_0}\int_\xi^\eta\frac{{\rm Ai}\left(\frac{\eta}{\eps^{2/3}}\right)^2}{{\rm Ai}\left(\frac{t}{\eps^{2/3}}\right)^2}dt\delta(\eta)(1+Q(\eta))d\eta,
\ee
where $F$ maps $\mathcal{C}^0([0,\xi_0])$ into itself. A change of variable provides
$$
F(Q)(\xi)=\eps^{2/3}\int_\xi^{\xi_0}
\left(\int_{\xi/\eps^{2/3}}^{\eta/\eps^{2/3}}{\rm Ai}(u)^{-2}du
  {\rm Ai}\left(\frac{\eta}{\eps^{2/3}}\right)^2\right)
\delta(\eta)(1+Q(\eta))d\eta.$$
Thanks to the asymptotic behavior (\ref{daaibi}), $f(x):=\int_0^x{\rm Ai}(y)^{-2}dy {\rm Ai}(x)^2\sim
\frac{1}{2\sqrt{x}}$ as $x\to+\infty$. In particular, $f$ is bounded on
$\R_+$. We deduce that for any $\xi\in (0,\xi_0)$,
$$\left|(F(Q))(\xi)\right|\leq\eps^{2/3}\|f\|_{L^\infty(\R_+)}
\int_\xi^{\xi_0}|\delta(\eta)|d\eta(1+\|Q\|_{L^\infty(0,\xi_0)}).$$
Since $\delta$ is clearly continuous on $(0,\xi_0]$ and
$$
\delta(\xi(x)) \longrightarrow \frac{9\cdot2^{2/3}}{560} \quad \mbox{as} \quad x \to 0,
$$
we deduce $\delta\in L^1(0,\xi_0)$. Thus, if $Q\in
\mathcal{C}^0([0,\xi_0])$, then
\be\label{ballstab}
\|F(Q)\|_{L^\infty(0,\xi_0)}\leq\eps^{2/3}\|f\|_{L^\infty(\R_+)}
  \|\delta\|_{L^1(0,\xi_0)}(1+\|Q\|_{L^\infty(0,\xi_0)}).
\ee
Moreover, if $Q_1,Q_2\in \mathcal{C}^0([0,\xi_0])$, we get similarly
\be\label{contrac}
\|F(Q_1)-F(Q_2)\|_{L^\infty(0,\xi_0)}\leq\eps^{2/3}
\|f\|_{L^\infty(\R_+)}\|\delta\|_{L^1(0,\xi_0)}\|Q_1-Q_2\|_{L^\infty(0,\xi_0)}.
\ee
From (\ref{ballstab}) and (\ref{contrac}) we infer that, if we take
$C:=2\|f\|_{L^\infty(\R_+)}
  \|\delta\|_{L^1(0,\xi_0)}$, for $\eps$
sufficiently small (namely $\eps^{2/3}<1/2C$), $F$ maps the ball of radius $C\eps^{2/3}$
in $\mathcal{C}^0([0,\xi_0])$ into itself, and is a contraction on that
ball. Then, $F$ has a unique fixed point $Q$ such that
$\|Q\|_{L^\infty(0,\xi_0)}\leq C\eps^{2/3}$. Such a fixed point
of $F$ gives a $\mathcal{C}^2$ solution of (\ref{eqQ}) on $(0,\xi_0)$. Defining $Q_A^\eps$ as
$Q_A^\eps(x):=Q(\xi(1-x))$ and applying the sequence of
substitutions backwards, we found a solution $\psi_A^\eps$ of the system
(\ref{eqinn}) with the required bounds.

For the existence of the solution $\psi_B^\eps$, we proceed
similarly. Namely, we look for a solution to (\ref{eqv}) in the
form $v(\xi)={\rm Bi}\left(\frac{\xi}{\eps^{2/3}}\right)(1+Q(\xi))$. It is equivalent for $v$ to solve (\ref{eqv}) or for $Q$ to solve
\be\label{eqQB}
\frac{d}{d\xi}\left[{\rm Bi}\left(\frac{\xi}{\eps^{2/3}}\right)^2Q'(\xi)\right]
& = & \delta(\xi){\rm Bi}\left(\frac{\xi}{\eps^{2/3}}\right)^2(1+Q(\xi)),
\quad \xi\in(0,\xi_0).
\ee
Since $g(x):={\rm Bi}(x)^2\int_x^{+\infty} {\rm Bi}(u)^{-2}du\sim
\frac{1}{2\sqrt{x}}$ as $x\to+\infty$ thanks to the asymptotic behavior (\ref{daaibi}) again, $g$ is
bounded on $\R_+$. It enables us to prove the existence of a fixed
point to the functional $G:\mathcal{C}^0([0,\xi_0]) \mapsto
  \mathcal{C}^0([0,\xi_0])$ defined by
\be\nonumber
G(Q)(\xi):=\int_0^{\xi}\int_\eta^\xi\frac{{\rm Bi}\left(\frac{\eta}{\eps^{2/3}}\right)^2}{{\rm Bi}\left(\frac{t}{\eps^{2/3}}\right)^2}dt\delta(\eta)(1+Q(\eta))d\eta,
\ee
similarly to what has been done for $F$.

The linear independence of $\psi_A^\eps$ and $\psi_B^\eps$ follows
from the linear independence of functions ${\rm Ai}$ and ${\rm
  Bi}$.\hfill $\blacksquare$

\paragraph{Proof of Lemma \ref{airy-back}.\ \ }
The proof is very similar to that of Lemma \ref{airy}, so that we will
only point out the differences. It is equivalent for $\psi$ to solve
(\ref{eqext}) on $(\sqrt{1+\eps\nu},+\infty)$ or for
$w(x):=\psi(\sqrt{1+\eps\nu}(1+x))$ to solve 
\be\label{eqww}
\tilde{\eps}^2w''(x)-x(x+2)w(x)=0
\ee
on $\R^+$, where $\tilde{\eps}:=\eps/\sqrt{1+\eps\nu}$. We look for
$w$ in the form $w(x)=a(x)v(\xi(x))$, where
$\xi(x)=\left(\frac{3}{2}\int_0^x\sqrt{t(2+t)}dt\right)^{2/3}$ and
$a(x)= (\xi'(x))^{-1/2}$. Then, it is equivalent for $w$ to solve (\ref{eqww})
on $\R^+$ or for $v$ to solve
\be\label{eqvv}
\tilde{\eps}^2v''(\xi)-\xi v(\xi)=\tilde{\eps}^2\delta(\xi)v(\xi)
\ee
on $\R^+$, where the function $\xi\mapsto \delta(\xi)$ is defined by
$\delta(\xi(x))=-a''(x)a(x)^3$. Since
$a\in\mathcal{C}^\infty([0,+\infty))$ and
$\delta(\xi)\underset{\xi\to\infty}\sim 7\xi^{-2}/1024$, we deduce
that $\delta\in L^1(\R^+)$. Then, the existence of
$Q\in\mathcal{C}_b^0(\R^+)$ with
$\|Q\|_{L^\infty(\R^+)}\lesssim\eps^{2/3}$, such that $v(\xi)={\rm
  Ai}(\xi/\tilde{\eps}^{2/3})(1+Q(\xi))$ solves (\ref{eqvv}), is
established like in the 
proof of Lemma \ref{airy}, applying the fixed point theorem to the
functional $F$ defined in (\ref{defF}), with
$\xi_0=+\infty$. Therefore, we obtain $\psi_A^{\nu,\eps}$. The
expression for $\psi_B^{\nu,\eps}$ is obtained similarly as in Lemma \ref{airy}. Next,
the expression of $\psi_A^{\nu,\eps}(x)$ at $x=\sqrt{1+\eps\nu}$
yields
\be\label{exp-psi-sqrt}
\psi_A^{\nu,\eps}(\sqrt{1+\eps\nu})=a(0){\rm
  Ai}(0)(1+Q_A^{\nu,\eps}(0))=a(0){\rm
  Ai}(0)(1+\mathcal{O}(\eps^{2/3})),
\ee
and similarly
\be\label{exp-psi'-sqrt}
\lefteqn{(\psi_A^{\nu,\eps})'(\sqrt{1+\eps\nu})}\nonumber\\ 
& = & a'(0){\rm
  Ai}(0)(1+\mathcal{O}(\eps^{2/3}))+a(0)\xi'(0){\rm
  Ai}'(0)\eps^{-2/3}(1+\mathcal{O}(\eps^{2/3}))
+a(0){\rm
  Ai}(0)\xi'(0)(Q_A^{\nu,\eps})'(0)\nonumber\\
&= & a(0)\xi'(0){\rm
  Ai}'(0)\eps^{-2/3}(1+\mathcal{O}(\eps^{2/3})),
\ee
where we have used that
$$\left|(Q_A^{\nu,\eps})'(0)\right|=\left|{\rm Ai}(0)^{-2}\int_0^{+\infty}{\rm
    Ai}(\eta/\tilde{\eps}^{2/3})^2\delta(\eta)(1+Q_A^{\nu,\eps}(\eta))d\eta \right|\leq \|\delta\|_{L^1(\R^+)}(1+{\cal O}(\eps^{2/3}))\lesssim 1.$$
At this point, the function $\psi_A^{\nu,\eps}$ has been defined on
the interval $[\sqrt{1+\eps\nu},+\infty)$. In the case $\nu>0$, we
extend into a solution of (\ref{eqext}) on the interval $[1,+\infty)$,
thanks to the Cauchy-Lipshitz Theorem. We denote
$I_\nu=[\sqrt{1+\eps\nu},1]$ if $\nu<0$, $I_\nu=[1,\sqrt{1+\eps\nu}]$
if $\nu>0$. Then, for any sign of $\nu$, we have
\be\label{loop1}
\left|\psi_A^{\nu,\eps}(1)-\psi_A^{\nu,\eps}(\sqrt{1+\eps\nu})\right|
&\lesssim & \eps\|(\psi_A^{\nu,\eps})'\|_{L^\infty(I_\nu)}\lesssim 
\eps\left|(\psi_A^{\nu,\eps})'(\sqrt{1+\eps\nu})\right|+\eps^2\|(\psi_A^{\nu,\eps})''\|_{L^\infty(I_\nu)}\nonumber\\
&\lesssim &\eps^{1/3}+\eps\|\psi_A^{\nu,\eps}\|_{L^\infty(I_\nu)}
\ee
and, thanks to (\ref{loop1})
\be
\|\psi_A^{\nu,\eps}\|_{L^\infty(I_\nu)} &\lesssim
&\left|\psi_A^{\nu,\eps}(\sqrt{1+\eps\nu})\right|+\eps\|(\psi_A^{\nu,\eps})'\|_{L^\infty(I_\nu)}\nonumber\\
&\lesssim
&1+\eps\|\psi_A^{\nu,\eps}\|_{L^\infty(I_\nu)},\nonumber
\ee
thus 
\be\label{loop}
\|\psi_A^{\nu,\eps}\|_{L^\infty(I_\nu)}&\lesssim &1.
\ee
From (\ref{loop1}), (\ref{loop}) and (\ref{exp-psi-sqrt}) it follows
that
\be\label{exp-psi-1}
\psi_A^{\nu,\eps}(1)=a(0){\rm Ai}(0)(1+{\cal O}(\eps^{1/3}).
\ee
Similarly,
\be\nonumber
\left|(\psi_A^{\nu,\eps})'(1)-(\psi_A^{\nu,\eps})'(\sqrt{1+\eps\nu})\right|
&\lesssim &
\eps\|(\psi_A^{\nu,\eps})''\|_{L^\infty(I_\nu)}\lesssim\|\psi_A^{\nu,\eps}\|_{L^\infty(I_\nu)}\lesssim
1,
\ee
and therefore thanks to (\ref{exp-psi'-sqrt}), we get
\be\label{exp-psi'-1}
(\psi_A^{\nu,\eps})'(1)=a(0)\xi'(0){\rm Ai}'(0)\eps^{-2/3}(1+{\cal O}(\eps^{2/3}).
\ee
The limit (\ref{asymptotic-limit}) follows from (\ref{exp-psi-1}) and
(\ref{exp-psi'-1}), since $\xi'(0)=2^{1/3}$, and because
$$
{\rm Ai}(0) = \frac{1}{3^{2/3} \Gamma(2/3)}, \quad
{\rm Ai}'(0) = -\frac{1}{3^{1/3} \Gamma(1/3)}.
$$
Notice that all the estimates we made in this proof are uniform in $\nu\in K$, 
for any fixed compact subset $K\subset \R$.
\hfill $\blacksquare$

\subsection{Proof of Lemma \ref{extension}}\label{Aextension}
If $f\in X'$ and $\phi\in D(L_X)$, we have
$$|\left<L_{X'}f,\phi\right>_{D(L_X)',D(L_X)}|\leq \|f\|_{X'}\|L_X\phi\|_X\leq
\|f\|_{X'}\|\phi\|_{D(L_X)},$$
which provides the continuity of $L_X$. If $f\in X'$ and $L_{X'}f=0$,
then for every $\phi\in D(L_X)$,
$\left<f|L_X\phi\right>_{X',X}=0$. We can apply this to $\phi=L_X^{-1}x$,
for any $x\in X$ and we get that $\left<f,x\right>_{X',X}=0$ for every $x\in
X$. Therefore $f=0$ and $L_{X'}$ is injective. Let us next prove the
surjectivity of $L_{X'}$. Let $T\in
D(L_X)'$. $f:x\mapsto \left<T,L_X^{-1}x\right>_{D(L_X)',D(L_X)}$
clearly defines a continuous
linear form on $X$, and for every $\phi\in D(L_X)$,
$$\left<L_{X'}f,\phi\right>_{D(L_X)',D(L_X)}=\left<f,L_X\phi\right>_{X',X}
=\left<T,L_X^{-1}L_X\phi\right>_{D(L_X)',D(L_X)}=\left<T,\phi\right>_{D(L_X)',D(L_X)},$$
which means that $T=L_{X'}f$. Moreover, the application
$L_{X'}^{-1}:D(L_X)'\mapsto X'$ we have just defined is
continuous. Indeed, if $T\in
D(L_X)'$ and $x\in X$,
\be
|\left<L_{X'}^{-1}T,x\right>_{X',X}| & = &
|\left<T,L_{X}^{-1}x\right>_{D(L_X)',D(L_X)}|\nonumber\\
&\leq & \|T\|_{D(L_X)'}\|L_{X}^{-1}x\|_{D(L_X)}\nonumber\\
&\lesssim & \|T\|_{D(L_X)'}(\|x\|_X+\|L^{-1}x\|_{D(L)})\nonumber\\
&\lesssim & \|T\|_{D(L_X)'}\|x\|_X, \nonumber
\ee
where we have used the continuous embeddings $D(L)\subset X\subset H$, as well
as the continuity of $L^{-1}\in \mathcal{L}(H)$.
Finally, we show that $L_{X'}$ is an extension of $L$. Here, we
classically identify elements of $H$ to elements of $X'$
(resp. $D(L_X)'$) as follows: if $f\in H$, $x\in X$ (resp. $T\in H$,
$\phi\in D(L_X)$), $\left<f,x\right>_{X',X}=(f|\overline{x})$
(resp.
$\left<T,\phi\right>_{D(L_X)',D(L_X)}=(T|\overline{\phi})$), where
$(\cdot|\cdot)$ denotes the scalar product in $H$. Thus, if
$f\in D(L)\subset X\subset X'$,
$$\left<L_{X'}f,\phi\right>_{D(L_X)',D(L_X)}
=\left<f,L\phi\right>_{X',X}=\left(f|\overline{L\phi}\right)
=\left(Lf|\overline{\phi}\right)=\left<Lf,\phi\right>_{D(L_X)',D(L_X)},$$
which means that $L_{X'}f=Lf$. \hfill $\blacksquare$

\subsection{Numerical methods for inner solutions}\label{Anumerics}

We rewrite the fourth--order equation (\ref{4-ode}) on $[0,1]$ in the form
$$
w''(x) = v(x), \quad \eps^2 v''(x) - 2(1-x^2) v(x) = \gamma w(x), \quad 0 < x < 1.
$$
Using the finite-difference approximation with the second--order
central differences \cite{GP}, the system of differential equations is
converted into the system of algebraic
equations
$$
A_1 {\bf w} = {\bf v}, \quad A_2 {\bf v} = \gamma {\bf w},
$$
where ${\bf v},{\bf w}$ are $n$-vectors of $v(x)$,$w(x)$ represented on a discrete grid
$\{ x_k \}_{k = 0}^{n-1} \subset [0,1]$ with $x_0 = 0$ and $x_0 < x_1 < ... < x_{n-1} < x_n = 1$.
Using an equally spaced grid with step size $h = 1/n$ and incorporating boundary conditions $w'(0) = 0$, $v'(0) = 0$,
we obtain $n \times n$ matrices $A_1$ and $A_2$ in the explicit form, where
$$
A_1 = \frac{1}{h^2} \left[ \begin{array}{cccccc} -2 & 2 & 0 & ... & 0 & 0 \\ 1 & -2 & 1 & ... & 0 & 0 \\
0 & 1 & -2 & ... & 0 & 0 \\ \vdots & \vdots & \vdots & \vdots & \vdots & \vdots  \\ 0 & 0 & 0 & ... & 1 & -2 \end{array} \right]
$$
and $A_2 = \eps^2 A_1 - 2 {\rm diag}(1-x^2)$. For the first solution
$w_1(x)$, with $w_n=1$ and $v_n=0$, we obtain solutions
of the finite-difference equations in the form
$$
{\bf w} = -\frac{1}{h^2} \left( A_1 - \gamma A_2^{-1} \right)^{-1} {\bf e}_n, \quad
{\bf v} = \gamma A_2^{-1} {\bf w},
$$
where ${\bf e}_n$ is the $n^{\rm th}$ unit vector in $\mathbb{R}^n$. For the second solution
$w_2(x)$, with  $w_n=0$ and $v_n=1$, the finite-difference equations are solved in the form
$$
{\bf w} = -\frac{\eps^2}{h^2} \left( A_1 - \gamma A_2^{-1} \right)^{-1} A_2^{-1} {\bf e}_n, \quad
{\bf v} = \gamma A_2^{-1} {\bf w} -\frac{\eps^2}{h^2} A_2^{-1} {\bf e}_n.
$$
The values of $w'(1)$ and $w'''(1)$ are obtained from the three-point finite-difference approximations
$$
w'(1) \approx \frac{3 w_n - 4 w_{n-1} + w_{n-2}}{2h}, \quad w'''(1) \approx \frac{3 v_n - 4 v_{n-1} + v_{n-2}}{2h},
$$
which preserves the second--order accuracy of the numerical method
\cite{GP}.

\end{document}